\documentclass[aps,prd,preprint,eqsecnum,nofootinbib,groupedaddress,showpacs,floatfix]{revtex4}
\usepackage[dvips]{graphicx}
\font\ttiny=cmr6 at 3.5pt

\newcommand{\BlackHat}{{\sc BlackHat}}
\newcommand{\SHERPA}{{\sc SHERPA}}
\newcommand{\AMEGIC}{{\sc AMEGIC++\/}}
\newcommand{\SISCone}{{\sc SISCone}}
\newcommand{\JETCLU}{{\sc JETCLU}}
\newcommand{\MCFM}{{\sc MCFM}}

\newcommand{\pt}{$p_T$}
\newcommand{\ptm}{p_T}
\newcommand{\ptjet}{$p_T^\jet$}
\newcommand{\ptjetmin}{p^\jet_{T\rm min}}
\newcommand{\ptv}{p_T^V}
\newcommand{\ptmax}{p_T^{\rm max}}
\newif\ifdraft
\drafttrue
\newif\ifpreprint
\preprinttrue

\def\fig#1{fig.~{\ref{#1}}}
\def\Fig#1{Fig.~{\ref{#1}}}
\def\figs#1#2{figs.~{\ref{#1}} and {\ref{#2}}}
\def\Figs#1#2{Figs.~{\ref{#1}} and {\ref{#2}}}
\def\sect#1{section~{\ref{#1}}}
\def\Sect#1{Section~{\ref{#1}}}

\def\eqn#1{eq.~(\ref{#1})}

\def\tab#1{table~{\ref{#1}}}

\def\tabs#1#2{tables~\ref{#1} and~\ref{#2}}
\def\Tabs#1#2{Tables~\ref{#1} and~\ref{#2}}

\def\qb{\bar q}
\def\pb{\bar p}

\def\Qb{\bar Q}
\def\nub{\bar \nu}
\def\e{\epsilon}

\def\HTpartonic{{\hat H}_T}

\def\Wjjj{$W\,\!+\,3$}

\def\Wjjja{$W\,\!+\,1,2,3$}

\def\Wjn{$W\,\!+\,n$}

\def\Zj{$Z\,\!+\,1$}

\def\Zjjj{$Z\,\!+\,3$}
\def\Zgamz{$Z,\gamma^*\, +\, 0$}
\def\Zgam{$Z,\gamma^*$}
\def\Zgamj{$Z,\gamma^*\,\!+\,1$}
\def\Zgamjj{$Z,\gamma^*\,\!+\,2$}
\def\Zgamjjj{$Z,\gamma^*\,\!+\,3$}
\def\Zgamjjjj{$Z,\gamma^*\,\!+\,4$}

\def\Znj{$Z\,\!+\,n$}

\def\Zgamjx{$Z,\gamma^*\,\!+\,1,2$}
\def\Zgamjz{$Z,\gamma^*\,\!+\,0,1,2$}
\def\Zgamjjjz{$Z,\gamma^*\,\!+\,0,1,2,3$}

\def\Zjjja{$Z\,\!+\,1,2,3$}
\def\Zgamjjja{$Z,\gamma^*\,\!+\,1,2,3$}
\def\Zgamjjjja{$Z,\gamma^*\,\!+\,1,2,3,4$}

\def\Zjn{$Z\,\!+\,n$}
\def\Zjnp1{$Z\,\!+\,(n+1)$}

\def\Zgamjn{$Z,\gamma^*\,\!+\,n$}
\def\Zgamjnp1{$Z,\gamma^*\,\!+\,(n+1)$}
\def\Zjnm1{$Z\,\!+\,(n-1)$}
\def\Zgamjnm1{$Z,\gamma^*\,\!+\,(n-1)$}
\def\Vj{$V\,\!+\,1$}
\def\Vjj{$V\,\!+\,2$}
\def\Vjjj{$V\,\!+\,3$}

\def\Vjn{$V\,\!+\,n$}

\def\jets{\,{\rm jets}\,}
\def\jet{{\rm jet}}
\def\eps{\epsilon}
\def\nn{\nonumber}
\def\MLO{d\sigma^{(0)}}
\def\MNLO{d\sigma_V^{(1)}}
\def\hatMNLO{\widehat{d\sigma}_V^{(1)}}
\def\HTpartonic{{\hat H}_T}
\def\HTp{{\hat H}_T}
 
\def\adipole{\alpha_{\rm dipole}} 

\def\Ord{{\cal O}}

\newbox\charbox
\newbox\slabox
\def\s#1{{      % Feynman slash
        \setbox\charbox=\hbox{$#1$}
        \setbox\slabox=\hbox{$/$}
        \dimen\charbox=\ht\slabox
        \advance\dimen\charbox by -\dp\slabox
        \advance\dimen\charbox by -\ht\charbox
        \advance\dimen\charbox by \dp\charbox
        \divide\dimen\charbox by 2
        \raise-\dimen\charbox\hbox to \wd\charbox{\hss/\hss}
        \llap{$#1$}
}}

\def\nuornub{{}^{\raise1.3pt\hbox{\ttiny(}}\hskip -0.2pt\overline{\kern 
-0.5pt \nu \kern -0.4pt}\hskip 0.3pt{}^{\raise1.3pt\hbox{\ttiny)}}}

\begin{document}

\hbox{
SLAC--PUB--13954$\null\hskip 1.125cm \null$
UCLA/10/TEP/101$\null\hskip 1.125cm \null$
MIT-CTP 4117$\null\hskip 1.125cm \null$
SB/F/379-10}

\hbox{
IPPP/10/13$\null\hskip 0.55cm \null$
Saclay IPhT--T10/040$\hskip 0.55cm \null$
NIKHEF-2010-004$\hskip 0.55cm \null$
CERN-PH-TH/2010-029}

\title{
Next-to-Leading Order QCD Predictions for $Z,\gamma^* +3$-Jet
Distributions at the Tevatron}

\author{C.~F.~Berger${}^{a}$,\, Z.~Bern${}^b$,\,
L.~J.~Dixon${}^c$,\, F.~Febres Cordero${}^d$,\, D.~Forde${}^{e,f}$,\,
T.~Gleisberg${}^c$,\,  H. Ita${}^b$,\,
D.~A.~Kosower${}^{g}$\,
and D.~Ma\^{\i}tre${}^{h}$ 
\\
$\null$
\\
${}^a$Center for Theoretical
Physics, MIT,
      Cambridge, MA 02139, USA \\
${}^b$Department of Physics and Astronomy, UCLA, Los Angeles, CA
90095-1547, USA \\
${}^c$SLAC National Accelerator Laboratory, Stanford University,
             Stanford, CA 94309, USA \\
${}^d$Universidad Sim\'on Bol\'{\i}var, Departamento de
F\'{\i}sica, Caracas 1080A, Venezuela\\
${}^e$Theory Division, Physics Department, CERN, CH--1211 Geneva 23, 
    Switzerland\\
${}^f$NIKHEF Theory Group, Science Park 105, NL--1098~XG
  Amsterdam, The Netherlands\\
${}^g$Institut de Physique Th\'eorique, CEA--Saclay,
          F--91191 Gif-sur-Yvette cedex, France\\
${}^h$Department of Physics, University of Durham,
          DH1 3LE, UK\\
}

\begin{abstract}
Using \BlackHat{} in conjunction with \SHERPA{}, we have computed
next-to-leading order QCD predictions for a variety of distributions
in \Zgamjjja-jet production at the Tevatron, where the $Z$ boson or
off-shell photon decays into an electron-positron pair.  We find good
agreement between the NLO results for jet \pt{} distributions and
measurements by CDF and D0.  We also present jet-production ratios, 
or probabilities of finding one additional jet.  As a function of
vector-boson \pt{}, the ratios have distinctive features which we describe
in terms of a simple model capturing leading logarithms and phase-space
and parton-distribution-function suppression.
\end{abstract}

\pacs{12.38.-t, 12.38.Bx, 13.87.-a, 14.70.-e \hspace{1cm}}

\maketitle

\section{Introduction}

The Large Hadron Collider (LHC) recently passed the milestone of first
collisions.  The start of the LHC era in particle physics opens new
opportunities to confront data with theoretical predictions for
Standard-Model scattering processes, at scales well beyond those
probed in previous colliders.  This confrontation
will be a key tool in the search for new physics beyond the Standard
Model.  Where new physics produces sharp peaks, Standard Model
backgrounds can be understood without much theoretical input.  For
many searches, however, the signals do not stand out so clearly, but
are excesses in broader distributions of jets accompanying missing
energy and charged leptons or photons.  Such searches require a much
finer theoretical understanding of the QCD backgrounds.

An important class of backgrounds is the production of multiple 
jets in association with a $Z$ boson.  If the $Z$ boson decays into
neutrinos, this process forms an irreducible background to LHC searches for
new physics, such as supersymmetry, that are based on missing
transverse energy and jets, as discussed in, {\it e.g.} ref.~\cite{MET}.
These processes can be calibrated experimentally
using events in which the $Z$ boson decays into a pair of charged
leptons, either an electron--positron pair or a di-muon pair.  The
latter samples are quite clean, and the QCD dynamics is of course
identical to the $Z\to\nu\nub$ mode.  The one experimental drawback is
the small branching ratio for $Z\to l^+ l^-$.  Nevertheless, there are
already results from the Tevatron on $Z$ production in association
with up to three jets~\cite{ZCDF,Z1jD0,ZD0,ZjanglesD0}, along with the
prospect of new analyses using larger data sets in the near
future.  Therefore in this paper we focus on the production of
\Zjjja{} jets at the Tevatron, which can also serve as a benchmark
for future LHC studies.

The first step toward theoretical control of QCD backgrounds at hadron
colliders is the evaluation of the cross section at leading order (LO)
in the strong coupling, $\alpha_S$. Several computer
codes~\cite{LOPrograms,HELAC,Amegic} are available for LO predictions.
These codes typically use matching (or merging) 
procedures~\cite{Matching,MLMSMPR} to incorporate higher-multiplicity
leading-order matrix elements into programs that shower and hadronize
partons~\cite{PYTHIA,HERWIG,Sherpa}.  Although such programs provide a
hadron-level description which has great utility, the LO approximation
suffers from large factorization- and renormalization-scale dependence,
which grows with increasing jet multiplicity.  This dependence is already up
to a factor of two in the processes we shall study; accordingly, LO
results do not generally provide a quantitatively reliable prediction.
The problems go beyond that of normalization of cross sections:
shapes of distributions may or may not be modeled correctly at
(matched) LO, and the results at this order can depend strongly
on the functional form and value chosen for the scale.
In order to resolve these problems, and provide
quantitatively reliable predictions, one must evaluate the
next-to-leading order (NLO) corrections to processes of interest.
Such computations are technically more challenging, but generically
yield results with a greatly reduced scale
dependence~\cite{LesHouches2007,LesHouches2009},
as well as displaying better agreement with 
measurements (see {\it e.g.} refs.~\cite{WCDF,ZCDF,PRLW3BH,ZD0}).

More generally, sufficiently accurate QCD predictions can provide
important theoretical input into experimentally-driven determinations
of backgrounds, by allowing measurements in one process to be
converted into a prediction for another, where theory is used only for
the ratio of the two processes.  For example, complete NLO
predictions for \Wjjj-jet production, followed by $W\to l\nu$, 
are already available at parton-level~\cite{W3jDistributions}, and the present
paper describes analogous predictions for \Zjjj-jet production,
followed by $Z\to l^+l^-$.
These results allow the ratio of $Z$ to $W$ production in
association with up to 
three jets to be computed at NLO.  Parton-level results do neglect
non-perturbative effects, such as hadronization and the underlying
event, which can contribute to both $Z$ and $W$ production.  However, 
as long as the experimental cuts on the jets are the same, 
the non-perturbative corrections should largely cancel in the
$Z$ to $W$ ratio.  Therefore, measurements of the (more copious) 
production of $W$ bosons in the presence of multiple jets can be 
extrapolated to the case of $Z$ bosons~\cite{CMSWZRatioNote}, 
using precise theoretical values for the ratio of $Z$ to $W$ events
with similar kinematics, and as a function of the kinematics.
Because leptonically decaying $Z$ bosons,
although rarer, are cleaner experimentally than $W$ bosons, it has also
been suggested to reverse the procedure and use $Z(\to l^+l^-)$ boson
samples to calibrate $W(\to l\nu)$ and $Z(\to\nu\nub)$
samples~\cite{ATLASZWRatioNote}.
However, this procedure requires more input from theory:
in order to have adequate statistics in the $Z\to l^+l^-$ channel,
less energetic kinematical configurations with larger cross sections
are measured, and then extrapolated to more energetic configurations
with smaller cross sections.

In the past, the bottleneck in computing NLO QCD corrections to processes
with large numbers of jets has been the evaluation of one-loop
amplitudes involving six or more partons~\cite{LesHouches2007}.  On-shell
methods~\cite{UnitarityMethod,Zqqgg,BCFUnitarity,Bootstrap,Genhel,%
OPP,Forde,EGK,GKM,OnShellReviews,LesHouches2009}
have successfully resolved
this bottleneck, by avoiding gauge-noninvariant intermediate steps
and reducing the problem to much smaller elements analogous to
tree-level amplitudes.  In this paper we evaluate the required
one-loop amplitudes with the \BlackHat{} program 
library~\cite{BlackHatI,ICHEPBH,PRLW3BH,W3jDistributions},
which implements on-shell methods numerically.
A number of programs based on on-shell techniques have been
constructed by other
groups~\cite{OtherOnShellPrograms,HPP,EGKMZ,EMZW3j,Czakon}.
Approaches based on Feynman diagrams have also led to new results
with six external partons, in particular the 
NLO cross section for producing four heavy quarks at hadron
colliders~\cite{BDDP,bbbb}.  The $t\bar{t}b\bar{b}$ case has also been 
computed via on-shell methods~\cite{Czakon}, and recently
the final state $t\bar{t}\,+\,$2~jets has been computed at NLO
in a similar way~\cite{ttjj}.  We
expect that on-shell methods will be especially advantageous for
processes involving many external gluons, which often dominate
multi-jet final states. 

NLO parton-level cross sections for the production of a $W$ or $Z$ boson in
association with one or two jets have long been available in the
\MCFM~\cite{MCFM} code, which utilizes the one-loop
amplitudes from ref.~\cite{Zqqgg} for the two-jet case.
More recently, complete NLO results have been obtained for
\Wjjj-jet production~\cite{W3jDistributions} using
\BlackHat{} in conjunction with the \SHERPA{} package~\cite{Sherpa}.
(Different leading-color approximations and ``adjustment procedures'' 
have also been applied to \Wjjj{} jets at NLO~\cite{PRLW3BH,EMZW3j}.)
In this work, we use the same basic calculational setup as in
ref.~\cite{W3jDistributions} to compute \Zjjj-jet production.
The real emission, dipole subtraction~\cite{CS} and integration over phase
space is handled by \AMEGIC{}~\cite{Amegic,AutomatedAmegic}, which is
part of the \SHERPA{} package~\cite{Sherpa}.  (Other automated
implementations of infrared subtraction methods~\cite{FKS,CS} have been
described elsewhere~\cite{AutomatedSubtractionOther}.)
\SHERPA{} is also used to perform the Monte Carlo integration over
phase space for all contributions.  One important improvement 
in the present study with respect to
ref.~\cite{W3jDistributions} is to increase the efficiency of the
phase-space integrator, making use of QCD antenna structures
along the lines of refs.~\cite{AntennaIntegrator,GleisbergIntegrator}.

In this article we present results for \Zgamjjj-jet production at the
Tevatron to NLO in QCD, at parton level, and with the vector boson decaying
to a lepton--anti-lepton pair.  We include off-shell photon exchange,
and $\gamma^*-Z$ interference, because the production of a charged-lepton 
pair by an off-shell photon is indistinguishable from the 
leptonic decay of a $Z$ boson.
Preliminary versions of some of the results presented here may be found in
ref.~\cite{Radcor09BH}, where slightly different jet cuts were
applied, along with a leading-color approximation.  

Here we present total production cross sections, with jet and
lepton cuts appropriate to existing CDF and D0
analyses~\cite{ZCDF,ZD0}, as well as a variety of distributions.  We
also study how \Zgamjjj-jet production at the Tevatron depends on a
common choice of renormalization and factorization scale $\mu$.  As
mentioned above, LO results are generically rather sensitive to the
scale choice.  This sensitivity usually is greatly reduced at NLO.
As an example, in \Zgamjjj-jet production, varying the scale by a
factor of two from our default central value causes nearly a 60\%
deviation from the central value. In contrast, 
at NLO this deviation drops to only
15--22\%.

Total cross sections with standard experimental jet cuts
are dominated by the production of jets with low transverse momentum,
$\ptm < M_V$, where the vector boson $V$ is a $W$ or $Z$.
Accordingly, scale choices such as the mass of the 
vector boson $\mu = M_V$ are reasonable for these quantities.
For the study of differential distributions, however, it is
better to choose a dynamical scale, event by event, in order to have
reasonable scales for each bin~\cite{DynamicalScaleChoice}.
This helps reduce the change in predicted shapes from LO to NLO,
and can improve the NLO prediction some too.
However, care is required in choosing the functional form of such scales.  
Greater care is required at the LHC than at the Tevatron, 
because the much larger dynamical range can ruin seemingly-reasonable
scale choices, such as the commonly-used vector-boson 
transverse energy, $\mu = E_T^V \equiv \sqrt{M_V^2+(p_T^V)^2}$
(see {\it e.g.} refs.~\cite{EarlyWplus2MP,ZCDF,WCDF,PRLW3BH,ZD0}).  
As noted in ref.~\cite{W3jDistributions}, a scale
choice of $\mu = E_T^V$ leads to negative cross
sections in tails of some NLO distributions, because typical energy
scales in the process are much larger than $E_T^V$.
In addition, as noted independently~\cite{Bauer}, the choice
$\mu = E_T^V$ leads to
undesirably large shape changes between LO and NLO.  As we discuss
here, for \Zjjj-jet production at the Tevatron, the scale $\mu = E_T^V$
is unsatisfactory, at least at LO: 
choosing it results in large shape changes in
distributions between LO and NLO.  The total partonic transverse
energy $\HTp$ (or a fixed fraction thereof), adopted in our previous
study~\cite{W3jDistributions}, is a satisfactory choice.  Other
choices, such as the combined invariant mass of the jets~\cite{Bauer},
should also be satisfactory.  In any case, the lesson is clear: the
vector-boson transverse energy is not satisfactory and should not be
used, especially for processes at the LHC.

At the Tevatron, both CDF and D0 have measured~\cite{ZCDF,ZD0}
jet-\pt{} distributions for \Zgamjx-jet production in the channel
$Z\to e^+e^-$.  The D0 measurements are not absolute cross sections,
but are normalized to the inclusive \Zgamz-jet cross section for the same
set of lepton cuts.  CDF and D0 have each
compared their data with NLO predictions from \MCFM~\cite{MCFM},
taking into account estimates of non-perturbative corrections.
For comparison we present our own NLO analysis of these processes.
We then turn to the more involved case of \Zgamjjj-jet production,
which we compare against the D0 experimental measurement.
A difference between our NLO study and the experimental
measurements is in our use of infrared-safe jet
algorithms (as reviewed in ref.~\cite{Jetography}).
Our default choice here is the
\SISCone{} jet algorithm~\cite{SISCONE}, although we present some
results using the anti-$k_T$ algorithm~\cite{antikT} as well.  (The
$k_T$ algorithm~\cite{KTAlgorithm,KTES} gives parton-level
results very similar to those for
the anti-$k_T$ one, so we do not present them.)
The algorithms used in the CDF and D0 analyses, which are in 
the ``midpoint'' class of iterative cone
algorithms~\cite{Midpoint,D0jetAlgorithm,CDFMidpoint},
are generically infrared unsafe, and cannot be used in an NLO
computation of \Vjjj-jet production.
Although a midpoint algorithm will
yield finite results at NLO for \Vjj-jet production, 
it suffers from uncontrolled non-perturbative corrections that 
are in principle of the same order as the NLO correction~\cite{Jetography}.

In comparing theory and experiment, the differing jet algorithms do
introduce an additional source of uncertainty.  Nevertheless, based
on our study of \Zgamjx-jet production we expect the NLO results to
match experiment reasonably well.  There have also been two
studies comparing inclusive jet cross sections for midpoint algorithms
with those for \SISCone~\cite{SISCONE,EHKetal}, using
{\sc Pythia}~\cite{PYTHIA}, which find that
the major non-perturbative differences between the algorithms are at the
level of the underlying event.  Although the kinematics of inclusive-jet
production differs somewhat from that of a vector boson plus multiple jets,
these results suggest that hadron-level data collected using \SISCone{}
would differ from that for midpoint algorithms primarily by the underlying
event correction (at least for larger cone sizes).  That is,
if the two measurements were corrected back to parton level, one
would expect them to have only percent-level differences.
(At the LHC, both ATLAS and CMS have adopted infrared-safe jet algorithms,
removing this important source of uncertainty.)

We shall present \Zgamjjja-jet production cross sections for the CDF
and D0 cuts, as well as jet \pt{} distributions.  For \Zgamjjj-jet
production, we show the \pt{} distributions for all three jets,
ordered in \pt.  In addition, we discuss ratios of cross sections.
Experimental and theoretical systematic uncertainties cancel to some degree
in such ratios.  Ratios of similar processes --- for example, the
ratio of \Wjjj-jet to \Zjjj-jet production --- are thus attractive
candidates for confronting experimental data with theoretical
predictions.  We shall not study this kind of ratio in the present
paper, but another kind, that of a production process to the same
process with one additional jet, sometimes known as the 
``jet-production ratio''~\cite{BerendsRatio,AbouzaidFrisch}.  (This ratio is
also called the ``Berends'' or ``staircase'' ratio.)  In particular,
we study the ratio of \Zgamjn-jet to \Zgamjnm1-jet production up to
$n=4$ at LO and $n=3$ at NLO. Its evaluation for different values of
$n$ can also test the lore, based on Tevatron studies, that the ratio
is roughly independent of $n$.  We find that for total cross sections
through $n=3$, with the experimental cuts used by CDF, this scaling is
valid to about 30\%, and for the D0 cuts, it is valid to about 15\%.

One may also ask whether not just total cross sections, but also
differential distributions, can be predicted for \Zgamjn-jet
production (at least approximately) by scaling results for
\Zgamjnm1-jet production.  We show explicitly for $n\le 4$, using the
example of the vector-boson \pt{} distributions, that shapes of
distributions {\it cannot} be reliably predicted by assuming a
constant factor between the $(n-1)$-jet and $n$-jet cases.  We
describe the nontrivial structure found in the jet-production ratios
for the vector-boson \pt{} distributions using a simple model that
incorporates leading-logarithmic behavior and suppression due to
phase-space and parton-distribution-function effects.  Related to 
the shape differences is a strong dependence of the jet-production
ratios on the experimental cuts.  In particular, there is about a 50\%
difference in the jet-production ratios for the CDF and D0 setups,
as shown in \tabs{CDFRatioTable}{D0RatioTable} of
\sect{JetProductionRatioSection}.

This paper is organized as follows.  In \sect{SetUpSection} we briefly
summarize our calculational setup, focusing on the differences from
our previous work on \Wjjj-jet
production~\cite{PRLW3BH,W3jDistributions}.
\Sect{CutsandScalesSection} records our choice of
couplings, renormalization and factorization scales, and cuts
matching those of the CDF and D0 measurements.  We also discuss issues
associated with the choice of scale. In \sect{CDFResultsSection}, we
present results for cross sections and for a variety of distributions,
matching CDF's cuts, and we compare with their measurements.  In
\sect{D0ResultsSection} we give distributions in the softest jet \pt{}
for \Zgamjjja-jet production, using D0's cuts and comparing with their
published data.  In \sect{JetProductionRatioSection} we present jet
production ratios for both the total cross section and the
vector-boson \pt{} distribution, and discuss a simple model for the
latter ratio.  We present our conclusions in \sect{ConclusionSection}.
We include one appendix defining observables, as well as one giving
values of the matrix elements at a single point in phase space.  The
latter appendix should aid future implementations of the \Zgamjjj-jet
one-loop matrix elements.

%%%%%%%%%%%%%%%%%%%%%%%%%%%%%%%%%%%%%%%%%%%%%%%%%%%%%%%%%%%%%%%%%%%%
\section{Calculational Setup}
\label{SetUpSection}

\subsection{Processes}

In this paper we calculate the inclusive processes,
\begin{equation}
 p\pb\ \ \rightarrow\ \ 
             Z,\gamma^* + n\, \jets\,+\,X 
\ \ \rightarrow\ \ e^{+} e^{-} + n\, \jets\,+\,X,
\label{BasicProcess}
\end{equation}
at $\sqrt{s}=1.96$~TeV to NLO accuracy, for $n=1,2,3$.  Both CDF and
D0 have measured production cross sections for all three of these
processes~\cite{ZCDF,ZD0} at the Tevatron, based on integrated
luminosities of 1.7~fb$^{-1}$ and 1.0~fb$^{-1}$, respectively. 
The D0 measurements, besides using slightly different cuts, use a
significantly smaller jet cone size, $R=0.5$ versus $R=0.7$ for CDF.
In addition, differential distributions have been provided for
\Zgamjx-jet production.  The set of available distributions is
particularly extensive in the case of one 
jet~\cite{Z1jD0,ZjanglesD0}.  The \Zgamjx-jet production measurements
have been compared to NLO predictions from \MCFM~\cite{MCFM}.  For the
case of three additional jets, the data sets analyzed are still small,
so that CDF has measured only a total cross section, while D0 has
provided three bins in the distribution of the transverse momentum of
the third jet ($p_T$ ordered).  The latter distribution was also
compared with a leading-order prediction computed using \MCFM.

The present article provides the first NLO predictions for
\Zgamjjj-jet production, allowing a comparison with both the CDF cross
section and the D0 $p_T$ distribution. We will provide a few other
distributions as well.  We hope that, as additional data is analyzed,
such distributions will be measured by CDF and D0.  A comparison between
NLO results and Tevatron data for various $W,Z + 3$-jet distributions
would provide a very important benchmark for future LHC studies of
these complex final states. 

%%%%%%%%%%%%% FIGURE %%%%%%%%%%%%%%%%%%
\begin{figure}[tbh]
\includegraphics[clip,scale=0.637]{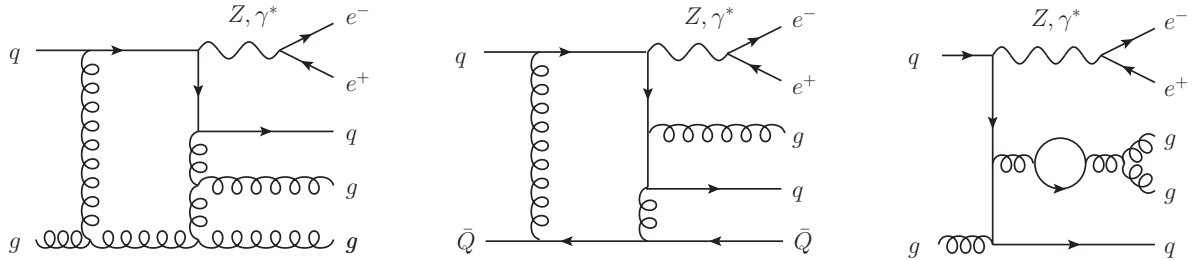}
\caption{A few representative diagrams contributing
to the $q g \rightarrow e^+ e^- \, q
g g$ and $q \bar Q \rightarrow e^+ e^- \, q g \bar Q$ one-loop
amplitudes.  The $e^+ e^-$ pair couples to the quarks via either a $Z$
boson or an off-shell photon.}
\label{LoopDiagramsLCFigure}
\end{figure}
%%%%%%%%%%%%%%%%%%%%%%%%%%%%%

%%%%%%%%%%%%% FIGURE %%%%%%%%%%%%%%%%%%
\begin{figure}[tbh]
\includegraphics[clip,scale=0.59]{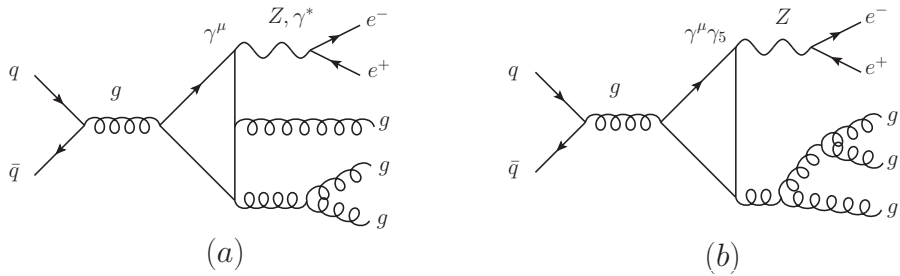}
\caption{Sample diagrams illustrating one-loop contributions 
to \Zgamjjj-jet production
where the vector boson couples directly to a quark loop, via either
(a) a vector coupling, or (b) an axial vector coupling.
These contributions are quite small for the corresponding
process with one parton less,
and therefore are not included in our calculation.}
\label{LoopDiagramsVAFigure}
\end{figure}
%%%%%%%%%%%%%%%%%%%%%%%%%%%%%

In more detail, the process under consideration (\ref{BasicProcess})
receives contributions from several partonic subprocesses.
At leading order, and in the virtual (one-loop) NLO contributions,
these subprocesses are all obtained from
\begin{eqnarray}
 && q \qb Q \Qb g\rightarrow Z, \gamma^* \rightarrow e^+\, e^- \,,
\label{Z3qqQQg} \\
 && q \qb g g g\rightarrow Z, \gamma^* \rightarrow e^+\, e^- \,,
\label{Z3qqggg}
\end{eqnarray}
by crossing three of the partons into the final state. 
The quarks are represented by $q$ and $Q$ and the gluons
by $g$.  The $Z$ or
photon couples to the quark line labeled $q$.
  Representative diagrams
for the virtual contributions are shown in
\fig{LoopDiagramsLCFigure}.  We include the
decay of the vector boson ($Z,\gamma^*$) into a lepton pair at the
amplitude level.  The photon is always off shell, and the $Z$ boson
can be as well.
For the $Z$ the lepton-pair invariant mass, $M_{e e}$, 
follows a relativistic
Breit-Wigner distribution whose width is determined by the $Z$ decay
rate $\Gamma_Z$.  We take the lepton decay products to be
massless.  Amplitudes containing identical quarks are generated by
antisymmetrizing in the exchange of appropriate $q$ and $Q$ labels.

The light quarks, $u,d,c,s,b$, are all treated as massless.  We do not
include contributions to the amplitudes from a real or virtual top
quark.  Nor do we include the pieces
in which the vector boson couples directly to a quark loop through
either a vector or axial coupling, as
illustrated in \fig{LoopDiagramsVAFigure}.  In \Zgamjj-jet production
these pieces affect the cross section by under 0.3\%.  We
therefore expect the omission of these pieces to have a small effect
on the \Zgamjjj-jet results presented here, well below the residual NLO
uncertainties of 10--20\%.

%%%%%%%%%%%%% FIGURE %%%%%%%%%%%%%%%%%%
\begin{figure}[tbh]
\includegraphics[clip,scale=0.495]{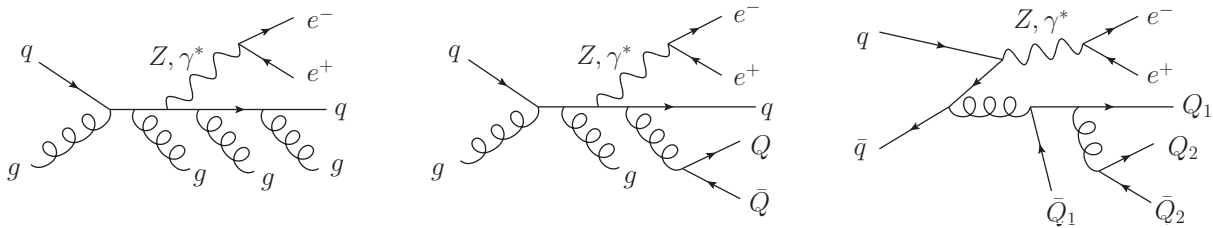}
\caption{Representative real-emission 
diagrams for the eight-point tree-level amplitudes,
$q g \rightarrow e^+ e^- q  g g g$,
$q g \rightarrow e^+ e^- q  g Q \bar Q$,
and $q \bar{q} \rightarrow e^+ e^- \, Q_1 \bar{Q}_1 Q_2 \bar{Q}_2$.}
\label{TreeDiagramsFigure}
\end{figure}
%%%%%%%%%%%%%%%%%%%%%%%%%%%%%

Besides the loop amplitudes, we need tree amplitudes for real 
emission contributions.  The relevant subprocesses are
\begin{eqnarray}
&& q \qb g g g g \rightarrow Z, \gamma^* \rightarrow e^+\, e^- \,, \\
&& q \qb Q \Qb g g \rightarrow Z, \gamma^* \rightarrow e^+\, e^- \,, \\
&& q \qb  Q_1 \bar{Q}_1 Q_2 \bar{Q}_2  \rightarrow Z, \gamma^*
      \rightarrow  e^+ e^- \,,
\end{eqnarray}
where all the physical processes are obtained by crossing four of the
partons into the final state.  Representative tree diagrams for
these contributions are given in \fig{TreeDiagramsFigure}.

To compute the NLO corrections we use \BlackHat{} and \SHERPA,
essentially following the same calculational setup described in
ref.~\cite{W3jDistributions} for the \Wjjj-jet process.  We
therefore discuss our setup only briefly, pointing out the 
few differences with ref.~\cite{W3jDistributions}.

\subsection{Setup}

The virtual contributions are evaluated with \BlackHat, which is based
on the unitarity method~\cite{UnitarityMethod}.  One-loop amplitudes
are expanded in terms of a basis set of scalar integrals composed of
box, triangle and bubble integrals, plus a rational remainder.  The
coefficients of box integrals are obtained from quadruple cuts by
solving the cut conditions~\cite{BCFUnitarity}.  Coefficients of
bubble and triangle functions are then obtained using a numerical
implementation of Forde's approach~\cite{Forde}.  In this
implementation, \BlackHat{} uses a procedure related to that of
Ossola, Papadopoulos and Pittau~\cite{OPP} to subtract box
contributions when determining triangle coefficients, and to subtract
box and triangle contributions when determining bubble
coefficients. The basis scalar integrals are evaluated numerically
using their known analytic expressions~\cite{IntegralsExplicit}. To
obtain rational terms, we have implemented both loop-level on-shell
recursion~\cite{Bootstrap,Genhel} and a numerical version of the
``massive continuation'' approach due to Badger~\cite{Badger}, which
is related to the $D$-dimensional generalized
unitarity~\cite{DdimUnitarity} approach of Giele, Kunszt and
Melnikov~\cite{GKM}.  The numerical version involves subtracting the
contributions of higher-point cuts rather than taking large-mass
limits.  It is similar to the numerical version of Forde's method~\cite{Forde}
for four-dimensional unitarity cuts, which is described in
ref.~\cite{W3jDistributions}.  In that paper we used on-shell recursion
for the leading-color terms, where speed is at a premium. For the
simplest helicity configurations, on-shell recursion is implemented
analytically and the results stored for numerical evaluation.  For
subleading color terms the massive continuation method was used
because it is presently more flexible.  For production runs in the
current study, we used the analytic formul\ae{} obtained via on-shell
recursion for the leading-color amplitudes, and the massive
continuation method for the remaining terms.

As discussed in ref.~\cite{W3jDistributions}, for efficiency purposes
it is useful to compute the leading-color parts of the virtual
contributions separately from the numerically much smaller, but
computationally more complicated, subleading-color contributions.  We
follow the same division of leading and subleading color as in
ref.~\cite{W3jDistributions}, except that here we assign the pieces
proportional to the number of quark flavors ($n_f$) to the
leading-color contributions instead of the subleading-color ones.
This has the effect of somewhat reducing the size of the (already very
small) subleading-color contributions, helping to reduce the number of
phase-space points at which they must be evaluated.  We add the
leading- and subleading-color contributions at the end of the
calculation to obtain the complete color-summed result. We refer the
reader to refs.~\cite{qqggg,Zqqgg} for detailed descriptions
of the primitive amplitude decomposition that we used.
Alternative organizations of color, within the context of the
unitarity method, may be found in refs.~\cite{HPP,ZoltanColor}.

%%%%%%%%%%%%% FIGURE %%%%%%%%%%%%%%%%%%
\begin{figure}[tbh]
\includegraphics[clip,scale=0.75]{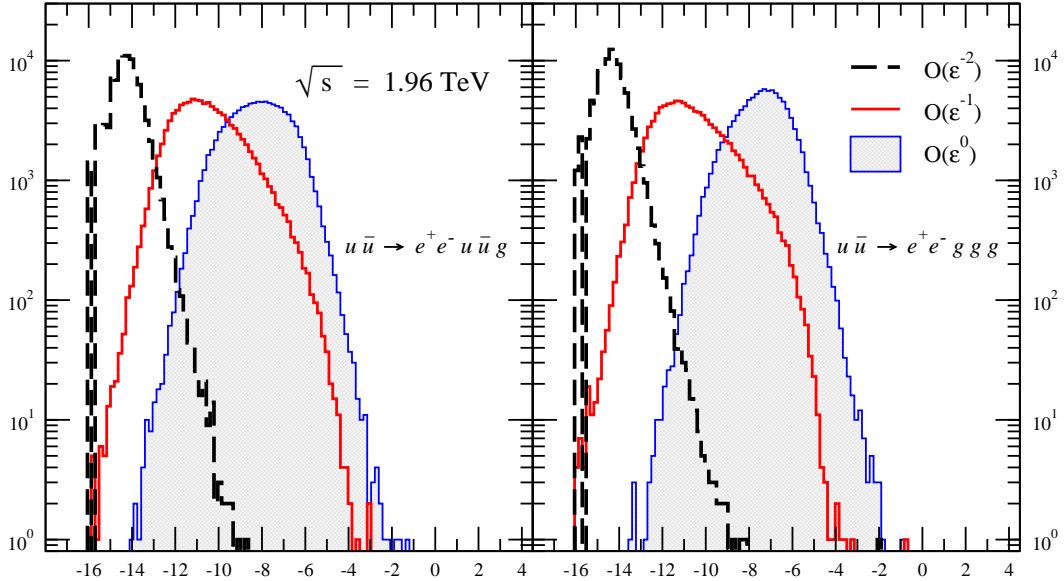}
\caption{The distribution of the relative error in the 
virtual cross section for the two subprocesses $u \bar u \rightarrow 
e^+ e^- u \bar u g$ and $u \bar u \rightarrow e^+ e^- g g g $.  
The horizontal axis is the logarithm of
the relative error~(\ref{RelError}) between an
evaluation by \BlackHat{}, running in production mode,
and a target expression evaluated using higher precision
with at least $32$ decimal digits (or up to $64$ decimal digits for
unstable points).  The vertical axis shows the number of phase-space
points out of 100,000 that have the corresponding error.
The dashed (black) line shows
the $1/\eps^2$ term; the solid (red) curve, the $1/\eps$ term; and the
shaded (blue) curve, the finite ($\eps^0$) term.
}
\label{StabilityFigure}
\end{figure}
%%%%%%%%%%%%%%%%%%%%%%%%%%%%%

An important issue is the numerical stability of the loop amplitudes.
In \fig{StabilityFigure}, we illustrate the stability of the
full-color virtual interference term (or squared matrix element),
$d\sigma_V$, summed over colors and over all helicity configurations
for the two subprocesses $u \bar u \rightarrow e^+ e^- u \bar u g$ and
$u \bar u \rightarrow e^+ e^- g g g$.  The horizontal axis of
\fig{StabilityFigure} shows the logarithmic error,
\begin{equation}
\log_{10}\left(\frac{|d \sigma_V^{\rm BH}- d \sigma_V^{\rm target}|}
          {| d \sigma_V^{\rm target}|} \right),
\label{RelError}
\end{equation}
for each of the three components: $1/\epsilon^2$, $1/\epsilon$ and
$\epsilon^0$, where $\eps = (4-D)/2$ is the dimensional regularization
parameter. In this expression $\sigma_V^{\rm BH}$ is the
cross section computed by \BlackHat{} as it normally operates for
production runs (switching from $16$ decimal digits 
to higher precision only when instabilities
are detected), whereas
$\sigma_V^{\rm target}$ is a target value computed by \BlackHat{}
using multiprecision arithmetic with at least $32$ digits, and
$64$ digits if the point is deemed unstable using the criteria described in 
refs.~\cite{BlackHatI,W3jDistributions}.
We use the {\sc QD} package~\cite{QD}
for higher-precision arithmetic. The phase-space points are
selected in the same way as those used to compute cross sections.  We
note that an overwhelming majority (above 99.9\%) of events are accurate
to better than one part in $10^3$ --- that is, to the left of the
`$-3$' mark on the horizontal axis.  Because we need only recompute parts
of amplitudes in most cases~\cite{W3jDistributions}, the extra time
spent in higher-precision operation is quite small, roughly 20\% more
than if only double precision had been used.

%\subsection{Real-emission contributions and infrared singularities}

In addition to the virtual corrections, the real-emission corrections
are also required.  These terms arise from
tree-level amplitudes with one additional parton: an additional gluon,
or a quark--antiquark pair replacing a gluon, as illustrated
in \fig{TreeDiagramsFigure}.  We use \SHERPA{} for these pieces.
The infrared singularities are canceled between 
real-emission and virtual contributions using the Catani--Seymour
dipole subtraction method~\cite{CS},
implemented~\cite{AutomatedAmegic} in the automated
program~\AMEGIC~\cite{Amegic}, which is part of the \SHERPA{}
framework~\cite{Sherpa}.  We follow the same setup described in
ref.~\cite{W3jDistributions}, taking $\adipole = 0.03$ as our default value.

The Monte Carlo integration over phase space of both the real-emission
and virtual pieces are carried out by \SHERPA{} using a
multichannel~\cite{MultiChannel} approach.  In this approach,
the integrand is not split up into pieces, but is sampled
differently in different channels.  For \Zgamjx-jet
production, we use \AMEGIC{}, and each channel generates a
momentum configuration based on the size of the denominators 
of the propagators of a tree-level Feynman diagram
(Born or real-emission, as appropriate).  For the more complicated case of 
\Zgamjjj-jet production, in order to improve the efficiency,
we have developed a specific phase-space generator, applicable
to $V\,+\,n$-jet production.
In this approach, a single channel generates
a momentum configuration for the partons that is based on the
size of denominators associated with a specific parton color
ordering (the color-ordered QCD antenna radiation pattern),
following the ideas of
refs.~\cite{AntennaIntegrator,GleisbergIntegrator}.
The lepton momenta are generated so that the invariant mass of the lepton
pair traces a Breit-Wigner distribution about the vector-boson mass.
For larger numbers of partons this
generator has a greatly reduced number of channels, compared to the number
of channels based on Feynman diagrams, so that it remains viable for
vector-boson production with up to five or six jets.

For the \Zgamjjj-jet process we integrate the
real-emission terms over about $4\times 10^7$ phase-space points, the
leading-color virtual parts over $7\times 10^5$ phase-space points and
the subleading-color virtual parts over $4 \times 10^4$ phase-space
points.  The LO and dipole-subtraction terms are run separately with
$10^7$ points each.  These numbers are chosen to achieve an
integration uncertainty of 0.5\% or less in the total cross section.

As a cross-check, we have compared our results for \Zgamjz-jet
production at NLO and \Zgamjjj-jet production at LO to those of
\MCFM{} and find agreement to
better than 1\%.  For \Zgamjj-jet production we used the same analytic 
one-loop matrix elements~\cite{Zqqgg} as used in \MCFM, with cross
checks against a purely numerical computation within \BlackHat{}.

%%%%%%%%%%%%%%%%%%%%%%%%%%%%%%%%%%%%%%%%%%%%%%%%%%%%%%%%%%%%%%%%%
\section{Couplings, Experimental Cuts and Scale Choices}
\label{CutsandScalesSection}

In this section we describe the basic parameters used in this work,
including couplings, experimental cuts and our 
choice of renormalization and factorization scales.
We also discuss the residual scale dependence remaining
in the NLO results.

\subsection{Couplings and parton distributions}
\label{CouplingsSection}

%%%%%%%%%%% TABLE %%%%%%%%%%%%%%%%%%%%%%%%%%
\begin{table}
\vskip .4 cm
\begin{tabular}{||c|c||}
\hline
parameter  & value  \\
\hline
$\alpha_{\rm QED}(M_Z)$ & $ 1/128.802 \;$ \\
\hline
$M_Z $  & $\;91.1876 $ GeV  $\;$  \\
\hline
$\sin^2\theta_W$ & $\; 0.230 \;$ \\
\hline
$\Gamma_Z$  & $2.49$ GeV  \\
%alpha_QED=1/128.802217
%$g^2_w$ &  0.4242 (calculated)  \\
\hline
\end{tabular}
\caption{Electroweak parameters used in this work.}
\label{ElectroWeakTable}
\end{table}
%%%%%%%%%%%%%%%%%%%%%%%%%%%%%

We express the $Z$-boson couplings to fermions using the
Standard Model input parameters shown in \tab{ElectroWeakTable}.
The parameter $g_w^2$ is derived from the others via,
\begin{equation}
g_w^2 = {4 \pi \alpha_{\rm QED}(M_Z) \over \sin^2\theta_W}\,.
\end{equation}
We use the CTEQ6M~\cite{CTEQ6M} parton distribution functions (PDFs)
at NLO and the CTEQ6L1 set at LO. The value of
the strong coupling constant is fixed to agree with the CTEQ choices,
so that $\alpha_S(M_Z)=0.118$ and $\alpha_S(M_Z)=0.130$
at NLO and LO respectively.  We evolve
$\alpha_S(\mu)$ using the QCD beta function for five massless quark
flavors for $\mu<m_t$, and six flavors for $\mu>m_t$.
(The CTEQ6 PDFs use a five-flavor scheme for all $\mu>m_b$, but we use the
SHERPA default of six-flavor running above the top-quark mass;
the effect on the cross section is very small, on the order of 1\%
at larger scales.)  At NLO we use two-loop running, and at LO,
one-loop running.

\subsection{Experimental cuts for CDF}
\label{CDFCutsSubsection}

To compare to CDF data we apply the same cuts as CDF~\cite{ZCDF},   
\begin{eqnarray}
&& p_T^\jet > 30~{\rm GeV}\,,\qquad|y^\jet| < 2.1\,,\nonumber\\
&& E_T^e > 25~{\rm GeV}\,,\qquad|\eta^{e_1}| < 1\,,\label{CDFCuts} \\
&&|\eta^{e_2}| < 1 \quad{\rm or}\quad 1.2 < |\eta^{e_2}| < 2.8\,,\nonumber\\
&& \Delta R_{e-\jet} > 0.7\,, 
\qquad  66 {\ \rm GeV} < M_{e e} < 116 {\ \rm GeV} \,. \nonumber
\end{eqnarray}
For any jet, \ptjet{} denotes the transverse momentum and $y^\jet$
the rapidity.  For the leptons, $E_T^e$ denotes the transverse energy
of either the electron or positron; $\eta^{e_1}$ refers to the 
pseudorapidity of either the electron or positron and $\eta^{e_2}$
refers to that of the other; $M_{e e}$ is the pair invariant mass.

In their study of $Z,\gamma^*$ production, CDF used a midpoint jet
algorithm~\cite{CDFMidpoint} with a cone size of $R=0.7$ and
a merging/splitting fraction of $f=0.75$.  We use instead three different
infrared-safe jet algorithms~\cite{SISCONE,antikT,KTAlgorithm}:
\SISCone{} ($f=0.75$), anti-$k_T$ and $k_T$, all with $R=0.7$.
\SISCone{} is our default choice for comparison to CDF.
(The $k_T$ algorithm gives very similar parton-level results as the
anti-$k_T$ algorithm, so we will not show those results explicitly.)

Our calculation is a parton-level one, and does not include corrections due
to non-perturbative effects, such as those induced by the underlying
event, induced for example by multiple parton interactions, 
or by fragmentation and hadronization of the outgoing partons.
In order to compare our parton-level results to data,
we require non-perturbative correction factors.  As discussed
further in \sect{CDFpTComparison}, for \Zgamjx-jet \pt{}
distributions, we adopt estimates of these correction factors made by
CDF~\cite{ZCDF}.

\subsection{Experimental cuts for D0}
\label{D0CutsSubsection}

To compare to D0 data we apply the jet cuts~\cite{ZD0},
\begin{eqnarray}
&& p_T^\jet > 20~{\rm GeV}\,, \qquad |\eta^\jet| < 2.5\,.
\label{D0jetcuts}
\end{eqnarray}
D0 defined jets using the D0 Run II midpoint jet
algorithm~\cite{D0jetAlgorithm}, with a cone size of $R=0.5$
and a merging/splitting fraction of $f=0.5$.  We use instead
the \SISCone{} algorithm, with $R=0.5$ and $f=0.5$.

D0 performed an analysis with two distinct sets of lepton cuts.
In their primary selection, 
which was compared directly to theory, only an invariant
mass cut was imposed on the electron--positron pair,
\begin{equation}
\hbox{(a): }\quad  65 {\ \rm GeV} < M_{e e} < 115 {\ \rm GeV} \,. \hskip 1 cm 
\label{D0LeptonAnalysisA}
\end{equation}
For the secondary selection, the lepton cuts were,
\begin{eqnarray}
\hbox{(b): }
           \begin{array}{ll}
        & 65 {\ \rm GeV} < M_{e e} < 115 {\ \rm GeV} \,, \\
        & E_T^e > 25~{\rm GeV}\,, \\
        & |\eta^e| < 1.1 \quad{\rm or}\quad 1.5 < |\eta^e| < 2.5 \,.
           \end{array}
\label{D0LeptonAnalysisB}
\end{eqnarray}
The latter ``(b)'' selection 
corresponds to the data D0
actually collected.  In their main selection (``(a)''), they
extrapolated to an ideal detector with full lepton coverage using
LO-matched parton-shower simulations.  This extrapolation
introduces an additional uncertainty and model dependence.
It more than doubles the absolute cross section, although
the quantities measured by D0, which are normalized by the inclusive
\Zgamz-jet cross section for the same lepton cuts, shift by much less.
(Comparing the (a) entry to the corresponding (b) entry in
\tab{Table-Zjets-D0-total-xs-HT} gives an estimate of the fraction of
cross section in selection (a) that comes from the extrapolation.)

We shall present NLO results corresponding to both selections, that is
with and without the lepton acceptance cuts in the secondary
selection~(\ref{D0LeptonAnalysisB}).  
Selection (b) allows us
to compare to unextrapolated data.  On the other hand,
D0 estimated the non-perturbative corrections, from hadronization
and the underlying event, for selection (a)~\cite{ZD0},
requiring us to extrapolate these corrections to selection (b) in
order to use them there, as we shall discuss further in
\sect{D0ResultsSection}.

\subsection{Scale dependence}
\label{ScaleDependenceSubsection}

%%%%%%%%%%%%% FIGURE %%%%%%%%%%%%%%%%%%
\begin{figure}[tbh]
\begin{center}
\includegraphics[clip,scale=0.55]{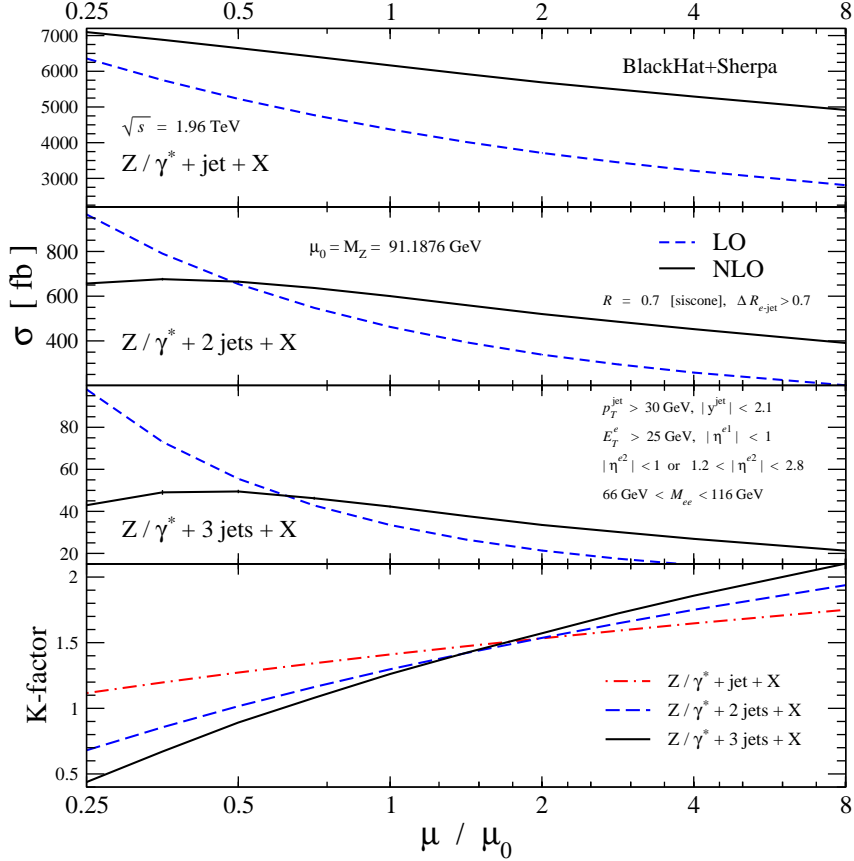}
\end{center}
\caption{The scale dependence of the LO (dashed blue) and NLO (solid black)
cross sections for \Zgamjjja-jet
production at the Tevatron, as a function of the common
renormalization and factorization scale $\mu$, with $\mu_0 = M_Z$.
Here the \SISCone{} jet algorithm is used; the lepton and jet cuts
match CDF~\cite{ZCDF}.
The bottom panels show the $K$ factor, or ratio between the NLO 
and LO result, for each of the three cases:  1 jet (dot-dashed red), 2 jets
(dashed blue), and 3 jets (solid black).}
\label{ZJetsScaleVariationSISConeFigure}
\end{figure}
%%%%%%%%%%%%%%%%%%%%%%%%%%%%%

%%%%%%%%%%%%% FIGURE %%%%%%%%%%%%%%%%%%
\begin{figure}[tbh]
\begin{center}
\includegraphics[clip,scale=0.55]{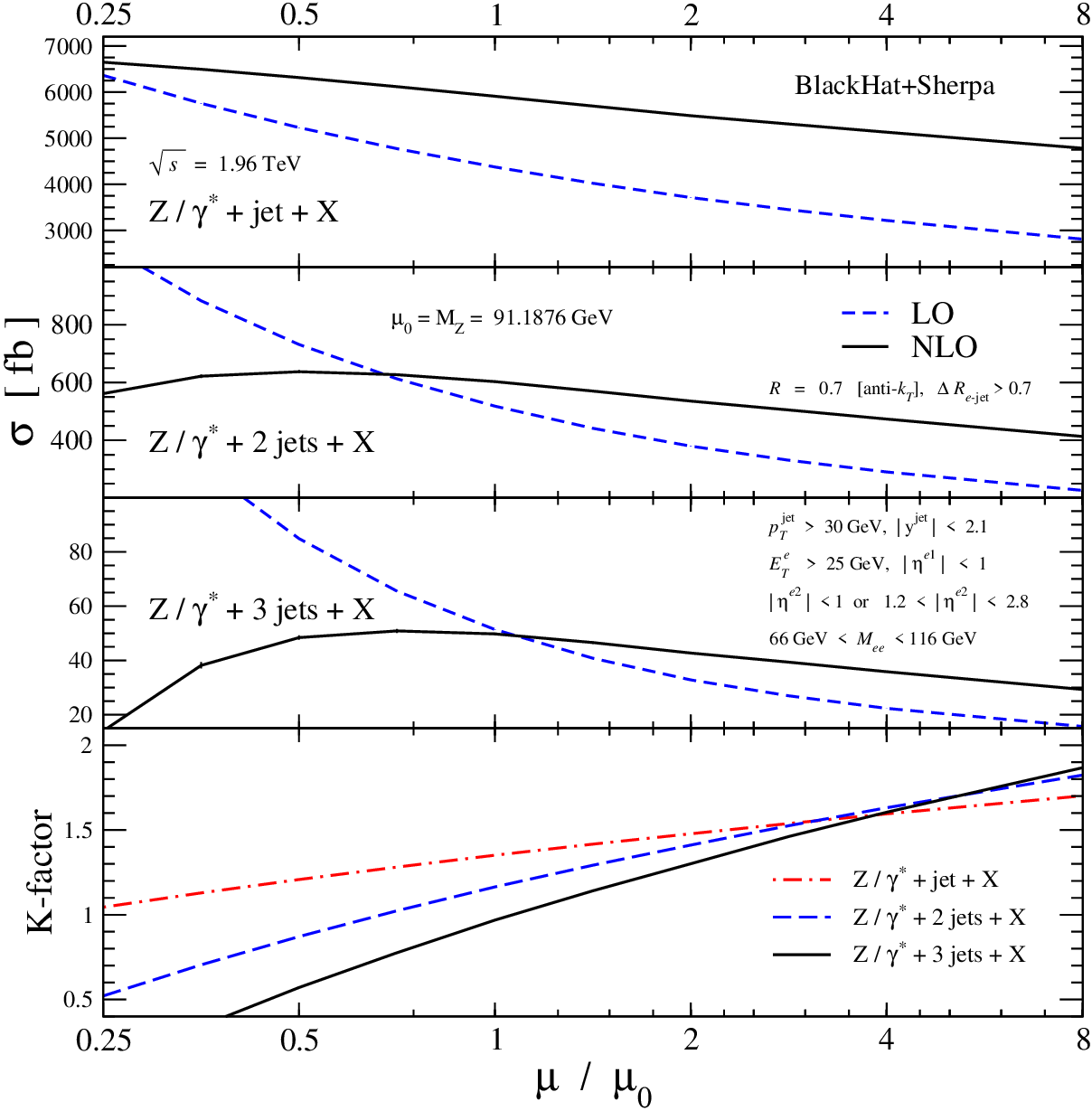}
\end{center}
\caption{The scale dependence for \Zgamjjja-jet
production at the Tevatron.
The plot is the same as \fig{ZJetsScaleVariationantiKTFigure}, 
except here the anti-$k_T$ jet algorithm is used.
}
\label{ZJetsScaleVariationantiKTFigure}
\end{figure}
%%%%%%%%%%%%%%%%%%%%%%%%%%%%%

Following the standard procedure, we test the stability of the
perturbative results by varying the renormalization and factorization
scales. In this article, we set the renormalization and factorization
scales equal, $\mu_R = \mu_F = \mu$.  In
\figs{ZJetsScaleVariationSISConeFigure}{ZJetsScaleVariationantiKTFigure},
we show the scale variation of the total cross section for the
\SISCone{} and anti-$k_T$ algorithms, respectively.  In both cases we
choose the central scale $\mu_0 = M_Z$ and then vary it down by a
factor of four and upwards by a factor of eight.  A fixed scale of
order the $Z$ mass is appropriate here, because the cross section is
dominated by low-\pt{} jets.  In both figures, the upper three panels
show the markedly reduced scale dependence at NLO compared to the
corresponding LO cross section in \Zgamj-, \Zgamjj-, and \Zgamjjj-jet
production, respectively.  The bottom panel combines the ratios of NLO
to LO predictions ($K$ factors) for all three cases, illustrating the
increasing sensitivity of the LO result with an increasing number of
jets.  This increase is expected, because there is an additional power
of $\alpha_s$ for every additional jet.  Accordingly, the reduction
in the scale dependence at NLO tends to become more significant
with an increasing number of jets.  The plots for the $k_T$ algorithm
are very similar to the ones for anti-$k_T$, so we do not
show them here.

\Figs{ZJetsScaleVariationSISConeFigure}{ZJetsScaleVariationantiKTFigure}
also reveal two further features.  First, for $n>1$ the cross
section for the anti-$k_T$ algorithm is significantly larger
than for \SISCone{} at the same value of $R$, especially at LO;
the difference lessens at NLO.  Second, the $K$ factor at $\mu=M_Z$
decreases significantly with the number of jets.  The first feature is due
to the smaller probability of two partons clustering into a jet
in the anti-$k_T$ algorithm.  In that algorithm (or in the $k_T$
algorithm), no clustering can take place unless the two partons are
separated by less than $R$ in the $(\eta,\phi)$ plane; whereas in \SISCone{}
they can be clustered out to a distance of $2R$.  Hence the effective
area of a cone algorithm, for the same value of $R$, is somewhat
larger (by a factor of about 1.35) than that of a cluster algorithm 
such as anti-$k_T$ or $k_T$~\cite{KTES,DMS,Jetography}.
At LO, clustering always causes a loss of events, and thus a decreased
cross section for \SISCone{}, relative to anti-$k_T$.  NLO corrections, 
however, tend to increase the cross section more for jet algorithms
with larger effective cone areas, because there is less chance of radiating
a parton out of the cone and thereby reducing the jet $p_T$ below the cut
threshold~\cite{DMS,Jetography}.
Hence the cross-section difference between the algorithms is lessened at NLO.
The differences between \SISCone{} and $k_T$ algorithms at LO and
NLO can also be examined as a function of the number of jets in
\Wjjja-jet production, using results for the LHC presented in
ref.~\cite{W3jDistributions}.  However, in this work $R=0.4$ was used,
resulting in far smaller perturbative differences between the algorithms.

The second feature, that the $K$ factors at $\mu=M_Z$
decrease with the number of jets, is not unrelated.
It was previously observed that \Wjjj-jet production for $R=0.4$
had quite a small $K$ factor~\cite{PRLW3BH,EMZW3j,W3jDistributions}.
The dependence on the number of jets was discussed in
ref.~\cite{HustonLONLOJets}, where it was attributed to the
LO cross section being ``too high'', in part because of collinear
enhancements associated with the small jet size. We can see from
\figs{ZJetsScaleVariationSISConeFigure}{ZJetsScaleVariationantiKTFigure}
that the trend of decreasing $K$ factor is stronger for the anti-$k_T$
algorithm than for \SISCone.  This feature is consistent with
the picture of ref.~\cite{HustonLONLOJets}, because the anti-$k_T$
effectively has a smaller jet size.  

For distributions, rather than total cross sections, we would like
to choose a characteristic renormalization and factorization scale
on an event-by-event basis, in particular to ensure that the tails
of distributions are described properly.
Previous studies (see {\it e.g.}
refs.~\cite{EarlyWplus2MP,ZCDF,WCDF,PRLW3BH,ZD0})
have used the transverse energy of the vector boson,
$E_T^V$, as a common renormalization and factorization scale.
As already argued in
refs.~\cite{Bauer,W3jDistributions} this choice is quite poor at LHC
energies. Indeed, because of the large dynamic range at the LHC, at
NLO the choice can go disastrously wrong for some distributions,
leading to negative cross sections~\cite{W3jDistributions}.  It also
causes large changes in shapes of generic distributions between LO and
NLO.  This behavior reflects the emergence
of large logarithms $\ln \mu/E$, which spoil
the validity of the perturbative expansion when $\mu$
does not match the characteristic energy scale $E$.
We note that without an NLO result for guidance, it may not be clear
that a given scale choice---such as $E_T^V$---is problematic.

%%%%%%%%%%%% FIGURE %%%%%%%%%%%%%%%%%%
\begin{figure}[tbh]
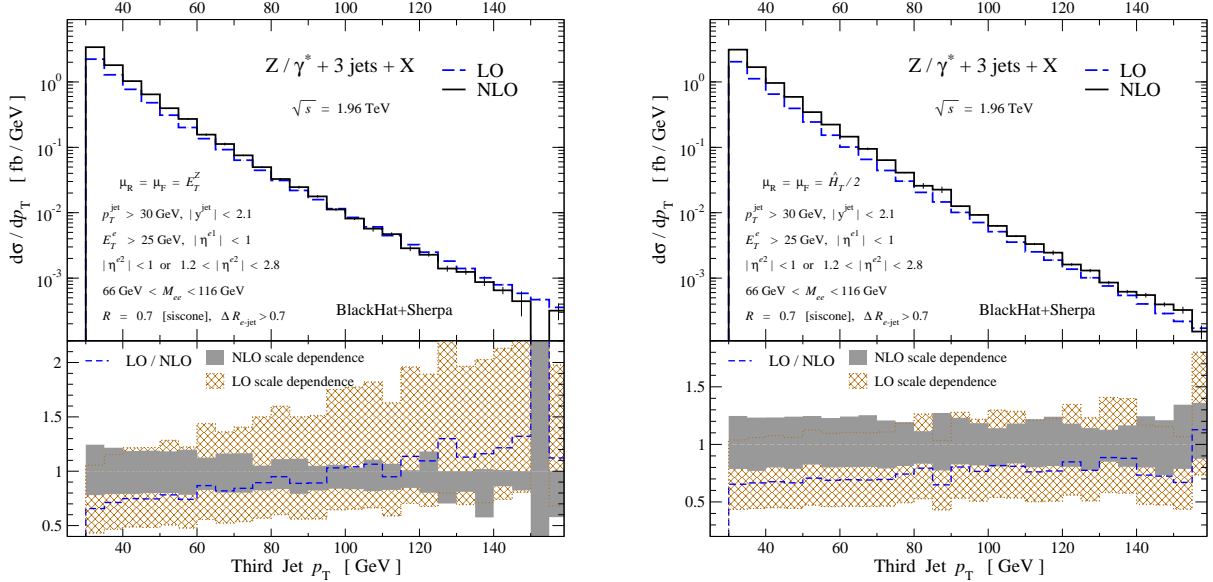

\begin{minipage}[b]{1.\linewidth}
\includegraphics[clip,scale=0.36]{figs/Z3j-CDF-ETZ_siscone-Pt30_jets_jet_1_1_pt_3.eps}
\hskip 1cm
\includegraphics[clip,scale=0.36]{figs/Z3j-CDF-HT_siscone-Pt30_jets_jet_1_1_pt_3.eps}
\end{minipage}
\caption{The NLO \pt{} distribution of the third jet in \Zgamjjj-jet
  production at the Tevatron, for the \SISCone{} algorithm. 
  In the left panel the scale choice $\mu= E_T^Z$
  is used and for the right panel $\mu = \HTp/2$.
  The thin vertical lines (where visible) indicate the numerical
  integration uncertainties.  The bottom part of each panel displays
  various ratios, where the denominator
  is always the NLO result at the reference scale choice, and the
  numerator is obtained by either evaluating the LO result at the same
  scale (dashed blue line), varying the LO scale by a factor of two in either
  direction (cross-hatched brown band), or varying the NLO scale in the
  same way (gray band). 
  Although the two NLO results are compatible, the LO results 
  have large shape differences, illustrating why $\mu
  = E_T^Z$ is not a good choice at LO.
  The jet and lepton cuts match those used by CDF~\cite{ZCDF}.
}
\label{ThirdJetTeVComparisonFigure}
\end{figure}
%%%%%%%%%%%%%%%%%%%%%%%%%%%%%

Even for the Tevatron, with its smaller dynamic range, the
commonly-used scale choice $\mu = E_T^V$ is not particularly good.  It
leads to a large change in shape between LO and NLO in the \pt{}
distribution of the third-hardest jet in \Zgamjjj-jet production, as
shown in the left panel of \fig{ThirdJetTeVComparisonFigure}. 
In contrast, the scale choice $\mu = \HTpartonic/2$,
where $\HTpartonic$ is the total partonic transverse
energy defined in \eqn{PartonicHTdef}, results in little change in
shape between LO and NLO.  This choice is shown in the right panel of
\fig{ThirdJetTeVComparisonFigure}.

%%%%%%%%%%%%% FIGURE %%%%%%%%%%%%%%%%%%
\begin{figure}[tbh]
\includegraphics[clip,scale=0.7]{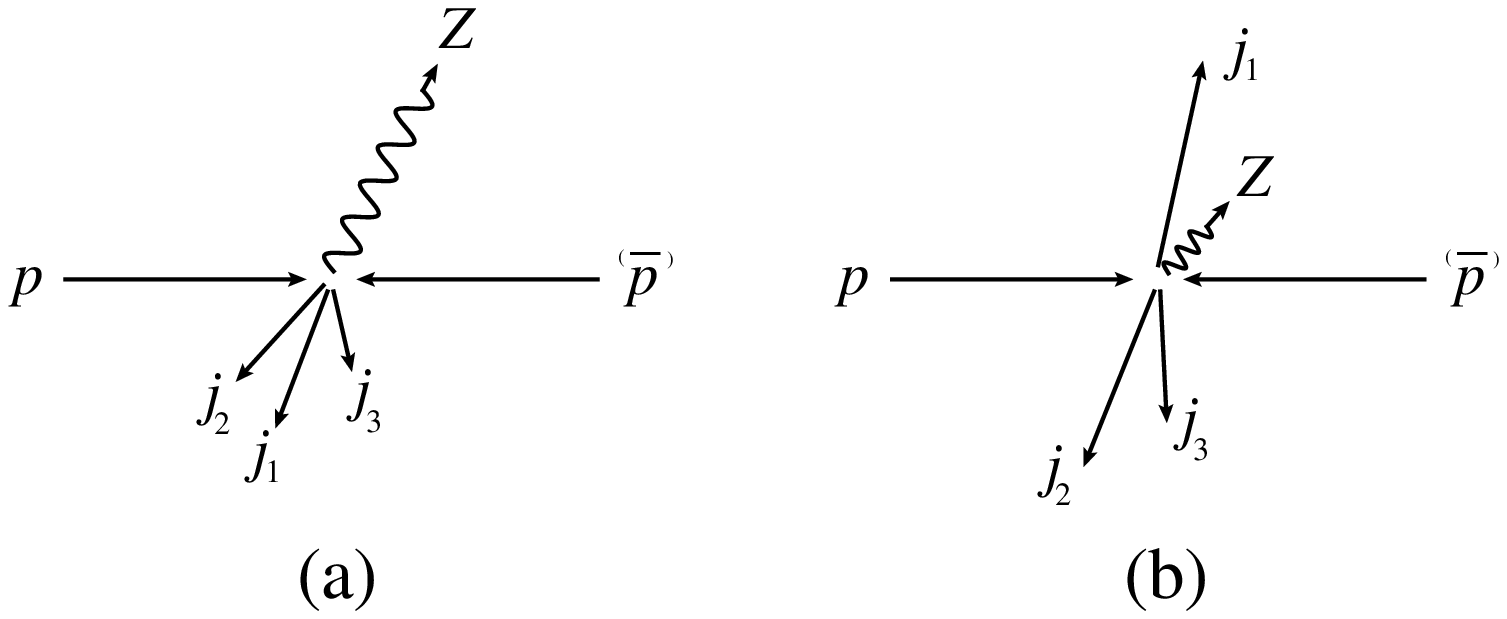}
\caption{Two distinct \Zjjj{} jet configurations with rather
different values for the $Z$ transverse energy.
In configuration (a) an energetic $Z$ balances the energy
of the jets, while in (b) the $Z$ is relatively soft.
Configuration (b) generally dominates over (a) when the transverse
energy of the third jet gets large. }
\label{TwoConfigurationsFigure}
\end{figure}
%%%%%%%%%%%%%%%%%%%%%%%%%%%%%

The difficulty with using the vector-boson transverse energy,
$\mu = E_T^Z$, as the scale can be exposed~\cite{W3jDistributions}
by considering the two configurations depicted in
\fig{TwoConfigurationsFigure}.  In configuration~(a), the $Z$ boson
has a transverse energy larger than that of the jets, and
sets the scale for the process.  In configuration~(b), the two leading
jets roughly balance in \pt, while the $Z$ has much lower transverse
energy.  Here, the scale $\mu = E_T^Z$ is too low, and not
characteristic of the process.  In the tails of
\fig{ThirdJetTeVComparisonFigure}, configuration~(b) dominates,
because it results in a larger third-jet \pt{} for fixed center-of-mass
partonic energy; contributions from higher center-of-mass energies which
might boost the $Z$-boson transverse energy are suppressed by the
fall-off of the parton distributions.  This explains the large deviation
between LO and NLO visible in the left panel of
\fig{ThirdJetTeVComparisonFigure}.

In contrast, $\HTpartonic$ (or some fixed fraction of it)
does properly capture both configurations
(a) and (b).  It is thus a much better choice of scale.
 (For the purposes of fixing the scale, we prefer
the partonic definition of the total transverse energy
over the experimental one in \eqn{PhysicalHTdef} because it is
independent of the experimental cuts and the jet
definitions~\cite{W3jDistributions}.)
In the remainder of this paper we take $\mu = \HTpartonic/2$ as our default
for both the renormalization and factorization scales,
except where noted.  To assess the remaining scale dependence
in the cross sections we evaluate them at five scales: $\mu/2,
\mu/\sqrt2, \mu, \sqrt2\mu, 2\mu$. We generate scale variation bands
using the minimum and maximum values. In our previous
analysis~\cite{W3jDistributions} of $W$ production in association with jets
at the LHC, we chose $\mu = \HTpartonic$.
Generally, $\HTpartonic$ tends to be on the
high side of typical energy scales, so here we divide by a factor of
two.  The difference between the two choices at NLO is not large, on the
order of 10\% in the normalization, and with very small effects on the
shapes of distributions. At LO the changes are, of course, larger,
with up to 40\% variations.

Although we adopt here $\mu = \HTpartonic/2$ as a good representative overall
scale for general distributions, other approaches may be superior
for particular distributions, or particular regions of phase space.
For example, it may be possible to resum large logarithms that appear
in particular corners of phase space, and match the resummed result to the
NLO one.  Even if that cannot be done, it is certainly possible that
choosing a scale that is a blend of different scales
(such as the different jet transverse momenta) is appropriate in
some cases.

In principle, perturbative approximations can break down in various 
kinematic regions, so it is important to check whether this can affect our
results.  The breakdown is often due to effects of soft-gluon
emission that in can be resummed in many cases.  Large soft-gluon effects
can be obtained when there are explicit vetoes on soft radiation,
or when such radiation is implicitly vetoed by fast-falling parton
distribution functions.  In this paper, we put no explicit vetoes on soft
radiation in any observable we consider.  However, there is an implicit
veto as one goes out in the tail of the $H_T$ distribution or
the third-jet $p_T$ distribution.  This implicit veto might lead to large
double Sudakov, or threshold, logarithms.

In order to investigate whether such logarithms might be large,
we take advantage of the fact that threshold logarithms in the
high-$p_T$ tail of the $p_T$ distribution for inclusive jet production
should be very similar to the tails we are looking at in \Vjn-jet
production, at comparable values of jet $p_T$.  In both cases there is 
comparable mix of partonic channels, and similar values of parton $x$.
Note, however, that one can reach much higher
$p_T$s experimentally in pure jet production because of the much
larger cross sections.  One recent resummation of threshold logarithms 
for inclusive jet production~\cite{deFV} shows that the effects are
quite modest.  For example, figure 6 of ref.~\cite{deFV} shows 
the ratio $K$ of the (matched) NLL result to the NLO result for
single-inclusive jet production at the Tevatron Run I $(\sqrt{s}=1.8$~TeV),
for $p_T$ from 50 to 500 GeV.  For various choices of the renormalization
and factorization scales, $K$ ranges from 0.98 to at most 1.14, as long as
$p_T < 300$~GeV.  Note that 300~GeV is well above the third-jet $p_T$s
shown in \fig{ThirdJetTeVComparisonFigure}.  (The relevant 
parton $x$ values probed at the Tevatron in figure 6 of ref.~\cite{deFV}
also correspond at the LHC to $p_T$s that are about seven times larger,
well above the range studied in
figure~9 of ref.~\cite{W3jDistributions}.  There, the NLO cross section
evaluated at $\mu=E_T^V$ became negative for a second-jet $E_T$ of
only 475~GeV.  Hence, even in this more extreme example, threshold
logarithms are very unlikely to play a role in this behavior.)

There is one other type of logarithm in \Vjn-jet production, which is
not present in inclusive jet production, and that is a (double)
logarithm of the form $\ln(p_{T,{\rm jet}}/M_V)$, due to emission of
electroweak vector bosons that are soft and collinear with respect to
the jets, as in the configuration shown in
\fig{TwoConfigurationsFigure}(b).  The importance of this logarithm
was emphasized very recently~\cite{RSS} for the case of \Zj-jet production.
Although the NLO correction to this process is enhanced by
$\alpha_s \, \ln^2(p_{T,{\rm jet}}/M_V)$ with respect to LO,
the effect is peculiar to \Vj-jet production.  It does not happen
when two or more final-state partons are present at LO, because then
the configuration in \fig{TwoConfigurationsFigure}(b) can already be 
reached at LO. Also, because it is associated with {\it electroweak}
boson emission, it does not represent a {\it QCD} double logarithm
that will reappear at higher orders in $\alpha_s$.

We conclude that there is no indication of a breakdown of fixed-order
perturbation theory for the ranges of observables studied in this paper
or in ref.~\cite{W3jDistributions}.

%%%%%%%%%%%%%%%%%%%%%%%%%%%%%%%%%%%%%%%%%%%%%%%%%%%%%%

\section{Results for CDF}
\label{CDFResultsSection}

In this section we present results for \Zgamjjja-jet production
(inclusive) at the Tevatron, and compare to data from CDF
that has been corrected back to hadron level~\cite{ZCDF}.
For jet \pt{} distributions in \Zgamjx-jet production, we 
use the (relatively large) non-perturbative corrections
estimated by CDF~\cite{ZCDF} to transform our parton-level results
to hadron-level ones.  For \Zgamjjj-jet production,
non-perturbative corrections were not explicitly presented by CDF, and
we give only parton-level results.

\subsection{Total cross sections}

%%%%%%%%%%%%%%%%%%%%%%% TABLE mu = HT/2, anti-kt, Siscone %%%%%%%%%%%%%%
\begin{table}[ht]
%\hspace{-2.5cm}
\begin{tabular}{|c||c||c|c||c|c|}
\hline
\# of jets  & CDF & LO parton & NLO parton & LO parton  & NLO parton \\[-.34cm] 
    $\,$  & midpoint  & \SISCone & \SISCone & anti-$k_T$ & anti-$k_T$ \\
\hline
\hline
1  & $\,7003 \pm 146 {}^{+483}_{-470} \pm 406 \, $ 
   & $4635(2)^{+928}_{-715}$
   & $6080(12)^{+354}_{-402}$
   & $4635(2)^{+928}_{-715}$
   & $5783(12)^{+257}_{-334}$ \\
\hline
2  & $ 695 \pm 37  {}^{ +59}_{ -60} \pm 40  $ 
   & $429.8(0.3)^{+171.7}_{-111.4}$ 
   & $564(2)^{+59}_{-70}$
   & $481.2(0.4)^{+191}_{-124}$ 
   & $567(2)^{+31}_{-57}$  \\
\hline
3  & $  60 \pm 11  {}^{  +8}_{  -8} \pm 3.5 $ 
   & $24.6(0.03)^{+14.5}_{-8.2}$ 
   & $36.8(0.2)^{+8.8}_{-7.8}$
   & $37.88(0.04)^{+22.2}_{-12.6}$ 
   & $44.7(0.24)^{+5.1}_{-6.8}$ 
    \\
\hline
\end{tabular}
\caption{\Zgamjjja-jet production (inclusive) cross section (in fb) at CDF.  
  The column labeled CDF gives the hadron-level results from ref.~\cite{ZCDF},
  using a midpoint jet algorithm.  The experimental uncertainties
  are statistical, systematics (upper and lower) and luminosity.
  The columns labeled by LO
  parton and NLO parton contain the parton-level results for the
  \SISCone{} and anti-$k_T$ jet algorithms.  The central scale choice
  for the theoretical prediction is $\mu=\hat{H}_T/2$, the numerical
  integration uncertainty is in parentheses, and the scale dependence
  is quoted in super- and subscripts.  Non-perturbative corrections
  should be accounted for prior to comparing the CDF measurement 
  to parton-level NLO theory.}
\label{Table-Zjets-Tev-total-xs-HT} 
\end{table}

%%%%%%%%%%%%%%%%%%%%%%%%%%%%%%%%%%%%%%%%%%%%%%%%%%%%%%%%%%%

%%%%%%%%%%%%%%%% TABLE mu = ETZ, combined anti-kt, Siscone %%%%%%%%%%%%%%
\begin{table}[ht]
%\hspace{-2.5cm}
\begin{tabular}{|c||c||c|c||c|c|}
\hline
\# of jets  & CDF & LO parton & NLO parton & LO parton  & NLO parton \\[-.34cm] 
    $\,$  & midpoint  & \SISCone & \SISCone & anti-$k_T$ & anti-$k_T$ \\
\hline
\hline
1  & $\, 7003 \pm 146 {}^{+483}_{-470} \pm 406 \, $ 
   & $4206(2)^{+801}_{-616}$ 
   & $6076(9)^{+501}_{-466}$
   & $4206(2)^{+801}_{-616}$
   & $5828(9)^{+425}_{-414}$ 
  \\
\hline
2  & $ 695 \pm 37  {}^{ +59}_{ -60} \pm 40  $ 
   & $422.2(0.3)^{+168}_{-109}$
   & $576(2)^{+72}_{-77}$
   & $469.4(0.4)^{+185}_{-120}$
   & $583(2)^{+51}_{-67}$ 
  \\
\hline
3  & $  60 \pm 11  {}^{  +8}_{  -8} \pm 3.5 $ 
   & $28.66(0.03)^{+17.9}_{-10.0}$ 
   & $40.28(0.2)^{+8.6}_{-8.5}$
   & $43.3(0.05)^{+26.6}_{-14.9}$
   & $48.7(0.3)^{+3.8}_{-7.9}$
    \\
\hline
\end{tabular}
\caption{\Zgamjjja-jet production cross section (in fb) at CDF.
   This table is similar to \tab{Table-Zjets-Tev-total-xs-HT},
  except that here the scale choice is $\mu  = E_T^Z$.}
\label{Table-Zjets-Tev-total-xs-ETZ} 
\end{table}

%%%%%%%%%%%%%%%%%%%%%%%%%%%%%%%%%%%%%%%%%%%%%%%%%%%%%%%%%%%

In \tab{Table-Zjets-Tev-total-xs-HT} we present the total inclusive
cross sections for \Zgamjjja-jet production, showing both the CDF
measurement and theoretical predictions, using our default scale
choice $\mu=\hat{H}_T/2$.  The theoretical results in the table are
given for both the \SISCone{} and anti-$k_T$ jet algorithms.  (The
$k_T$ algorithm gives identical results as anti-$k_T$ at LO, and is
within 1\% at NLO.)  In the second column we give the CDF measurement, 
for its midpoint jet algorithm and corrected to hadron level,
along with the experimental uncertainties.  The statistical,
systematic (upper and lower) and luminosity uncertainties are
given after the central values.  The third and fourth
columns present the LO and NLO parton-level predictions. Here 
we quote the uncertainties from integration statistics in parentheses,
 and the scale dependence in super- and sub-scripts
(upper and lower).  The scale
dependence is determined following the traditional prescription of
varying the scale by a factor of two around the central choice
$\mu = \HTp/2$, as described above.

To assess the effect of changing the jet algorithm, we compare the
\SISCone{} and anti-$k_T$ results in \tab{Table-Zjets-Tev-total-xs-HT},
which further quantifies the differences that were visible in 
\figs{ZJetsScaleVariationSISConeFigure}{ZJetsScaleVariationantiKTFigure},
which used a fixed scale $\mu$.  Although the \SISCone{} algorithm
gives noticeably different results from the anti-$k_T$ algorithm, the
variations are similar in magnitude to the residual scale dependence.
The reasons for the perturbative differences between \SISCone{} and anti-$k_T$
algorithms were outlined in section~\ref{ScaleDependenceSubsection}.

It is also interesting to compare the results of
\tab{Table-Zjets-Tev-total-xs-HT} to cross sections obtained with the
widely-used scale $\mu = E_T^Z$ instead of our default choice $\mu =
\HTpartonic/2$.  In \tab{Table-Zjets-Tev-total-xs-ETZ} we give cross
sections with this scale choice for the \SISCone{} and anti-$k_T$ jet
algorithms.  Comparing these results to those of
\tab{Table-Zjets-Tev-total-xs-HT}, we see that, at least for
\Zgamjx{} jets, the $K$ factor (ratio of NLO to LO) is much closer to unity
for the choice $\mu = \HTpartonic/2$, than for $\mu = E_T^Z$.
Although the $\mu = E_T^Z$ choice is problematic in general, as
already noted in \sect{CutsandScalesSection}, at NLO it gives results
for the total cross section that are similar to those from our default
choice of $\mu = \HTpartonic/2$.

In order to compare parton-level results to the experimental
measurement we must account for non-perturbative corrections,
using estimates from CDF~\cite{ZCDF}.  These corrections are
sizable, increasing the total cross section by a factor between 1.1 and 1.4
as the number of jets increases from one to three.  
As we will see below in \fig{Figure-Z12jet-CDFdata-siscone-HT}
for the jet $p_T$ distributions, these estimated correction factors 
align NLO theory with the measurement within uncertainties, although a 
much more careful study of the non-perturbative corrections 
and the differences in jet algorithms
is needed.  It is
interesting to note that in the CDF measurement of \Wjn-jet
production~\cite{WCDF}, the corrections are significantly smaller.
That measurement used a jet cone size of $R=0.4$
(with the \JETCLU{} algorithm~\cite{JETCLU}).  
There the hadronization and
underlying event corrections were under 10\% below 50 GeV and under
5\% at higher $E_T$.  The CDF study may
also be contrasted with the D0 study~\cite{ZD0} discussed below, in which
the cone size of $R=0.5$ leads to non-perturbative corrections on the
order of 15\%.  From the perspective of maintaining the precision of
NLO predictions, it is advantageous to choose jet-cone sizes which
minimize non-perturbative corrections, while not increasing the size
of ($\ln R$-enhanced) higher-order perturbative corrections too much.
As discussed in {\it e.g.} refs.~\cite{DMS,Jetography},
there is a tradeoff between the underlying event correction 
(increases as $R$ increases) and splash-out (increases as $R$ decreases),
and a careful study would be needed to find the best choice.

\subsection{Comparison to CDF jet \pt{} distributions}
\label{CDFpTComparison}

%%%%%%%%%%%%%%%%%%%%%%%%%%% FIGURE %%%%%%%%%%%%%%%%%%%%%%%%%%%%%%%%
\begin{figure}[tbh]
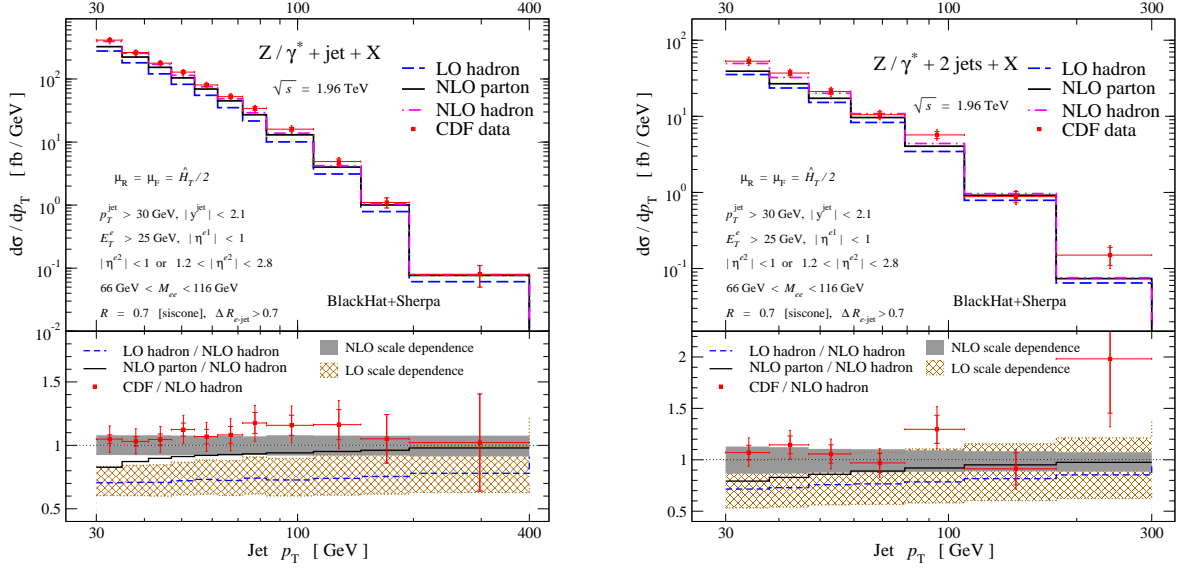

\begin{center}
\begin{minipage}[b]{1.\linewidth}
\includegraphics[clip,scale=0.35]{figs/Z1j-CDF-HT_siscone-Pt30_jets_jet_1_1_pt_0_with_CDF_data.eps}
\hskip 1 cm 
\includegraphics[clip,scale=0.35]{figs/Z2j-CDF-HT_siscone-Pt30_jets_jet_1_1_pt_0_with_CDF_data.eps}
\end{minipage}
\end{center}
\caption{Jet \pt{} distributions for \Zgamjx-jet production at the
  Tevatron with CDF's cuts.  The theoretical predictions use the
  \SISCone{} algorithm and the scale choice $\mu=\hat{H}_T/2$.
  In the upper panels the parton-level
  NLO distribution are the solid (black) histograms, and the NLO
  distributions corrected to hadron level are given by dash-dot
  (magenta) curves.   The CDF data are the (red)
  points, whose inner and outer error bars denote respectively the
  statistical and total uncertainties
  on the measurements (the latter obtained by adding separate uncertainties
  in quadrature).
  The LO predictions have been corrected to hadron level and are
  shown as dashed (blue) lines.
  The lower panels show the distributions normalized to
  the full hadron-level NLO prediction for $\mu=\hat{H}_T/2$. 
  The scale-dependence bands in the lower
  panels are shaded (gray) for the NLO prediction corrected to
  hadron level and cross-hatched (brown) for LO corrected to hadron level.
}
\label{Figure-Z12jet-CDFdata-siscone-HT}
\end{figure}
%%%%%%%%%%%%%%%%%%%%%%%%%%%%%%%%%%%%%%%%%%%%%%%%%%%%%%%%%%%%%%%%%%%%%5

In this subsection, we compare our results with CDF data for jet \pt{}
distributions in \Zgamj-jet and \Zgamjj-jet production.
In the observables used by CDF, sometimes referred to as 
inclusive jet \pt{} distributions,
all jets passing the cuts are included in the
distributions. That is, if $n$ jets pass the cuts, the event is
counted $n$ times, with contributions to each of the $n$ bins containing
the transverse energy of one of the jets.  By definition, for
inclusive \Znj-jet production, at least $n$ jets pass the cuts, and
periodically additional jets can also pass the cuts. At NLO these
extra jets are modeled by a single extra jet from the real-emission
contribution.  This causes the area under the curve to be slightly
more than $n$ times the total cross section.
In contrast, the \Wjn-jet production distributions
measured in ref.~\cite{WCDF} and the \Zjn-jet production distributions
measured in ref.~\cite{ZD0} are differential in the transverse energy
(or momentum) of the $n^{\rm th}$ jet, and each event is counted only
once, so they integrate to give the total cross section for \Vjn-jet
production.

The left and right panels of
\fig{Figure-Z12jet-CDFdata-siscone-HT} are for \Zgamj-jet and
\Zgamjj-jet production, respectively.
The upper part of each panel compares the LO and NLO results
against CDF data from ref.~\cite{ZCDF}.  \BlackHat{}+\SHERPA{}
produces NLO parton-level predictions.  To compare to the
CDF measurement we need to account for non-perturbative corrections.
We use the last column of Table~I of ref.~\cite{ZCDF} as an estimate
of their size.  This table of corrections was determined for
the CDF midpoint jet algorithm using {\sc Pythia}~\cite{PYTHIA}, 
an LO-based parton-shower, hadronization and underlying-event program.
Because we used the (infrared-safe) \SISCone{} algorithm, the possible
algorithm-dependence of the non-perturbative corrections
introduces additional uncertainty into the comparison.
As mentioned in the introduction, studies~\cite{SISCONE,EHKetal}
of inclusive-jet cross section differences between midpoint algorithms 
and \SISCone{} (which were also performed for $R=0.7$) suggest
relatively small ``parton-level'' differences between the algorithms,
which in turn suggests that applying the CDF non-perturbative
corrections to our \SISCone{} perturbative prediction is not unreasonable. 
The size of the corrections can be seen in the upper
panels of \fig{Figure-Z12jet-CDFdata-siscone-HT} by comparing the
curves labeled ``NLO parton'', which are the parton-level predictions,
to the ones labeled ``NLO hadron'', which are the hadron-level ones.
It is easier to judge the size in the lower panels, using the
solid (black)
curves which give the ratios of the two predictions.  For example, for
\Zgamjj-jet production, non-perturbative corrections are significant
for low \pt, on the order of 20\% at 30 GeV, and gradually drop to
under 5\% at larger jet transverse momenta.  Uncertainties in the
non-perturbative corrections are not included in the plots.

The bottom panels shows various ratios, normalized to the NLO
hadron-level prediction for the central scale $\mu = \HTpartonic/2$. 
We include scale-dependence bands, as described above, for the
predictions corrected to hadron level.
As expected, for NLO the scale dependence is greatly reduced when
compared to LO.  We note that for both \Zgamjx-jet production, the NLO
hadron-level jet \pt{} distributions match the CDF results quite well,
noticeably better than the hadron-level LO distributions or
parton-level NLO distributions.  A similar comparison of the
experimental data to NLO predictions was given in the CDF study, using
\MCFM~\cite{MCFM}.  The ratios of data to NLO presented there differ by
up to 10\% from those shown in \fig{Figure-Z12jet-CDFdata-siscone-HT}.
Most of the difference can be attributed to the choice of central scale
in the NLO result, $\mu=E_T^Z$ versus $\mu = \HTpartonic/2$.
CDF also assessed the uncertainties on the NLO
predictions arising from the parton distribution functions. They
found them to vary from 4\% at low jet $p_T$ to 10\% at high $p_T$,
which is generally smaller than the NLO scale variation.

\subsection{Predictions for \Zgamjjj-jet distributions at CDF}

%%%%%%%%%%%%%%%%%%%%%%%%%%% FIGURE %%%%%%%%%%%%%%%%%%%%%%%%%%%%%%%%
\begin{figure}[ht]
\begin{center}
\includegraphics*[scale=0.45]{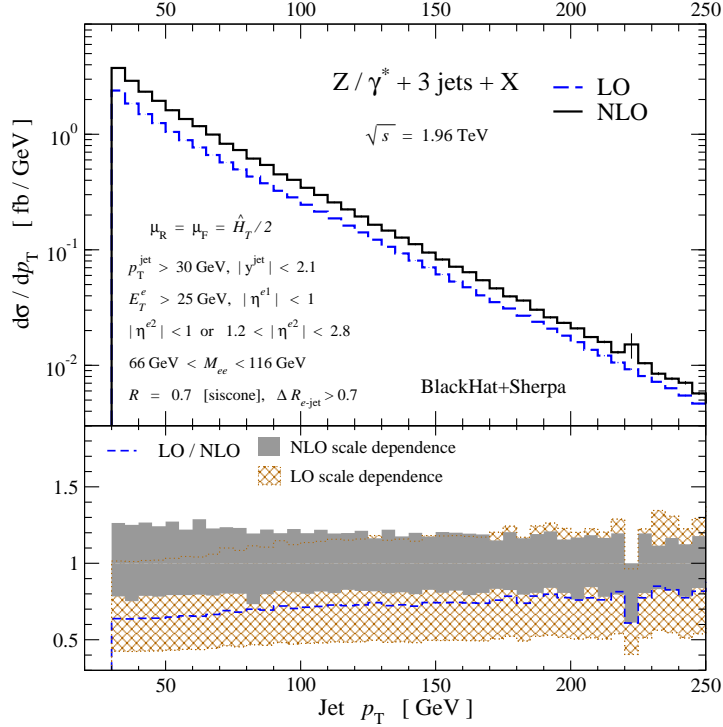}
\end{center}
\caption{\pt{} of all jets for \Zgamjjj-jet production with the CDF setup,
  using the \SISCone{} jet algorithm and the scale choice
  $\mu=\hat{H}_T/2$, for LO and NLO at parton level.  The thin vertical
  lines (where visible) indicate the numerical integration uncertainties.
  The lower panel bands are normalized to the central NLO prediction, as in 
  \fig{Figure-Z12jet-CDFdata-siscone-HT}.}
\label{Figure-Z3jets-jets-pt-siscone-HT}
\end{figure}
%%%%%%%%%%%%%%%%%%%%%%%%%%%%%%%%%%%%%%%%%%%%%%%%%%%%%%%%%%%%%%%%%%%%%

%%%%%%%%%%%%%%%%%%%%%%%%%%% FIGURE %%%%%%%%%%%%%%%%%%%%%%%%%%%%%%%%
\begin{figure}[ht]
\begin{center}
\includegraphics*[scale=0.65]{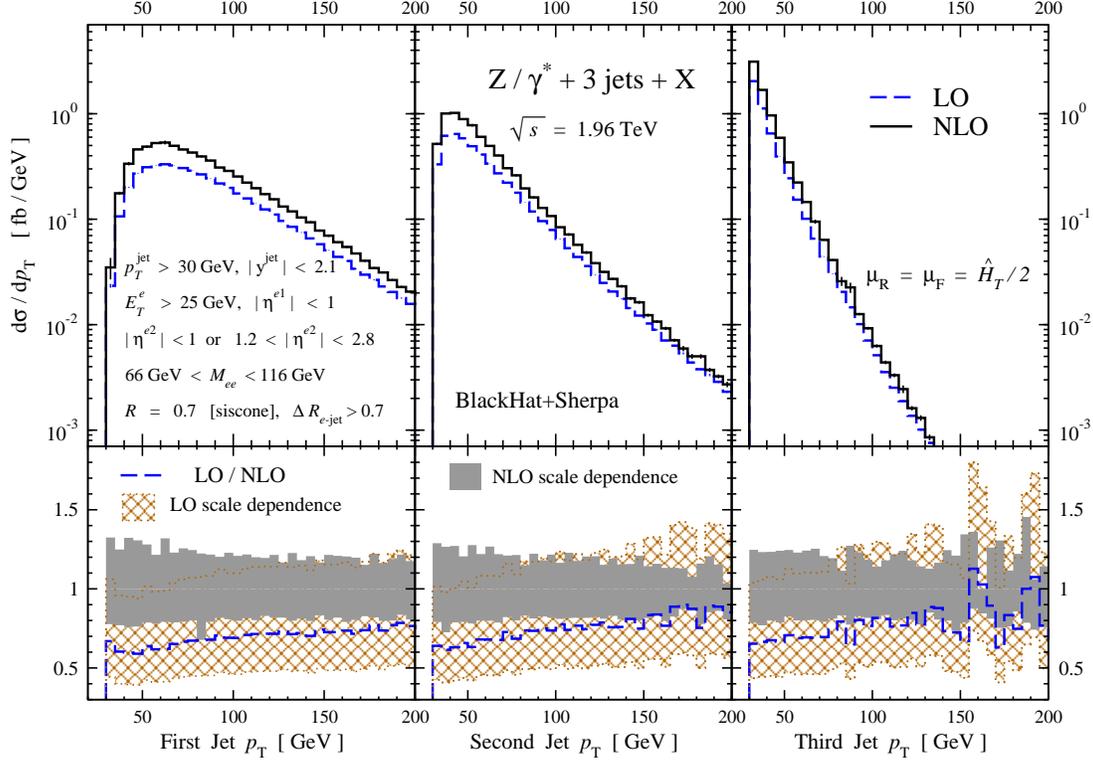}
\end{center}
\caption{First, second and third jet \pt{} distributions for 
\Zgamjjj-jet production.  The dashed (blue) lines are LO
predictions and the solid (black) lines are NLO predictions.  The
\SISCone{} jet algorithm and a scale choice of $\mu = \HTpartonic/2$ are
used for these plots.}
\label{Figure-Z3jets-jets-all-pt-siscone-HT}
\end{figure}
%%%%%%%%%%%%%%%%%%%%%%%%%%%%%%%%%%%%%%%%%%%%%%%%%%%%%%%%%%%%%%%%%%%%%

%%%%%%%%%%%%%%%%%%%%%%%%%%% FIGURE %%%%%%%%%%%%%%%%%%%%%%%%%%%%%%%%
\begin{figure}[ht]
\begin{center}
\includegraphics[clip,scale=0.733]{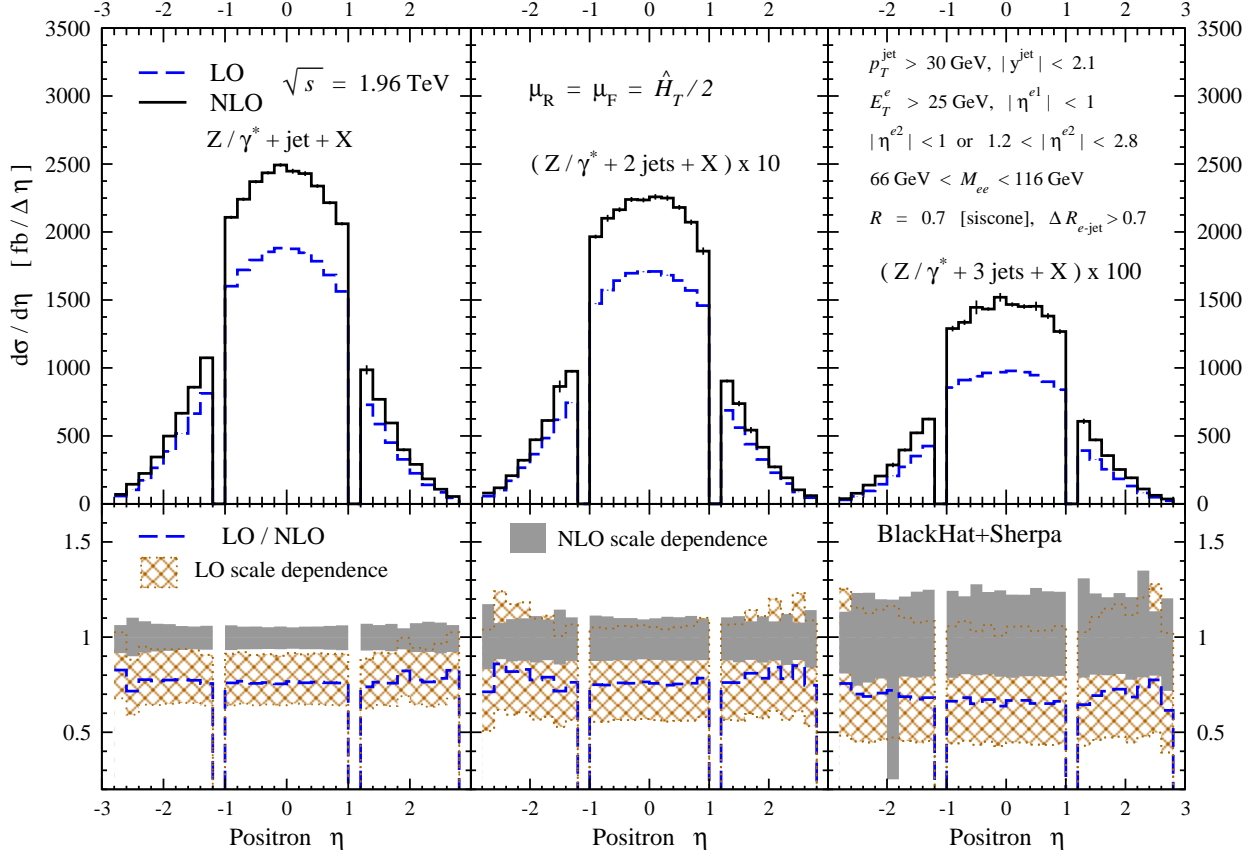}
\end{center}
\caption{The $\eta$ distribution of the positron, for \Zgamjjja-jet
  production. The discontinuities in the plots are due to the experimental
  cuts~(\ref{CDFCuts}). The cross sections in the second and third panels 
  are multiplied by factors of 10 and 100, respectively.}
\label{Figure-Z3jets-jets-eta-siscone-HT}
\end{figure}
%%%%%%%%%%%%%%%%%%%%%%%%%%%%%%%%%%%%%%%%%%%%%%%%%%%%%%%%%%%%%%%%%%%%%

In \fig{Figure-Z3jets-jets-pt-siscone-HT}, we show the combined
distribution of all jet \pt{}s in \Zgamjjj-jet production.  It would
be very interesting to compare this prediction to CDF data, after
accounting for non-perturbative effects. As discussed above, the
integral under the curve gives a bit more than three times the total
cross section. As can be seen in the plot, with the scale choice $\mu
= \HTpartonic/2$ there is only a modest change in shape between LO
and NLO, especially at higher jet \pt.  This is similar to the
parton-level results at \Zgamjx-jet production shown in
\fig{Figure-Z12jet-CDFdata-siscone-HT}.  We expect
non-perturbative corrections to lead to larger shape changes at lower
\pt.

The separate distributions for the hardest, second-hardest, and third-hardest
jet are shown in \fig{Figure-Z3jets-jets-all-pt-siscone-HT}.  The shapes
of the LO and NLO distributions are again similar, with our default
scale choice.  As in \Wjjj-jet production~\cite{W3jDistributions}, 
successive jets have increasingly steeply falling distributions.

In \fig{Figure-Z3jets-jets-eta-siscone-HT} we show the $\eta$
distribution of the positron for \Zgamjjja-jet production.  The
discontinuity and gap between $\eta= \pm 1$ and $\eta = \pm 1.2$ result
from the discontinuity in the charged-lepton cuts in
\eqn{CDFCuts}.  A careful inspection reveals a small forward-backward
asymmetry, which can be traced to the left-right asymmetry in the 
$Z$ boson couplings to fermions.  (Similar asymmetries are discussed
in {\it e.g.} ref.~\cite{ESW}.)  Once again, there is only a modest
shape change between LO and NLO.

%%%%%%%%%%%%%%%%%%%%%%%%%%%%%%%%%%%%%%%%%%%%%%%%%%%%%%%%%%%%
\section{Results for D0}
\label{D0ResultsSection}

%%%%%%%%%%%%%%%%%%%%%%% TABLE D0 mu = HT/2, Siscone %%%%%%%%%%%%%%
\begin{table}[th]
%\hspace{-2.5cm}
\begin{tabular}{|c||c|c||c|c|}
\hline
\# of jets  & LO parton & NLO parton & LO parton  & NLO parton \\[-.34cm] 
   $\,$    & selection (a) &  selection (a)  & selection (b) & selection (b) \\
\hline
\hline
0 [pb]
 % & ($179008(18)^{+0}_{-649} )  $ 
  & $179.01(0.02)^{+0.}_{-0.649} $ 
%   & $236957(75)^{+3745}_{-2478}$
   & $236.96(0.08)^{+3.75}_{-2.48} $
%   & $84088(30)^{+78}_{-643}$
   & $84.08(0.03)^{+0.78}_{-0.64} $
%   & $106812(148)^{+879}_{-401}$ \\
   & $106.81(0.15)^{+0.88}_{-0.40}$  \\
\hline
1 [fb] & $25223(9)^{+5011}_{-3877}$ 
   & $30230 (55)^{+1212}_{-1667}$
   & $10083(6)^{+1927}_{-1501}$ 
   & $12537(44)^{+580}_{-721} $  \\
\hline
2 [fb] & $3787(3)^{+1539}_{-999} $ 
   & $4415(14)^{+260}_{-476}$ 
   & $1538(2)^{+608}_{-398}$ 
   & $1848(10)^{+127}_{-201}$  \\
\hline
3 [fb] & $462(0)^{+280}_{-158}$ 
   & $553(3)^{+70}_{-92} $ 
   & $190(0)^{+113}_{-64} $ 
   & $236(1)^{+30}_{-42} $
    \\
\hline
\end{tabular}
\caption{NLO parton-level \Zgamjjjz-jet production cross sections
  corresponding to D0 selections (a) and (b). The columns labeled by 
  LO parton and NLO parton correspond to the parton-level results
  for the \SISCone{} algorithm.  The central scale used for one or
  more jets is $\mu=\hat{H}_T/2$.  
  The numerical integration uncertainties are in parentheses and the 
  scale dependence in super- and subscripts. 
}
\label{Table-Zjets-D0-total-xs-HT} 
\end{table}

The D0 collaboration has studied jet \pt{} distributions
in inclusive \Zgamjn-jet production for up to three jets~\cite{ZD0},
using the D0 Run II midpoint jet algorithm.  Here we present the
corresponding NLO parton-level results.  To compare NLO theory and
experiment we again need to account for non-perturbative corrections.
D0 has provided estimates of non-perturbative corrections due to the
underlying event and hadronization effects for their study using the
lepton cuts (a) of \eqn{D0LeptonAnalysisA}. With the smaller cone size
used by D0, $R=0.5$, the net non-perturbative corrections turn out to
be no larger than about 15\%, significantly smaller than in the CDF
analysis for $R=0.7$.  As described further below, we will use these
correction factors as a rough estimate of the non-perturbative
corrections for selection (b) as well.

\subsection{Total cross section}

The theoretical predictions for the total cross sections for 
selection 
(a), with lepton cuts~(\ref{D0LeptonAnalysisA}), and
for selection  (b), with lepton cuts~(\ref{D0LeptonAnalysisB}),
are given in \tab{Table-Zjets-D0-total-xs-HT}.  The LO and NLO parton-level
cross-sections are for the \SISCone{} algorithm, with the central
scale choice $\mu = \HTpartonic/2$, and the scale dependence determined
as before.  For the case of \Zgamz-jet production we use
$\mu = E_T^Z$, because  $\mu = \HTpartonic/2$ can vanish.
As seen from \tab{Table-Zjets-D0-total-xs-HT}, for
\Zgamjjja-jet production the LO scale dependence is quite large, but
is substantially reduced at NLO. In particular, for \Zgamjjj-jet
production a shift in the scale by a factor of two causes a variation of
up to 60\% at LO and under 20\% at NLO.  For the case of
\Zgamz-jet production, the scale dependence of the LO result
is anomalously small~\cite{ADMP}.  This scale independence is due to
the absence of factors of the strong coupling $\alpha_s$
in the LO matrix elements, along with the mild dependence of the quark
distribution functions $q(x,\mu_F)$ on $\mu_F$ at values of $x$
that are relevant at the Tevatron.\footnote{The smaller scale
variation of inclusive \Zgam{} production at LO, with respect to NLO,
does not, of course, imply that the LO prediction is more accurate;
indeed, the NLO result is much closer to the NNLO one~\cite{ADMP}.}

In ref.~\cite{ZD0}, D0 provided jet \pt{} distributions rather than summed
cross sections.  We could, of course, use the differential measurement
to obtain the total cross section for \Zgamjjja-jet production (as
a fraction of the inclusive $Z$-boson production cross section), but as
the systematic error correlations were not specified, we instead turn
to a direct comparison of theory and experiment for the \pt{}
distributions.

\subsection{Comparison to D0 jet $p_T$ distributions}

We now compare our predictions for jet \pt{} distributions to the D0
measurements.  From our vantage point, selection  (a) 
allows a direct comparison of our NLO
predictions to the plots presented in ref.~\cite{ZD0}, which also
include results from various LO and parton-shower based programs.  On
the other hand, as noted earlier, in selection  (a) more than
half the events are extrapolated from the actual measurement using
the cuts~(\ref{D0LeptonAnalysisB}) of selection~(b).
This extrapolation introduces additional uncertainty
into the measurement.  Thus selection~(b) would normally be preferred
for comparison to theory.  However, because the non-perturbative
corrections determined in ref.~\cite{ZD0} were for selection~(a),
they do not directly apply to selection~(b).  
Selection (b) corresponds
to a subset of (a), which could have somewhat different average values
for the $Z$ boson rapidity and transverse momenta; these values could
in turn affect the jet kinematics and therefore the non-perturbative
corrections.  Because the correlation between lepton cuts
and jet kinematics is at second order, and because the quoted 
non-perturbative corrections for (a) are no larger than about 15\%,
one may hope that the corrections for (b) are not too different.
For either selection, we face the same issue as with CDF, that the
non-perturbative corrections were estimated for a midpoint jet
algorithm rather than \SISCone.
In light of these issues, we
present NLO parton-level results with the \SISCone{} algorithm as our
primary predictions.  We then adopt the non-perturbative corrections
given in ref.~\cite{ZD0} for both selections~(a) and~(b),
leaving possible (significant) improvements to future studies.

In the comparison, we follow D0 in normalizing the \Zgamjjja-jet
\pt{} distributions by dividing by the inclusive \Zgamz-jet
cross section.  The latter cross section is defined using the same set
of lepton cuts as applied in the numerator, for both the 
(a) and (b) selections.  When assessing the scale dependence,
we vary the scale by a factor of two in each direction in the numerator,
but in the denominator for simplicity we always take the central-scale 
values in the first row of \tab{Table-Zjets-D0-total-xs-HT}.
Because both the LO and NLO inclusive \Zgam{} cross
sections vary by under 2\% from their central values, this procedure
modifies the scale variation band only slightly, with respect to
varying scales in the denominator as well.

%%%%%%%%%%%%%%%%%%%%%%%%%%% FIGURE %%%%%%%%%%%%%%%%%%%%%%%%%%%%%%%%
\begin{figure}[tbh]
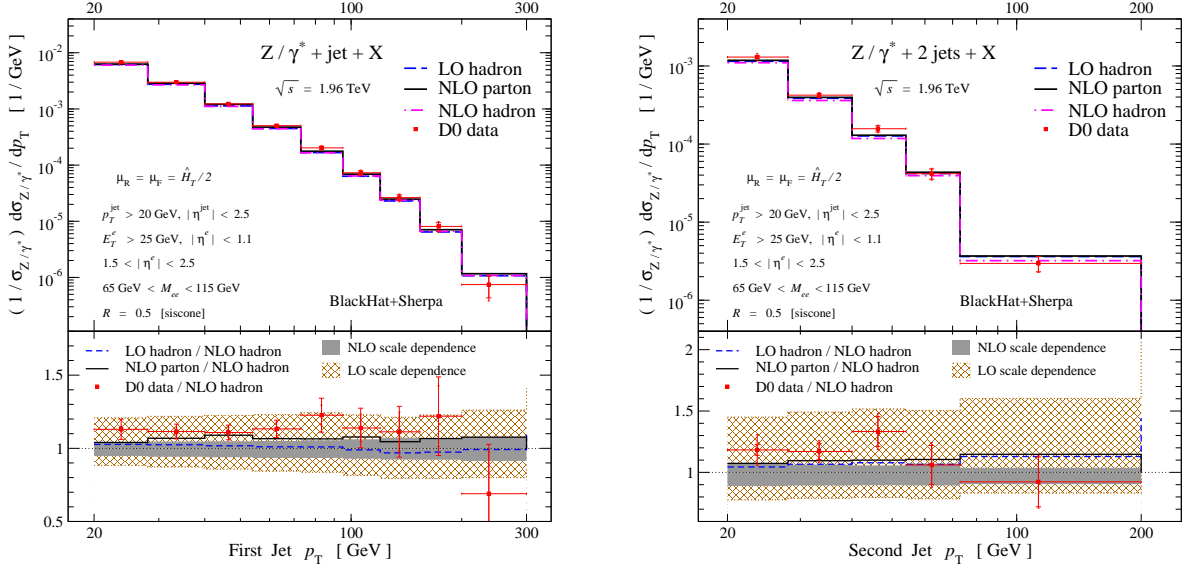

\begin{center}
\begin{minipage}[b]{1.\linewidth}
\includegraphics[clip,scale=0.35]{figs/Z1j-D0-HT_siscone-Pt20_jets_blc_jet_1_1_pt_1_with_D0_data.eps}
\hskip 1 cm 
\includegraphics[clip,scale=0.35]{figs/Z2j-D0-HT_siscone-Pt20_jets_blc_jet_1_2_pt_2_with_D0_data.eps}
\end{minipage}
\end{center}
\caption{Normalized jet \pt{} distributions for D0 
  selection
  (b).  The left plot shows $1/\sigma_{Z,\gamma^*}
  \times d \sigma/d p_T$ for the leading jet in \Zgamj-jet production.  The
  right plot shows the distribution for the second-hardest jet in
  \Zgamjj-jet production.  The \SISCone{} algorithm and scale
  choice $\mu=\hat{H}_T/2$ are used in the theoretical predictions.  In
  the upper panels the parton-level NLO distributions are the solid
  (black) histograms, while the NLO distributions corrected to hadron
  level are given by dash-dot (magenta) histograms.  
  The D0 data is indicated
  by the (red) points; the inner and outer error bars 
  denote respectively the statistical and total experimental 
  uncertainties on the measurements.  The LO predictions corrected
  to hadron level are shown as dashed (blue) lines.
  Each lower panel shows the
  distribution normalized to the full hadron-level NLO prediction. The
  scale-dependence bands in the lower panels are shaded (gray) for NLO
  and cross-hatched (brown) for LO.  }
\label{Figure-Z12jet-D0data-siscone-HT}
\end{figure}
%%%%%%%%%%%%%%%%%%%%%%%%%%%%%%%%%%%%%%%%%%%%%%%%%%%%%%%%%%%%%%%%%%%%%5

We have compared D0 data for \Zgamjjja-jet production to our NLO
prediction for both selections~(a) and~(b).  Ref.~\cite{ZD0} already
showed a comparison between selection~(a) for inclusive \Zgamjx-jet production
and NLO predictions using \MCFM~\cite{MCFM}, so for these processes,
we show instead a comparison to selection~(b).  The comparison
is displayed in \fig{Figure-Z12jet-D0data-siscone-HT}.  
As explained above, we use tables~IV and V of ref.~\cite{ZD0} to
convert our parton-level results to hadron-level ones.
Ref.~\cite{ZD0} studied the effects of parton distribution
function uncertainties on the perturbative predictions
for selection~(a), and found them to be 5--10\% for the leading two jets,
and 5--15\% for the third jet.
We have compared our results for both selections~(a) and~(b) 
to \MCFM, using the $k_T$ algorithm with scale choice
$\mu = E_T^Z$; we find agreement, with the total cross section
agreeing to better than 0.5\%.  (As was the case for the CDF study, 
the ratio of data to NLO presented by D0 differs by up to 10\% from our
corresponding results, though for selection (b), 
in \fig{Figure-Z12jet-D0data-siscone-HT}; again the difference is largely
due to the different choice of central scale in the NLO result,
$\mu=E_T^Z$ versus $\mu = \HTpartonic/2$.)

%%%%%%%%%%%%%%%%%%%%%%%%%%% FIGURE %%%%%%%%%%%%%%%%%%%%%%%%%%%%%%%%
\begin{figure}[tbh]
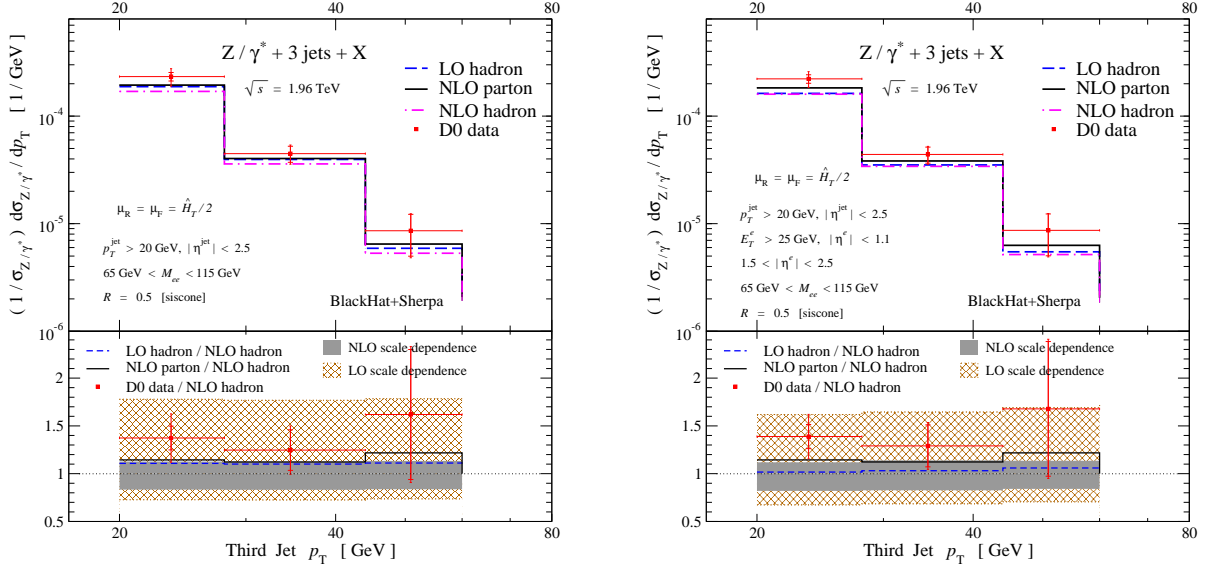

\begin{center}
\begin{minipage}[b]{1.\linewidth}
\includegraphics[clip,scale=0.35]{figs/Z3j-D0-HT_siscone-Pt20_jets_jet_1_3_pt_3_with_D0_data.eps}
\hskip 1 cm 
\includegraphics[clip,scale=0.35]{figs/Z3j-D0-HT_siscone-Pt20_jets_blc_jet_1_3_pt_3_with_D0_data.eps}
\end{minipage}
\end{center}
\caption{ Comparison of NLO theory to the D0 result for
the distribution of
$1/\sigma_{Z,\gamma^*} \times d \sigma/d p_T$ for the third-hardest
jet in inclusive \Zgamjjj-jet production. The left plot shows the
comparison for selection~(a), and the right plot for selection~(b).
The quoted~\cite{ZD0} non-perturbative corrections for selection~(a)
were used as estimated for both selections. The \SISCone{} jet algorithm
and the scale $\mu=\hat{H}_T/2$ were used in the theoretical predictions.
The labeling is as in \fig{Figure-Z12jet-D0data-siscone-HT}. }
\label{Figure-Z3jet-D0data-siscone-HTAB}
\end{figure}
%%%%%%%%%%%%%%%%%%%%%%%%%%%%%%%%%%%%%%%%%%%%%%%%%%%%%%%%%%%%%%%%%%%%%5

%%%%%%%%%%%%%%%%%%%%%%% TABLE  Analysis (a) %%%%%%%%%%%%%%
\begin{table}[ht]
%\hspace{-2.5cm}
\begin{tabular}{|c|c||c||c|c||c|}
\hline
\pt{} bin & \, LO parton \, & \, NLO parton \, & \, LO hadron \, & 
             \, NLO hadron \, & \, D0 \,   \\[-.30cm] 
 [GeV] \,   & [$10^{-6}$/GeV] & [$10^{-6}$/GeV] & [$10^{-6}$/GeV]  & [$10^{-6}$/GeV] & [$10^{-6}$/GeV]\\
\hline\hline
20-28 
& $\,  215(0.3)^{+130}_{-74} \,$ 
& $\,  194(1.5)^{+25}_{-32} \,$
& $\,  188(0.4)^{+114}_{-65} \,$ 
& $\,  170(1.3)^{+22}_{-28}  \,$
& $\,  233 \pm  21 \pm 37 \,$ 
 \\ 
\hline
28-44
   & $\, 44.4(0.06)^{+26.9}_{-15.1} \,$
   & $\, 40.3(0.2)^{+5.0}_{ -1.05} \,$
   & $\, 39.5(0.05)^{+24.0}_{-13.5} \,$ 
   & $\, 35.9(0.2)^{+4.3}_{-0.9} \,$ 
   & $\, 44.8 \pm  7.6 \pm  4.9  \, $
  \\
\hline
44-60    & $ \, 7.19(0.02)^{+4.35}_{-2.43} \,$
   & $\, 6.47(0.09)^{+0.86}_{-0.11} \,$
   & $\, 5.90(0.01)^{+3.57}_{-1.99} \,$ 
   & $\, 5.31(0.08)^{+0.71}_{-0.86} \,$  
   & $\, 8.60 \pm 3.61 \pm  1.12 \, \,$
 \\
\hline
\end{tabular}
\caption{\Zgamjjj-jet production for D0 selection (a),
$1/\sigma_{Z,\gamma^*} \times d\sigma/d p_T$ for the third jet
(\pt{} ordered) in inclusive \Zgamjjj-jet production, 
with the lepton cuts of \eqn{D0LeptonAnalysisA}.  The LO and NLO
hadron columns include hadronization and underlying-event corrections
from ref.~\cite{ZD0}.  The hadron-level columns do not include
the uncertainties from non-perturbative corrections; estimates
of these may be found in table~VI of ref.~\cite{ZD0}.
}
\label{Table-Z3jets-D0-PTdist-HT-a} 
\end{table}

%%%%%%%%%%%%%%%%%%%%%%% TABLE Analysis (b) %%%%%%%%%%%%%%
\begin{table}[ht]
%\hspace{-2.5cm}
\begin{tabular}{|c||c|c||c|c||c|}
\hline
\pt{} bin  &\,  LO parton\,  &\,  NLO parton\,  & \, LO hadron \, & \, NLO hadron \, & \,D0 \, \\[-.30cm] 
 [GeV] \,   & [$10^{-6}$/GeV] & [$10^{-6}$/GeV] & [$10^{-6}$/GeV]  & [$10^{-6}$/GeV]  & [$10^{-6}$/GeV] \\
\hline
20-28 
& $\,  186(0.3)^{+23}_{-63}\, $ 
& $\,  183(1.3)^{+22}_{-32}\, $ 
& $\,  162(0.3)^{+21}_{-55}\, $ 
& $\,  160(1.1)^{+19}_{-29}\, $ 
& $\,  222 \pm 20 \pm 31\, $ 
\\ 
\hline
28-44
   & $\, 39.4(0.07)^{+23}_{-13} \,$ 
   & $\, 38.3(0.3)^{+5.5}_{-6.7} \,$ 
   & $\, 35.2(0.07)^{+20.9}_{-11.8} \,$ 
   & $\, 34.1(0.3)^{+6.6}_{-5.9} \,$ 
   & $\, 44.0 \pm   7.5 \pm 3.7 \,$
 \\
\hline
44-60 
   & $\, 6.66(0.02)^{+3.97}_{-2.23} \,$ 
   & $\, 6.29(0.09)^{+ 0.85}_{-1.00} \,$ 
   & $\, 5.47(0.02)^{+3.26}_{-1.83} \,$
   & $\, 5.17(0.08)^{+0.70}_{-0.82} \,$  
   & $\, 8.67 \pm 3.64  \pm  0.95 \,$
 \\
\hline
\end{tabular}
\caption{\Zgamjjj-jet production. $1/\sigma_{Z,\gamma^*} \times d
\sigma/d p_T$ for the third jet (\pt{} ordered) in inclusive
\Zgamjjj-jet production.  The setup is the same as in
\tab{Table-Z3jets-D0-PTdist-HT-a}, except the additional lepton cuts
given in \eqn{D0LeptonAnalysisB} for selection (b) are imposed.  In
the columns labeled by ``hadron'' we again multiplied by the 
non-perturbative corrections (computed for (a)) from Table VI of 
ref.~\cite{ZD0}. }
\label{Table-Z3jets-D0-PTdist-HT-b} 
\end{table}

\Fig{Figure-Z3jet-D0data-siscone-HTAB} compares our theoretical
predictions for the third-jet \pt{} distribution in \Zgamjjj-jet
production to both selections~(a) and~(b).
As expected, the scale dependence of the NLO predictions is quite
a bit smaller than for LO, throughout the distribution, with only
a 15\% deviation from the central value.  
For reference, we also present the results shown in
\fig{Figure-Z3jet-D0data-siscone-HTAB} in tabular form,
in \tabs{Table-Z3jets-D0-PTdist-HT-a}{Table-Z3jets-D0-PTdist-HT-b}.
The columns labeled ``LO parton'' and ``NLO parton''
give the parton-level results.  These are the primary results of
our D0 study, and would be the key input to any future analyses. 
The columns labeled  ``LO/NLO hadron'' give these predictions
multiplied by the non-perturbative corrections given in table~VI
of ref.~\cite{ZD0} and represent a rough estimate.
As in \fig{Figure-Z3jet-D0data-siscone-HTAB} we have
not included the uncertainties in the non-perturbative corrections.
Finally, the column labeled ``D0'' gives the D0 measurement, followed by
its statistical and systematic uncertainties. 
\Fig{Figure-Z3jet-D0data-siscone-HTAB} and
\tabs{Table-Z3jets-D0-PTdist-HT-a}{Table-Z3jets-D0-PTdist-HT-b}
show that, although the experimental central values are always
a bit above the theoretical bands, the agreement between theory 
and experiment is reasonable, given the experimental statistical
uncertainties, and despite the various unquantified uncertainties
discussed above.  For both theory and experiment, the shift in values
between selections (a) and (b) is quite
small, significantly smaller than the respective uncertainties.

%%%%%%%%%%%%%%%%%%%%%%%%%%%%%%%%%%%%%%%%%%%%%%%%%%%%%%%%%%%%%%%%%%%%%%%%%%%%

\section{Jet-Production Ratios}
\label{JetProductionRatioSection}

The measurement of jet cross sections is sensitive to a number of
experimental and non-perturbative issues, in particular the measurement
of the jet energy and contributions due to the underlying event.
The latter is not
modeled in perturbative predictions, and accordingly introduces a
systematic uncertainty.  Jet \pt{} distributions fall rapidly: the
distribution of the third-hardest jet in \Zjjj-jet production
in \fig{Figure-Z3jets-jets-all-pt-siscone-HT}, for example,
falls by two orders of magnitude over a factor-of-two range in \pt.
Thus small errors in \pt{} measurements or non-perturbative shifts in
jet \pt{} can have important effects on distributions.
Furthermore, the cross section for an additional jet
is roughly an order of magnitude smaller, so a misidentification of an
$(n-1)$-jet process as an $n$-jet process can cause a significant
error in the measured cross section for the process with the extra
jet.  These difficulties increase as the number of jets accompanying a
vector boson grows.

%%%%%%%%%%%%%%%%%%%%% RATIO TABLE CDF SISCone %%%%%%%%%%%%%%%%%%%%%%%%%%
\begin{table}[tbh]
%\hspace{-2.5cm}
\begin{tabular}{|c||c||c|c|}
\hline
jet ratio   
 &   CDF
 &   LO
 &   NLO
 \\
\hline
2/1    & $ 0.099 \pm 0.012$
       & $  0.093^{+0.015}_{-0.012}$
       & $  0.093^{+ 0.004}_{-0.006}$
  \\
\hline
3/2    & $ 0.086 \pm 0.021 $
       & $ 0.057^{+0.008 }_{-0.006} $
       & $ 0.065^{+0.008}_{-0.007}$
  \\
\hline
4/3    & ---
       & $ 0.040^{+0.005}_{-0.004}$
       & --- 
  \\
\hline
\end{tabular}
\caption{ The ratios of the \Zgamjn-jet to \Zgamjnm1-jet
  cross sections for CDF's cuts, using the \SISCone{} algorithm and
  $\mu = \HTpartonic/2$.  For the
  experimental uncertainties we removed the luminosity errors, as
  these cancel; we treated all remaining uncertainties as uncorrelated. For
  the theoretical ratios we varied the scale in the same way in the 
  numerator and the denominator.
  The integration uncertainties are small compared to the 
  remaining scale dependence.}
\label{CDFRatioTable} 
\end{table}
%%%%%%%%%%%%%%%%%%%%%%%%%%%%%%%%%%%%%%%%%%%%%%%%%%%%%%%%%%%%%%%%%%

%%%%%%%%%%%%%%%%%%%%% RATIO TABLE D0 SISCone %%%%%%%%%%%%%%%%%%%%%%%%%%%%
\begin{table}[tbh]
%\hspace{-2.5cm}
\begin{tabular}{|c||c|| c|c|}
\hline
jet ratio   
 &   D0
 &   LO
 &   NLO
 \\
\hline
2/1    & $  0.151 \pm  0.005$
       & $  0.153^{+0.026}_{-0.020}$
       & $  0.147^{+0.003}_{-0.008} $
  \\
\hline
3/2    & $ 0.139 \pm 0.012$
       & $ 0.124^{+0.018}_{-0.013} $ 
       & $ 0.128^{+0.007}_{-0.010} $
  \\
\hline
4/3    & ---
       & $0.104^{+0.013}_{-0.010} $
       & ---
  \\
\hline
\end{tabular}
\caption{ The ratios of the \Zgamjn-jet to \Zgamjnm1-jet cross
  sections for D0's selection (b), using the \SISCone{} algorithm and
  $\mu = \HTpartonic/2$.  Here
  we keep only the experimental statistical uncertainties and drop the
  systematic ones.  The scale dependence is treated as in
  the previous table. }
\label{D0RatioTable} 
\end{table}
%%%%%%%%%%%%%%%%%%%%%%%%%%%%%%%%%%%%%%%%%%%%%%%%%%%%%%%%%%%%%%%%%%

A simple way to control some of the systematic uncertainties is to
consider instead ratios of cross
sections~\cite{BerendsRatio,AbouzaidFrisch, OtherBerendsRatio}.
As an example, consider the
jet-production ratio (also known as the ``Berends'' or ``staircase''
ratio), the ratio of \Zgamjn- to \Zgamjnm1-jet production, displayed
in \tab{CDFRatioTable} for the CDF cuts (\ref{CDFCuts}) and in
\tab{D0RatioTable} for D0 type (b) cuts (\ref{D0LeptonAnalysisB}).
We expect such ratios to be much less sensitive to
experimental systematic measurement uncertainties than the individual
cross sections.\footnote{Note that the jet-energy scale uncertainty
will not completely cancel, however, because the $p_T$ distribution
in the $n^{\rm th}$ jet is steeper than that in the $(n-1)^{\rm st}$ jet.}
Ratios also mitigate various theoretical uncertainties,
such as uncertainties in the non-perturbative corrections.
The theoretical values in \tab{CDFRatioTable} and \tab{D0RatioTable}
are given at parton level.  In assessing the scale dependence of the 
ratio, we varied the renormalization 
and factorization scales (by a factor of two) in the same way for both
numerator and denominator. (Of course, one could consider alternative
schemes to estimate scale dependence.)
Lacking knowledge of the correlations between the experimental
systematic uncertainties, we took them to be uncorrelated.
(We also took the larger of the upper and
lower uncertainties, as they are quite close for all cases.)  
Both the NLO and LO results are compatible with the experimental
ratios, within the estimated uncertainties.  It would be
interesting to re-examine these ratios using the larger data set 
collected more recently at the Tevatron, and properly incorporating all
correlations in the experimental systematic uncertainties.

Previous studies have noted that the jet-production ratios are roughly
independent of the base number of jets.  
With the choice of cuts
used by CDF, \eqn{CDFCuts}, our NLO theoretical results for $n=2,3$ 
are similar, within 30\% of each other.  With D0's choice
of cuts, the agreement is better, to within 15\%.  The predicted
ratio for $n=4$ is significantly lower, but this prediction is still
only at LO, because the NLO results for \Zgamjjjj-jet production are
not yet available.
\Tabs{CDFRatioTable}{D0RatioTable} also show that the NLO predictions
for the jet-production ratios are quite close to the LO central values,
but tend to have less scale variation.  

%%%%%%%%%%%%% FIGURE %%%%%%%%%%%%%%%%%%
\begin{figure}[tbh]
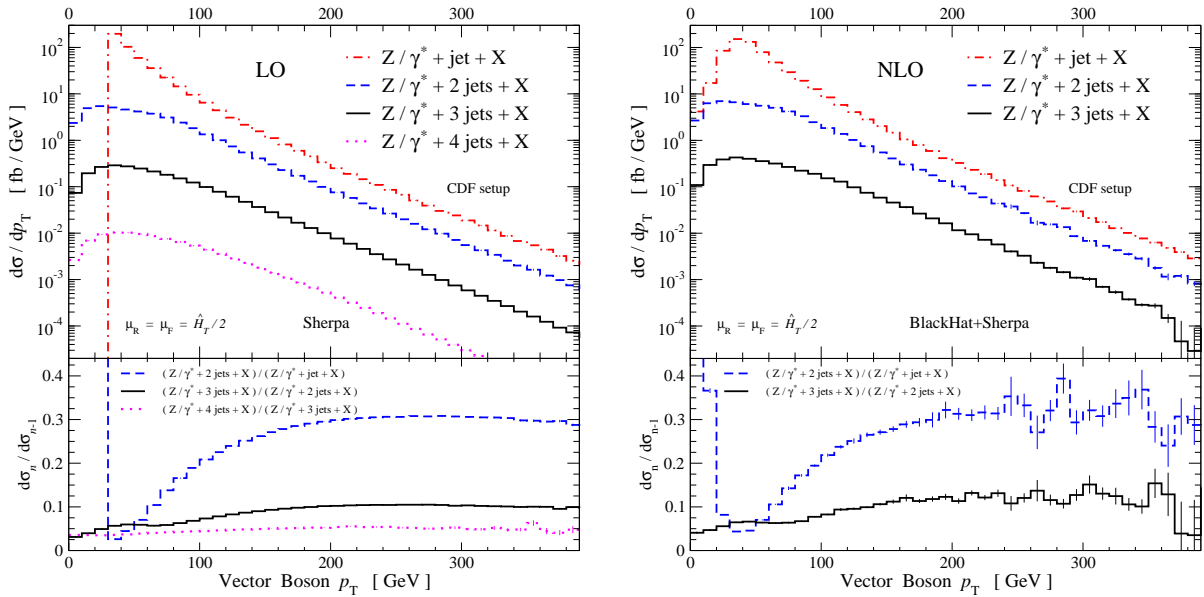

\begin{center}
\begin{minipage}[b]{1.\linewidth}
\includegraphics[clip,scale=0.37]{figs/Z1j-Z2j-Z3j-Z4j-LO-CDF-HT_siscone-Pt30_PTe-e+.eps} \hskip .5 cm 
\includegraphics[clip,scale=0.37]{figs/Z1j-Z2j-Z3j-NLO-CDF-HT_siscone-Pt30_PTe-e+.eps}
\end{minipage}
\end{center}
\caption{The LO and NLO vector-boson \pt{} distributions
for \Zgamjn-jet production at CDF. In the upper panels, the
top distribution is for one-jet inclusive production, the one
underneath it is for two-jet inclusive production, and the next one is for
three-jet inclusive production.  At LO, in the left plot the bottom
curve is for four-jet production.  The lower panel gives the jet-production 
ratios as a function of vector-boson \pt.  In the bottom
panel the upper curve is the 2/1-jet ratio and the one underneath it
the 3/2-jet ratio. The 4/3-jet ratio at LO is displayed as well
(magenta dotted curve).
}
\label{BerendsRatioZBosonPTFigure}
\end{figure}
%%%%%%%%%%%%%%%%%%%%%%%%%%%%%%%%%%%%%%%

%%%%%%%%%%%%% FIGURE %%%%%%%%%%%%%%%%%%
\begin{figure}[tbh]
\begin{center}
\includegraphics[clip,scale=0.5]{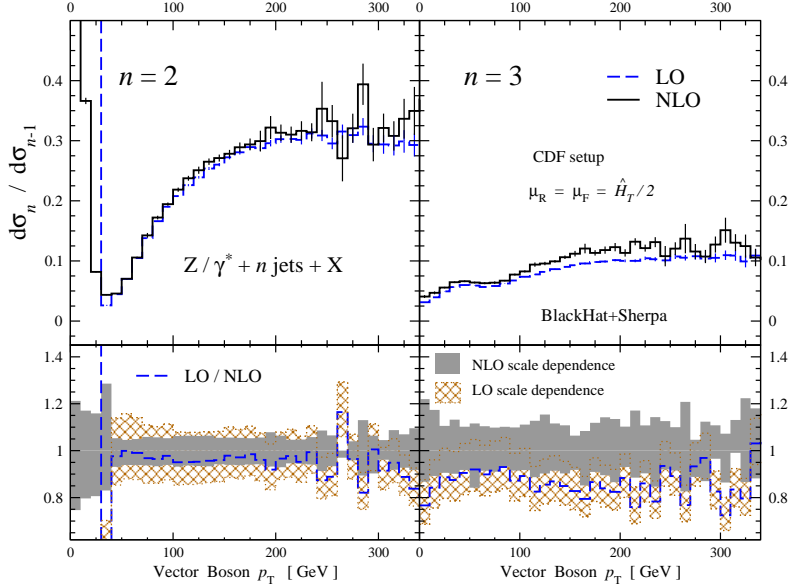} 
\end{center}
\caption{Comparison of jet-production ratios, differential in the
vector-boson \pt, at LO and NLO. The lower
panel shows these ratios, divided by the NLO ratio evaluated at 
the default central scale choice, and including scale variation bands
(computed as for the total cross section ratios). }
\label{RatioOfRatiosFigure}
\end{figure}
%%%%%%%%%%%%%%%%%%%%%%%%%%%%%%%%%%%%%%%

This approximate independence, however, hides a great deal of
variation in differential distributions.  In
\fig{BerendsRatioZBosonPTFigure}, we show the differential
distributions in the vector-boson transverse momentum ($\ptv$)
 for inclusive \Zgamjjjja-jet production
at LO (left panel) and for inclusive \Zgamjjja-jet production at NLO (right
panel).  The lower panes show the corresponding ratios:
(\Zgamjj-jet)/(\Zgamj-jet) (2/1), (\Zgamjjj-jet)/(\Zgamjj-jet) (3/2),
(\Zgamjjjj-jet)/(\Zgamjjj-jet) (4/3) at LO; 2/1 and 3/2 at NLO.  These
ratios have shapes that are quite stable in going from LO to NLO, as
shown in \fig{RatioOfRatiosFigure}, with the exception of the 2/1-jet
ratio at low \pt{} and small differences in the 3/2 ratio at
higher \pt.  In these plots, we use parton-level results, without
any corrections for hadronization or the underlying event.  We expect
substantial cancellations of these non-perturbative corrections
in the ratios.

This figure shows that the jet-production ratios depend strongly on
the \pt{} of the vector boson, and that the $3/2$-jet and $2/1$-jet
ratios are rather different beyond low \pt.  This means that their
putative independence of the base number of jets is illusory: in
reality, they depend sensitively on the cuts applied.  For example, a
\pt~$> 70$~GeV cut on the vector-boson transverse momentum would
result in a rather sizable difference between the 3/2-jet and 2/1-jet
ratios of total cross sections.

How does the non-trivial dependence of these ratios on $\ptv$ arise?
At LO, the 2/1-jet ratio is undefined (infinite) at low $\ptv$;
it rises smoothly from a very small value at $\ptv = \ptjetmin$,
where $\ptjetmin$ is the minimum jet $\ptv$ set by the experimental
cut (30~GeV in this case).  
In contrast, at NLO the 2/1-jet ratio rises to a finite value
as $\ptv\rightarrow 0$.
The other ratios have no structure at low \pt.  All of the
ratios rise noticeably beyond $\ptv$ of 70 GeV or so, a rise which
continues up to around~200 GeV, where the ratios flatten out or start to 
decline somewhat.

The behavior of the ratios at low \pt{} is relatively easy to understand.
For $\ptv$ below the minimal jet \pt, the vector's 
transverse momentum cannot be balanced by a lone parton, so the leading-order
contribution to \Zgamj-jet production vanishes.  Accordingly, the 2/1
ratio is infinite at LO for $\ptv<\ptjetmin$, and rises from a very
small value just above $\ptv=\ptjetmin$.
For \Zgamj-jet production at NLO, as $\ptv\to0$, the only contribution
is from real-emission configurations with two partons which
are roughly balanced in transverse momentum.  It is reasonably likely
that the second hardest parton also has $\ptv>\ptjetmin$.
Therefore the differential cross section in this region
is of the same order in $\alpha_s$, $\Ord(\alpha_s^2)$,
as the leading contribution to \Zgamjj-jet production.
Hence the NLO 2/1 ratio rises as $\ptv\to0$,
to a number independent of $\alpha_s$ (and ``of order unity'').
No such kinematic constraints arise in vector-boson production 
accompanied by more than one jet, even at LO, so the 3/2 and 4/3
ratios remain small as $\ptv\to0$.

What about the rise at larger vector-boson transverse momentum?  For a
given large $\ptv$, we expect that the matrix element is maximized
for an asymmetric configuration of jets, corresponding to a
near-singular configuration of the partons.  A typical configuration,
for example, would have one hard jet recoiling against the vector, and
additional jets (if any) with transverse momenta down near the \pt{}
cut.  In these configurations, the short-distance matrix element will
factorize into a matrix element for production of one hard gluon, and
a singular factor: either a splitting function in collinear limits, or
an eikonal one in soft limits.  The phase-space integrals over these
near-singular configurations give rise to potentially large
logarithms.  Because the minimum $\Delta R$ is relatively large,
collinear logarithms should not play an important role; on the other
hand, $\ptv/\ptjetmin$ can become large, so its logarithm will play a
role.

The approximate factorization suggests that we can model the 
differential cross sections shown in \fig{BerendsRatioZBosonPTFigure}
by the following forms,
\begin{eqnarray}
\sigma_1 &=& a_s f(\ptv)\,,\nn\\
\sigma_2 &=& a_s^2 \bigr(b_0 + b_1 \ln \rho \bigl) f(\ptv)
            (1-\ptv/\ptmax)^{\gamma_2}\,,\nn\\
\sigma_3 &=& a_s^3 \bigr(c_0 + c_1 \ln\rho  
                        +c_2 \ln^2\rho\bigl) f(\ptv)
            (1-\ptv/\ptmax)^{\gamma_3} \,,
\label{JetRatioModel}\\
\sigma_4 &=& a_s^4 \bigr(d_0 + d_1 \ln\rho
                        +d_2 \ln^2\rho
                        +d_3 \ln^3\rho
                    \bigl) f(\ptv)
            (1-\ptv/\ptmax)^{\gamma_4} \,.\nn
\end{eqnarray}
where $a_s \equiv \alpha_s(\ptv) N_c/(2 \pi)$, 
$\rho \equiv (\ptv/\ptjetmin)^2$ and
$\ptmax = 980$~GeV.  The additional factors of $(1-\ptv/\ptmax)^\gamma$
take into account the limits of the different-dimension phase spaces
and possibly different parton-distribution-function suppression in the
four cases.  The function $f(\ptv)$, which describes the 
overall rapidly-falling form of the distribution, will cancel in the
ratios, leaving us with three parameters for the 2/1 ratio, four
additional ones for the 3/2 ratio, and five further parameters
for the 4/3 ratio.

%%%%%%%%%%%%% FIGURE %%%%%%%%%%%%%%%%%%
\begin{figure}[tbh]
\begin{center}
\begin{minipage}[b]{1.1\linewidth}\leftskip -2mm
\includegraphics[clip,scale=0.8]{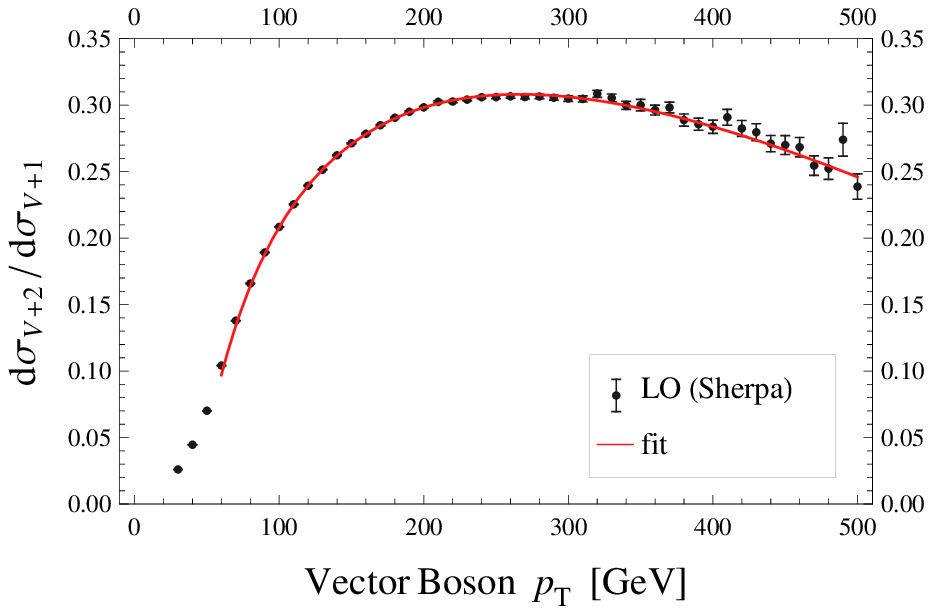} \hskip .5 cm 
\includegraphics[clip,scale=0.8]{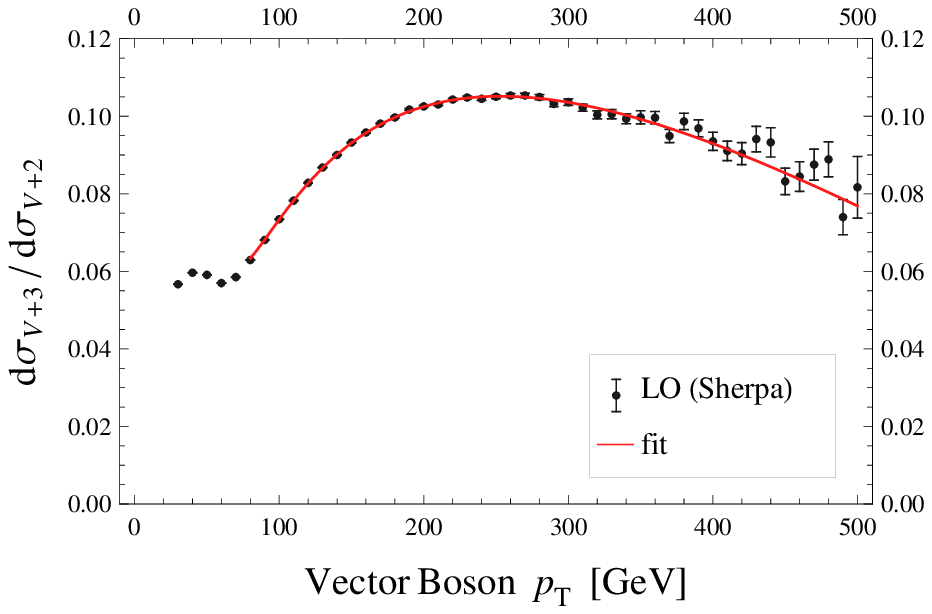} 
\end{minipage}
\includegraphics[clip,scale=0.8]{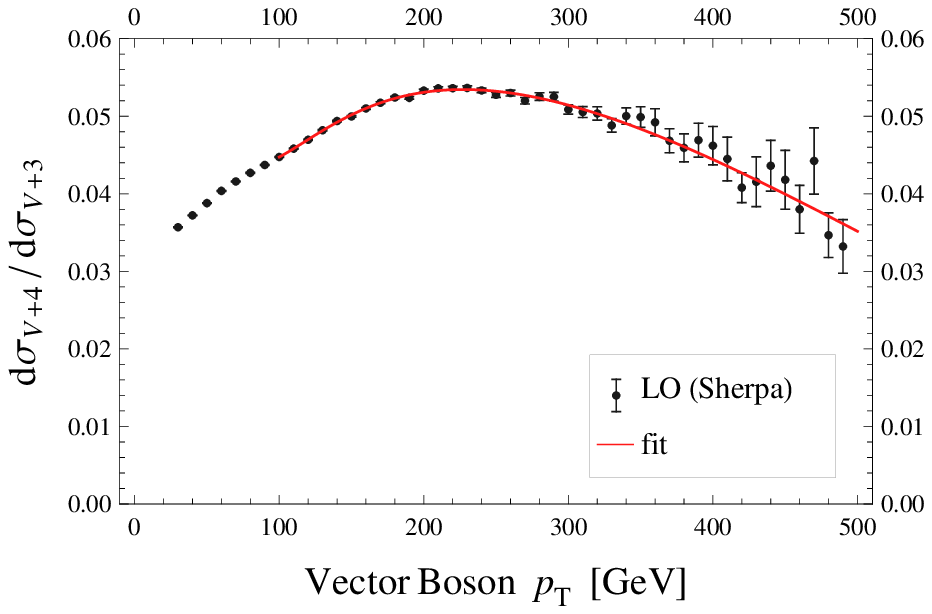} 
\end{center}
\caption{Fits of the LO jet-production ratios (using CDF cuts) 
as a function of
\pt{} to forms derived from \eqn{JetRatioModel}.  Top left, the 2/1 ratio;
top right, the 3/2 ratio; bottom, the 4/3 ratio.  The vertical scales
are different.
}
\label{BerendsRatiosFitFigure}
\end{figure}
%%%%%%%%%%%%%%%%%%%%%%%%%%%%%%%%%%%%%%%

We can determine these parameters for the CDF cuts
by fitting the ratios to our LO results (where much smaller
statistical uncertainties are easier to achieve).  We evaluate
the LO distributions at the central $\mu$ value, and fit in the range
$100~{\rm GeV}\le\ptv\le500~{\rm GeV}$.  The result is
\begin{eqnarray}
b_0 &=& -1.642\,,\qquad b_1 = 2.437\,,\qquad \gamma_2 = 1.118\,,\nn\\
c_0 &=& 5.081\,,\qquad c_1 = -5.812\,,\qquad c_2 = 2.658\,,\qquad 
 \gamma_3 = 2.539\,,\\
d_0 &=& -5.728\,,\qquad d_1 = 10.945\,,\qquad
 d_2 = -6.4\,,\qquad d_3 = 1.628\,,\qquad \gamma_4 = 4.195\,.\nn
\end{eqnarray}
The separate fits to the LO ratios are shown in
\fig{BerendsRatiosFitFigure}.
In spite of the limited number of parameters, the model gives an
excellent approximation to the LO ratios, even down to $\ptv$
of 70~GeV for the 2/1 ratio.  
(For the 2/1 ratio, the model's predictions are within 0.7\% of
the numerical results for 
two-thirds of the points in the fit range, and all points are within 
twice the Monte Carlo statistical error.  For the 3/2 ratio, two-thirds
of the points in the fit range are within 1\%.)
Although we did not attempt a fit to the NLO results for the 2/1 or 3/2
ratios, due to larger statistical uncertainties, \fig{RatioOfRatiosFigure}
indicates that the parameters entering the 2/1 ratio would be essentially
unchanged, and the 3/2 ratio parameters would only change modestly.
We stress that the model is a purely phenomenological one, with some
physically motivated input as to its form.  It is not intended to supply
a perfect fit, but it is remarkable that it does so well with so few
parameters.

The overall conclusion of this section is that the jet-production
ratio is {\it not\/} solely a measure of $\alpha_s$, but also depends
sensitively on kinematical ratios.

%%%%%%%%%%%%%%%%%%%%%%%%%%%%%%%%%%%%%%%%%%%%%%%%%%%%%%

\section{Conclusions}
\label{ConclusionSection}

In this paper we have presented the first full NLO results for 
\Zgamjjj-jet production at a hadron collider.  
(Ref.~\cite{Radcor09BH} contains a preliminary version of some of
these results, based on a leading-color approximation.)  
We chose to present results for the Tevatron, so that we could compare to
existing data from CDF and D0~\cite{ZCDF,ZD0}.  Our study should
also serve as a benchmark for future NLO studies of $Z$ boson 
production in association with multiple jets at the LHC.

In line with expectations, we found that NLO results for
\Zgamjjja-jet production are much less sensitive than their LO
counterparts to the choice of 
a common renormalization and factorization scale $\mu$.
The deviation in the cross section, when varying $\mu$ around our
default central scale choice by a factor of two, drops from 60\%
at LO to 15--22\% at NLO.  As noted in earlier 
studies~\cite{Bauer,W3jDistributions}, the
commonly-used scale choice of the vector-boson transverse energy is not 
particularly good at LHC energies, as it leads to large shape changes in
distributions between LO and NLO. At LHC energies, this choice can
even lead to negative cross sections at NLO, in
tails of various distributions~\cite{W3jDistributions}.
As we have shown in the present paper, this choice is also a poor one
for higher-multiplicity \Zgamjn-jet production,
even at Tevatron energies, because it generically results in large
shape changes from LO to NLO.  We used half the total transverse energy,
$\mu = \HTpartonic/2$, as our default scale choice instead.

The NLO results presented in this paper are parton-level results. To
compare to the measurements, non-perturbative corrections need to be
taken into account.
We used the non-perturbative corrections
tabulated by CDF and D0 to estimate the corrections to our parton-level
predictions.  These corrections were computed for different cone algorithms,
and in the case of D0 selection (b), for a slightly different setup.
We appealed to inclusive-jet studies~\cite{SISCONE,EHKetal} of algorithm
differences to suggest that applying these corrections is reasonable;
clearly, further study of this issue is desirable.
Applying the corrections, we found good agreement between the NLO results
and the CDF and D0 measurements of cross sections and jet \pt{}
distributions.  For \Zgamjjj-jet production, both CDF and D0 data
lie somewhat higher than the theoretical scale-dependence band, although
the statistical uncertainties are still large.
Obviously, from the perspective of comparing to
NLO results, it is helpful to choose infrared-safe jet algorithms,
as well as cone sizes for which the non-perturbative corrections
and their associated uncertainties are small.

For the cuts used by CDF, we presented results for two algorithms,
\SISCone{} and anti-$k_T$.  The differences in cross sections for
the two algorithms can be explained qualitatively by the larger effective
cone size of \SISCone~\cite{KTES,DMS,Jetography}, for a given cone-size
parameter $R$.

We presented new results on jet-production ratios.  We confirmed that
for total cross sections the ratio of $n$-jet production to
$(n-1)$-jet production is roughly
constant~\cite{BerendsRatio,AbouzaidFrisch}.  For the cuts used by
CDF, the NLO prediction gives ratios for $n=2,3$ that are within about
30\% of each other, whereas for the cuts used by D0 the corresponding
ratios are within about 15\%.  
(We note that the ratios for D0's setup are about 50\% higher than 
for CDF's setup.)
In both cases, the NLO corrections to
the LO ratios are quite small, 12\% or less.  This suggests that the
LO prediction of the ratio should also be fairly reliable for $n=4$,
though confirmation awaits an NLO computation.  In this case, however,
the predicted ratio for $n=4$ would lie significantly below the ratios
for $n=2,3$.  We expect that non-perturbative effects will partly
cancel in the ratios, making them theoretically more robust.  Although
the jet-production ratios can be used to give a rough estimate of
higher-multiplicity jet total cross sections, for differential cross
sections there is a strong dependence on the cuts and the number of
jets.  In particular, we found that the ratios depend strongly on the
vector-boson \pt.  These ratios of distributions have generic features
which we described using a simple model capturing leading logarithms
along with phase-space and parton-distribution-function suppression
factors.  The fits based on this model are surprisingly good and offer
a simple parametrization of the theoretical predictions.

Eventually, we would like to match the NLO results to parton showers
and hadronization models, allowing non-perturbative effects to be
modeled directly in an NLO program, instead of relying on LO-based
tools to model them.  This has been done for a variety of processes
within the {\sc MC@NLO} program~\cite{MCNLO} and the {\sc POWHEG}
method~\cite{POWHEG}.  It would be desirable to extend this matching to
higher-multiplicity processes such as those presented in this
present paper.  A first step in this direction, linking \BlackHat{} 
to an automated implementation of the FKS subtraction formalism used
in {\sc MC@NLO}, has been reported
recently~\cite{RikkertRadcor,LesHouches2009}.

With the on-shell methods as implemented in \BlackHat{}, we expect the
computation of 
virtual corrections to cease presenting a bottleneck to obtaining new
NLO results.  A publicly available version of \BlackHat{} is in
preparation and is currently being tested in a variety of projects
(see {\it e.g.}~ref.~\cite{RikkertRadcor}).  This version uses the
proposed Binoth Les Houches interface for one-loop matrix
elements~\cite{BinothInterface}. It has been tested with both C++ and
Fortran clients.  We intend the public version to provide all
processes that have been carefully tested with the full \BlackHat{}
code.  The dipole-subtraction implementation that we used is
available in the latest release of \SHERPA~\cite{SHERPAWebPage}.

NLO results will provide unprecedented precision for studies of
vector bosons in association with many jets.  They should prove useful
for experimentally-driven determinations of backgrounds such as the
invisible $Z\to\nu\nub$ background to missing-energy-plus-jets
signatures. By measuring the corresponding $Z\to l^+ l^-$ or $W\to l\nu$
processes, with NLO predictions for cross-section ratios
providing the necessary conversion factors, the $Z\to\nu\nub$
background can be determined
precisely~\cite{CMSWZRatioNote,ATLASZWRatioNote}.
The results presented in this paper are
examples of the predictions that can be obtained using \BlackHat{}
in conjunction with \SHERPA.  We look forward to applying these tools
to a wide range of studies of the forthcoming LHC data.

\section*{Acknowledgments}

\vskip -.3 cm 
We thank Joey Huston, Sabine Lammers, Michael Peskin,
Gavin Salam, Peter Skands and Rainer Wallny for helpful
conversations, and in particular Henrik Nilsen for useful
discussions and comments on the manuscript. 
This research was supported by the US Department of Energy under
contracts DE--FG03--91ER40662, DE--AC02--76SF00515 and
DE--FC02--94ER40818.  DAK's research is supported by the European
Research Council under Advanced Investigator Grant ERC--AdG--228301.
HI's work is supported by a grant from the US
LHC Theory Initiative through NSF contract PHY-0705682.
This research used resources of Academic Technology Services at UCLA,
PhenoGrid using the GridPP infrastructure, and the National Energy
Research Scientific Computing Center, which is supported by the Office
of Science of the U.S. Department of Energy under Contract
No. DE--AC02--05CH11231. 

%%%%%%%%%%%%%%%%%%%%%%%%%%%%%%%%%%%%%%%%%%

\appendix

\section{Kinematics and Observables}
\label{KinematicsandObservables}

In this appendix we give our definitions of standard
kinematic variables used to characterize scattering
events. The angular separation of two objects (partons, jets
or leptons) is denoted by 
\begin{eqnarray}
\Delta R= \sqrt{(\Delta \phi)^2 + (\Delta y)^2}\,,
\label{DeltaRphiy}
\end{eqnarray}
with $\Delta\phi$ the difference in the azimuthal angles, and
$\Delta y$ the difference in the rapidities.  The rapidity is
defined to be
\begin{equation}
y = {1\over 2} \ln\left({E+ p_L} \over {E-p_L} \right)\,,
\end{equation}
where $E$ is the energy and $p_L$ is the component of the momentum
along the beam axis (the $z$ axis). The pseudorapidity $\eta$ is given by
\begin{equation}
\eta= -\ln\left(\tan \frac{\theta}{2}\right) = {1\over 2} 
\ln\left({|\vec p\,|+ p_L} \over {|\vec p\,|-p_L} \right)\,,
\end{equation}
where $\theta$ is the polar angle with respect to the beam axis.

Jets are formed using cluster or cone algorithms based on the angular
separation~(\ref{DeltaRphiy}).  It is also possible to use the
pseudorapidity in place of the rapidity.  We have checked, for the
production of \Zgamjjj{} jets at NLO, that using $\eta$ instead of $y$
for the anti-$k_T$ and $k_T$ algorithms with $R=0.5$ makes no
discernible difference on the cross section.

The transverse energies of massless outgoing partons and leptons,
$E_T=\sqrt{p_x^2+p_y^2}$, can be summed to give the total partonic
transverse energy, $\HTpartonic$, of the scattering process,
\begin{equation}
\HTpartonic = \sum_p E_T^p + E_T^{e^+} + E_T^{e^-} \,.
\label{PartonicHTdef}
\end{equation}
All final-state partons $p$ and leptons are included in $\HTpartonic$,
whether or not they are inside jets that pass the cuts.  As discussed
in \sect{ScaleDependenceSubsection}, the variable $\HTpartonic$
represents a good choice for the renormalization and factorization
scale of a given event.  Although the partonic version is not directly
measurable, for practical purposes as a scale choice, it is
essentially equivalent (and identical at LO) to the more usual
jet-based total transverse energy,
\begin{equation}
H_T = \sum_j E_{T,j}^\jet + E_T^{e^+} + E_T^{e^-} \,.
\label{PhysicalHTdef}
\end{equation}

The jet four-momenta are computed by summing the four-momenta
of all partons that are clustered into them,
\begin{equation}
p^\jet_\mu = \sum_{i\in \jet} p_{i\mu}\,.
\end{equation}
The jet transverse momentum is then defined in the usual way,
\begin{equation}
p_T^\jet = \sqrt{(p^\jet_x)^2 + (p^\jet_y)^2}\,.
\end{equation}
%

%%%%%%%%%%%%%%%%%%%%%%%%%%%%%%%%%%%%%%%%%%%%%%%%%%%%%%%%%%%%%%%%

\section{Squared Matrix Elements at One Point in Phase Space}
\label{MatrixElementAppendix}

As an aid to future implementation of \Zgamjjj-jet production in other
numerical codes, we present values of the one-loop virtual
corrections to the squared matrix elements, $\MNLO$, at one point in
phase space.  These contributions arise from the interference between the
tree and one-loop amplitudes, summed over all colors and helicities,
for $N_c=3$ and $n_f=5$ massless quark flavors.  As discussed in 
the text, we do not include the small effects from the top quark, or 
from axial or vectorial loop contributions (see \fig{LoopDiagramsVAFigure}).

In \tab{ME2Table} we present numerical values for three representative
subprocesses.  The other subprocesses are related to these by crossing
symmetry or by change of coupling constants.  In the second line of
\tab{ME2Table}, the presence of two identical quarks (after crossing
all particles into the final state) implies that amplitudes are
antisymmetrized under exchange of the two.

%%%%%%%%%%% TABLE %%%%%%%%%%%%%%%%%%%%%%%%%%
\begin{table}[htp]
\vskip .4 cm
\begin{tabular}{|c||c|c|r|}
\hline
$\hatMNLO$ &$1/\epsilon^2$&$1/\epsilon$ &
 $\epsilon^0 \hskip .95 cm $ \\\hline\hline
%ME2 distinct up/down-type quarks 4q1g2l 
$(1_{\bar u} 2_d \rightarrow 3_d 4_{\bar u}  5_g 6_{e^-} 7_{e^+})$  & $ -8.3333333333 $ & $ -32.3745606495 $ & $ 5.2374716277 $\\ \hline 
%ME2 ident up-quarks 4q1g2l 
$(1_{\bar u} 2_u \rightarrow 3_u 4_{\bar u} 5_g 6_{e^-} 7_{e^+})$  & $ -8.3333333333 $ & $ -32.5180902998 $ & $ 0.4387412994 $\\ \hline 
%ME2 2q3g2l up-quarks
$(1_{\bar u} 2_g \rightarrow 3_g 4_g 5_{\bar u} 6_{e^-} 7_{e^+})$  & $ -11.6666666667 $ & $ -42.3279266518 $ & $ -15.2326082853 $\\ \hline
\end{tabular}
\caption{
Numerical values of the normalized virtual correction to
the squared matrix elements,
$\hatMNLO$, at the phase-space point
given in \eqn{SevenPointKinematics}, for three basic
partonic subprocesses for \Zjjj-jet production at a hadron collider.
We give both the finite parts and the coefficients
of the poles in $\epsilon$.
}
\label{ME2Table}
\end{table}
%%%%%%%%%%%%%%%%%%%%%%%%%%%%%

We quote numerical results for the ultraviolet-renormalized virtual
corrections in the 't~Hooft-Veltman variant of dimensional
regularization~\cite{HV}.  The remaining singularities in the
dimensional regularization parameter $\epsilon = (4-D)/2$ arise from
the virtual soft and collinear singularities in the one-loop
amplitudes.

The quoted values are for the ratio of the virtual corrections to the
tree-level squared matrix element $\MLO$.  Explicitly, we define the
ratio,
\begin{equation}
  \hatMNLO
\equiv
\frac{1}{8\pi\alpha_S \, c_\Gamma(\e)}
\frac{\MNLO}{\MLO}\,,
  \label{ME2normalization}
\end{equation}
where we have also separated out the dependence on the
strong coupling $\alpha_S$ and the overall
factor $c_\Gamma(\e)$, defined by
\begin{equation}
c_\Gamma(\e)
= \frac{1}{(4\pi)^{2-\epsilon}}
\frac{\Gamma(1+\epsilon)\Gamma^2(1-\epsilon)}
 {\Gamma(1-2\epsilon)}\,.
  \label{cGammaDef}
\end{equation}
The coupling constants, mass and width of the $Z$ boson are given in
\sect{CouplingsSection}.  

We choose the phase-space point given in eqs. (9.3) and (9.4) of
ref.~\cite{Genhel},
\begin{eqnarray}
k_1 &=&  {\mu \over 2}\, (1, -\sin \theta,
           -\cos \theta \sin \phi, -\cos \theta \cos\phi)\,, \nn\\
k_2 &=&  {\mu \over 2}\, (1,  \sin \theta,
          \cos \theta \sin \phi,  \cos \theta \cos \phi)\,, \nn \\
k_3 &=&  {\mu\over 3} (1,1,0,0) \,,\nn \\
k_4 &=&  {\mu\over 8} (1, \cos\beta, \sin \beta,0) \,, \nn \\
k_5 &=&  {\mu \over 10} (1, \cos\alpha \cos\beta, \cos \alpha \sin \beta,
                                 \sin \alpha )\,, \nn \\
k_6 &=& {\mu \over 12} (1, \cos\gamma \cos\beta,
            \cos \gamma \sin \beta, \sin \gamma)\, \nn, \\
k_7 &=& k_1+k_2-k_3-k_4-k_5-k_6\,,
\label{SevenPointKinematics}
\end{eqnarray}
where
\begin{eqnarray}
 \theta = {\pi\over 4}\,,\hskip 1 cm
 \phi   =  {\pi\over 6}\,,\hskip 1 cm
 \alpha = {\pi \over 3}\,,\hskip 1 cm
 \gamma = {2 \pi \over 3} \,, \hskip 1 cm
 \cos \beta = - {37\over 128} \,,
\end{eqnarray}
and the renormalization scale $\mu_R$ is set to $\mu_R = 7$~GeV.
We have flipped the signs of $k_1$ and $k_2$ compared to
ref.~\cite{Genhel}, to correspond to $2\rightarrow 5$
kinematics, instead of $0\rightarrow 7$ kinematics.
The labeling of the parton and lepton momenta is indicated
explicitly in the first column of Table~\ref{ME2Table}.

%%%%%%%%%%%%%%%%%%%%%%%%%%%%%%%%%%%%%%%


\begin{thebibliography}{99}

%+% 1 ref
\bibitem{MET}
S.~D.~Ellis, R.~Kleiss and W.~J.~Stirling,
%``Missing Transverse Energy Events And The Standard Model,''
Phys.\ Lett.\  B {\bf 158}, 341 (1985);\\
%%CITATION = PHLTA,B158,341;%%
%
M.~L.~Mangano,
%``Standard Model backgrounds to supersymmetry searches,''
Eur.\ Phys.\ J.\  C {\bf 59}, 373 (2009)
[0809.1567 [hep-ph]].
%%CITATION = EPHJA,C59,373;%%

%+% 18 refs
\bibitem{ZCDF}
T.~Aaltonen {\em et al.} [CDF Collaboration],
%``Measurement of inclusive jet cross-sections in Z/gamma*(---> $e^{+} e^{-)}$
%+ jets production in $p \bar{p}$ collisions at $\sqrt{s}$ = 1.96-TeV,''
Phys.\ Rev.\ Lett.\  {\bf 100}, 102001 (2008)
[0711.3717 [hep-ex]].
%%CITATION = PRLTA,100,102001;%%

%+% 2 refs
\bibitem{Z1jD0}
V.~M.~Abazov {\it et al.}  [D0 Collaboration],
%``Measurement of differential $Z / \gamma^{*}$ + jet + $X$ cross sections in
%$p \bar{p}$ collisions at $\sqrt{s}$ = 1.96-TeV,''
Phys.\ Lett.\  B {\bf 669}, 278 (2008)
[0808.1296 [hep-ex]].
%%CITATION = PHLTA,B669,278;%%

%+% 25 refs
\bibitem{ZD0}
V.~M.~Abazov {\it et al.}  [D0 Collaboration],
%``Measurements of differential cross sections of $Z /\gamma^\ast$+jets+X
%events in proton anti-proton collisions at $\sqrt{s}$=1.96 TeV,''
Phys.\ Lett.\  B {\bf 678}, 45 (2009)
[0903.1748 [hep-ex]].
%%CITATION = PHLTA,B678,45;%%

%+% 2 refs
\bibitem{ZjanglesD0}
V.~M.~Abazov {\it et al.}  [D0 Collaboration],
%``Measurement of $Z / \gamma^\ast +jet+X$ angular distributions in $p
%\bar{p}$ collisions at $\sqrt{s}=1.96$ TeV,''
Phys.\ Lett.\  B {\bf 682}, 370 (2010)
[0907.4286 [hep-ex]].
%%CITATION = PHLTA,B682,370;%%

%+% 1 ref
\bibitem{LOPrograms}
T.~Stelzer and W.~F.~Long,
%``Automatic generation of tree level helicity amplitudes,''
Comput.\ Phys.\ Commun.\  {\bf 81}, 357 (1994)
[hep-ph/9401258];\\
%%CITATION = HEP-PH 9401258;%%
%
A.~Pukhov {\it et al.},
%``CompHEP: A package for evaluation of Feynman diagrams and integration  over
%multi-particle phase space. User's manual for version 33,''
hep-ph/9908288;\\
%%CITATION = HEP-PH/9908288;%%
%
M.~L.~Mangano, M.~Moretti, F.~Piccinini, R.~Pittau and A.~D.~Polosa,
%``ALPGEN, a generator for hard multiparton processes in hadronic
%collisions,''
JHEP {\bf 0307}, 001 (2003)
[hep-ph/0206293].
%%CITATION = JHEPA,0307,001;%%
%

%+% 1 ref
\bibitem{HELAC}
A.~Kanaki and C.~G.~Papadopoulos,
%``HELAC: A package to compute electroweak helicity amplitudes,''
Comput.\ Phys.\ Commun.\  {\bf 132}, 306 (2000)
[hep-ph/0002082].
%%CITATION = CPHCB,132,306;%%

%+% 3 refs
\bibitem{Amegic}
F.~Krauss, R.~Kuhn and G.~Soff,
%``AMEGIC++ 1.0: A Matrix element generator in C++,''
JHEP {\bf 0202}, 044 (2002)
[hep-ph/0109036].
%%CITATION = JHEPA,0202,044;%%

%+% 1 ref
\bibitem{Matching}
S.~Catani, F.~Krauss, R.~Kuhn and B.~R.~Webber,
%``QCD matrix elements + parton showers,''
JHEP {\bf 0111}, 063 (2001)
[hep-ph/0109231];\\
%%CITATION = JHEPA,0111,063;%%
M.~Mangano,
%``The so-called MLM prescription for ME/PS matching'' (2004),
presented at the Fermilab ME/MC TuningWorkshop, October 4,
2004.

%+% 1 ref
\bibitem{MLMSMPR}
M.~L.~Mangano, M.~Moretti, F.~Piccinini and M.~Treccani,
%``Matching matrix elements and shower evolution for top-quark production in
%hadronic collisions,''
JHEP {\bf 0701}, 013 (2007)
[hep-ph/0611129];\\
%%CITATION = JHEPA,0701,013;%%
%
S.~Mrenna and P.~Richardson,
%``Matching matrix elements and parton showers with HERWIG and PYTHIA,''
JHEP {\bf 0405}, 040 (2004)
[hep-ph/0312274].
%%CITATION = JHEPA,0405,040;%%

%+% 3 refs
\bibitem{PYTHIA}
H.~U.~Bengtsson and T.~Sj\"ostrand,
%``The Lund Monte Carlo for Hadronic Processes: Pythia Version 4.8,''
Comput.\ Phys.\ Commun.\  {\bf 46}, 43 (1987);\\
%%CITATION = CPHCB,46,43;%%
T.~Sj\"ostrand, P.~Eden, C.~Friberg, L.~L\"onnblad, G.~Miu,
S.~Mrenna and E.~Norrbin,
%``High-energy-physics event generation with PYTHIA 6.1,''
Comput.\ Phys.\ Commun.\ {\bf 135}, 238 (2001)
[hep-ph/0010017];\\
%%CITATION = CPHCB,135,238;%%
T.~Sj\"ostrand, L.~L\"onnblad, S.~Mrenna and P.~Skands,
%``PYTHIA 6.3: Physics and manual,''
hep-ph/0308153;\\
%%CITATION = HEP-PH/0308153;%%
T.~Sj\"ostrand, S.~Mrenna, P.~Z.~Skands,
%``PYTHIA 6.4 Physics and Manual,''
JHEP {\bf 0605}, 026 (2006).
[hep-ph/0603175].
%%CITATION = HEP-PH/0603175;%%

%+% 1 ref
\bibitem{HERWIG}
G.~Marchesini and B.~R.~Webber,
%``HERWIG: A NEW MONTE CARLO EVENT GENERATOR FOR SIMULATING HADRON EMISSION
%REACTIONS WITH INTERFERING GLUONS,''
Cavendish-HEP-87/9;\\
%%CITATION = CAVENDISH-HEP-87/9;%%
%
G.~Marchesini, B.~R.~Webber, G.~Abbiendi, I.~G.~Knowles, M.~H.~Seymour
and L.~Stanco,
%``HERWIG: A Monte Carlo event generator for simulating hadron emission
%reactions with interfering gluons. Version 5.1 - April 1991,''
Comput.\ Phys.\ Commun.\  {\bf 67}, 465 (1992);\\
%%CITATION = CPHCB,67,465;%%
%
G.~Corcella {\it et al.},
%``HERWIG 6.5 release note,''
hep-ph/0210213;\\
%%CITATION = HEP-PH/0210213;%%
%
M.~Bahr, S.~Gieseke, M.~A.~Gigg {\it et al.},
%``Herwig++ Physics and Manual,''
Eur.\ Phys.\ J.\  {\bf C58}, 639-707 (2008).
[0803.0883 [hep-ph]].

%+% 4 refs
\bibitem{Sherpa}
T.~Gleisberg,
S.~H\"{o}che, F.~Krauss, M.~Sch\"{o}nherr, S.~Schumann,
F.~Siegert and J.~Winter,
%``Event generation with SHERPA 1.1,''
JHEP {\bf 0902}, 007 (2009)
[0811.4622 [hep-ph]].
%%CITATION = JHEPA,0902,007;%%

%+% 2 refs
\bibitem{LesHouches2007}
Z.~Bern {\it et al.},
%``The NLO multileg working group: summary report,''
0803.0494 [hep-ph].
%%CITATION = ARXIV:0803.0494;%%

%+% 3 refs
\bibitem{LesHouches2009}
T.~Binoth {\it et al.},
%``The SM and NLO multileg working group: Summary report,''
1003.1241 [hep-ph].
%%CITATION = ARXIV:1003.1241;%%

%+% 5 refs
\bibitem{WCDF}
T.~Aaltonen {\it et al.} [CDF Collaboration],
%``Measurement of the cross section for W-boson production in association with
%jets in $p\bar{p}$ collisions at $\sqrt{s}=1.96$ TeV,''
Phys.\ Rev.\  D {\bf 77}, 011108 (2008)
[0711.4044 [hep-ex]].
%%CITATION = PHRVA,D77,011108;%%

%+% 7 refs
\bibitem{PRLW3BH}
C.~F.~Berger {\it et al.},
%``Precise Predictions for $W$ + 3 Jet Production at Hadron Colliders,''
Phys.\ Rev.\ Lett.\ {\bf 102}, 222001 (2009)
[0902.2760 [hep-ph]].
%%CITATION = PRLTA,102,222001;%%

%+% 24 refs
\bibitem{W3jDistributions}
C.~F.~Berger {\it et al.},
%``Next-to-Leading Order QCD Predictions for W+3-Jet Distributions at Hadron
%Colliders,''
Phys.\ Rev.\  D {\bf 80}, 074036 (2009)
[0907.1984 [hep-ph]].
%%CITATION = PHRVA,D80,074036;%%

%+% 2 refs
\bibitem{CMSWZRatioNote}
CMS collaboration, 
http://cdsweb.cern.ch/record/1194471/files/SUS-08-002-pas.pdf

%+% 2 refs
\bibitem{ATLASZWRatioNote}
G.~Aad {\it et al.}  [The ATLAS Collaboration],
%``Expected Performance of the ATLAS Experiment - Detector, Trigger and
%Physics,''
0901.0512 [hep-ex],
p. 1552.
%%CITATION = ARXIV:0901.0512;%%

%+% 2 refs
\bibitem{UnitarityMethod}
Z.~Bern, L.~J.~Dixon, D.~C.~Dunbar and D.~A.~Kosower,
%``One-loop $n$-point gauge theory amplitudes, unitarity
% and collinear limits,''
Nucl.\ Phys.\ B {\bf 425}, 217 (1994)
[hep-ph/9403226];
%%CITATION = HEP-PH 9403226;%%
%
% Z.~Bern, L.~J.~Dixon, D.~C.~Dunbar and D.~A.~Kosower,
%``Fusing gauge theory tree amplitudes into loop amplitudes,''
Nucl.\ Phys.\ B {\bf 435}, 59 (1995)
[hep-ph/9409265];\\
%%CITATION = HEP-PH 9409265;%%
%
Z.\ Bern, L.\ J.\ Dixon and D.\ A.\ Kosower,
%``Progress in one-loop QCD computations,''
Ann.\ Rev.\ Nucl.\ Part.\ Sci.\  {\bf 46}, 109 (1996)
[hep-ph/9602280].
%%CITATION = HEP-PH 9602280;%%

%+% 4 refs
\bibitem{Zqqgg}
Z.~Bern, L.~J.~Dixon and D.~A.~Kosower,
%``One-loop amplitudes for e+ e- to four partons,''
Nucl.\ Phys.\  B {\bf 513}, 3 (1998)
[hep-ph/9708239].
%%CITATION = NUPHA,B513,3;%%

%+% 2 refs
\bibitem{BCFUnitarity}
R.~Britto, F.~Cachazo and B.~Feng,
%``Generalized unitarity and one-loop amplitudes in N = 4  super-Yang-Mills,''
Nucl.\ Phys.\  B {\bf 725}, 275 (2005)
[hep-th/0412103].
%%CITATION = NUPHA,B725,275;%%

%+% 2 refs
\bibitem{Bootstrap}
R.~Britto, F.~Cachazo, B.~Feng and E.~Witten,
%``Direct proof of tree-level recursion relation in Yang-Mills theory,''
Phys.\ Rev.\ Lett.\ {\bf 94}, 181602 (2005)
[hep-th/0501052];\\
%%CITATION = HEP-TH 0501052;%%
%
Z.~Bern, L.~J.~Dixon and D.~A.~Kosower,
%``On-shell recurrence relations for one-loop QCD amplitudes,''
Phys.\ Rev.\  D {\bf 71}, 105013 (2005)
[hep-th/0501240];
%%CITATION = PHRVA,D71,105013;%%
%Z.~Bern, L.~J.~Dixon and D.~A.~Kosower,
%``The last of the finite loop amplitudes in QCD,''
Phys.\ Rev.\  D {\bf 72}, 125003 (2005)
[hep-ph/0505055];
%%CITATION = PHRVA,D72,125003;%%
%Z.~Bern, L.~J.~Dixon and D.~A.~Kosower,
%``Bootstrapping multi-parton loop amplitudes in QCD,''
Phys.\ Rev.\  D {\bf 73}, 065013 (2006)
[hep-ph/0507005];\\
%%CITATION = PHRVA,D73,065013;%%
%
D.~Forde and D.~A.~Kosower,
%``All-multiplicity amplitudes with massive scalars,''
Phys.\ Rev.\  D {\bf 73}, 065007 (2006)
[hep-th/0507292];
%%CITATION = PHRVA,D73,065007;%%
%
% D.~Forde and D.~A.~Kosower,
%``All-multiplicity one-loop corrections to MHV amplitudes in QCD,''
Phys.\ Rev.\  D {\bf 73}, 061701 (2006)
[hep-ph/0509358];\\
%%CITATION = PHRVA,D73,061701;%%
%
C.~F.~Berger, Z.~Bern, L.~J.~Dixon, D.~Forde and D.~A.~Kosower,
%``All One-loop Maximally Helicity Violating Gluonic Amplitudes in QCD,''
Phys.\ Rev.\  D {\bf 75}, 016006 (2007)
[hep-ph/0607014].
%%CITATION = PHRVA,D75,016006;%%

%+% 4 refs
\bibitem{Genhel}
C.~F.~Berger, Z.~Bern, L.~J.~Dixon, D.~Forde and D.~A.~Kosower,
%``Bootstrapping one-loop QCD amplitudes with general helicities,''
Phys.\ Rev.\ D {\bf 74}, 036009 (2006)
[hep-ph/0604195].
%%CITATION = HEP-PH 0604195;%%

%+% 2 refs
\bibitem{OPP}
G.~Ossola, C.~G.~Papadopoulos and R.~Pittau,
%``Reducing full one-loop amplitudes to scalar integrals at the integrand
%level,''
Nucl.\ Phys.\  B {\bf 763}, 147 (2007)
[hep-ph/0609007].
%%CITATION = NUPHA,B763,147;%%

%+% 2 refs
\bibitem{Forde}
D.~Forde,
%``Direct extraction of one-loop integral coefficients,''
Phys.\ Rev.\  D {\bf 75}, 125019 (2007)
[0704.1835 [hep-ph]].
%%CITATION = PHRVA,D75,125019;%%

%+% 1 ref
\bibitem{EGK}
R.~K.~Ellis, W.~T.~Giele and Z.~Kunszt,
%``A Numerical Unitarity Formalism for Evaluating One-Loop Amplitudes,''
JHEP {\bf 0803}, 003 (2008)
[0708.2398 [hep-ph]].
%%CITATION = JHEPA,0803,003;%%

%+% 2 refs
\bibitem{GKM}
W.~T.~Giele, Z.~Kunszt and K.~Melnikov,
%``Full one-loop amplitudes from tree amplitudes,''
JHEP {\bf 0804}, 049 (2008)
[0801.2237 [hep-ph]].
%%CITATION = JHEPA,0804,049;%%

%+% 1 ref
\bibitem{OnShellReviews}
Z.~Bern, L.~J.~Dixon and D.~A.~Kosower,
%``On-Shell Methods in Perturbative QCD,''
Annals Phys.\  {\bf 322}, 1587 (2007)
[0704.2798 [hep-ph]];\\
%%CITATION = APNYA,322,1587;%%
C.~F.~Berger and D.~Forde,
%``Multi-Parton Scattering Amplitudes via On-Shell Methods,''
0912.3534 [hep-ph].
%%CITATION = ARXIV:0912.3534;%%

%+% 2 refs
\bibitem{BlackHatI}
C.~F.~Berger,
Z.~Bern, L.~J.~Dixon, F.~Febres Cordero, D.~Forde, H.~Ita,
D.~A.~Kosower and D.~Ma\^{\i}tre,
%``An Automated Implementation of On-Shell Methods for One-Loop
%Amplitudes,''
Phys.\ Rev.\ D {\bf 78}, 036003 (2008)
[0803.4180 [hep-ph]].
%%CITATION = PHRVA,D78,036003;%%

%+% 1 ref
\bibitem{ICHEPBH}
C.~F.~Berger, Z.~Bern, L.~J.~Dixon, F.~Febres Cordero,
D.~Forde, H.~Ita, D.~A.~Kosower and D.~Ma\^{\i}tre,
%``One-Loop Multi-Parton Amplitudes with a Vector Boson for the LHC,''
0808.0941 [hep-ph].
%%CITATION = ARXIV:0808.0941;%%

%+% 1 ref
\bibitem{OtherOnShellPrograms}
G.~Ossola, C.~G.~Papadopoulos and R.~Pittau,
%``CutTools: a program implementing the OPP reduction method to compute
%one-loop amplitudes,''
JHEP {\bf 0803}, 042 (2008)
[0711.3596 [hep-ph]];\\
%%CITATION = JHEPA,0803,042;%%
%
W.~T.~Giele and G.~Zanderighi,
%``On the Numerical Evaluation of One-Loop Amplitudes: The Gluonic Case,''
JHEP {\bf 0806}, 038 (2008)
[0805.2152 [hep-ph]];\\
%%CITATION = JHEPA,0806,038;%%
%
A.~Lazopoulos,
%``Multi-gluon one-loop amplitudes numerically,''
0812.2998 [hep-ph];\\
%%CITATION = ARXIV:0812.2998;%%
%
J.-C.~Winter and W.~T.~Giele,
%``Calculating gluon one-loop amplitudes numerically,''
0902.0094 [hep-ph].
%%CITATION = ARXIV:0902.0094;%%

%+% 2 refs
\bibitem{HPP}
A.~van Hameren, C.~G.~Papadopoulos and R.~Pittau,
%``Automated one-loop calculations: a proof of concept,''
JHEP {\bf 0909}, 106 (2009).
%[0903.4665 [hep-ph]].
%%CITATION = JHEPA,0909,106;%%

%+% 1 ref
\bibitem{EGKMZ}
R.~K.~Ellis, W.~T.~Giele, Z.~Kunszt, K.~Melnikov and G.~Zanderighi,
%``One-loop amplitudes for $W^+$ 3 jet production in hadron collisions,''
JHEP {\bf 0901}, 012 (2009)
[0810.2762 [hep-ph]].
%%CITATION = JHEPA,0901,012;%%

%+% 3 refs
\bibitem{EMZW3j}
R.~K.~Ellis, K.~Melnikov and G.~Zanderighi,
%``Generalized unitarity at work: first NLO QCD results for hadronic $W^+$
%3jet production,''
JHEP {\bf 0904}, 077 (2009)
[0901.4101 [hep-ph]];
%%CITATION = JHEPA,0904,077;%%
% R.~K.~Ellis, K.~Melnikov and G.~Zanderighi,
%``W+3 jet production at the Tevatron,''
Phys.\ Rev.\  D {\bf 80}, 094002 (2009)
[0906.1445 [hep-ph]];\\
%%CITATION = PHRVA,D80,094002;%%
%
K.~Melnikov and G.~Zanderighi,
%``W+3 jet production at the LHC as a signal or background,''
Phys.\ Rev.\  D {\bf 81}, 074025 (2010)
[0910.3671 [hep-ph]].
%%CITATION = PHRVA,D81,074025;%%

%+% 2 refs
\bibitem{Czakon}
G.~Bevilacqua, M.~Czakon, C.~G.~Papadopoulos, R.~Pittau and M.~Worek,
%``Assault on the NLO Wishlist: pp -> tt bb,''
JHEP {\bf 0909}, 109 (2009)
[0907.4723 [hep-ph]].
%%CITATION = JHEPA,0909,109;%%

%+% 1 ref
\bibitem{BDDP}
A.~Bredenstein, A.~Denner, S.~Dittmaier and S.~Pozzorini,
%``NLO QCD corrections to top anti-top bottom anti-bottom production at the
% LHC: 1. quark-antiquark annihilation,''
JHEP {\bf 0808}, 108 (2008)
[0807.1248 [hep-ph]];
%%CITATION = JHEPA,0808,108;%%
%A.~Bredenstein, A.~Denner, S.~Dittmaier and S.~Pozzorini,
%``NLO QCD corrections to pp -> t anti-t b anti-b + X at the LHC,''
Phys.\ Rev.\ Lett.\  {\bf 103}, 012002 (2009)
[0905.0110 [hep-ph]];
%%CITATION = PRLTA,103,012002;%%
%A.~Bredenstein, A.~Denner, S.~Dittmaier and S.~Pozzorini,
%``NLO QCD corrections to top anti-top bottom anti-bottom production at the
%LHC: 2. full hadronic results,''
JHEP {\bf 1003}, 021 (2010)
[1001.4006 [hep-ph]].
%%CITATION = JHEPA,1003,021;%%

%+% 1 ref
\bibitem{bbbb}
T.~Binoth, N.~Greiner, A.~Guffanti, J.~P.~Guillet, T.~Reiter and J.~Reuter,
%``Next-to-leading order QCD corrections to pp --> b b_bar b b_bar + X at the
%LHC: the quark induced case,''
Phys.\ Lett.\  B {\bf 685}, 293 (2010)
[0910.4379 [hep-ph]].
%%CITATION = PHLTA,B685,293;%%

%+% 1 ref
\bibitem{ttjj}
G.~Bevilacqua, M.~Czakon, C.~G.~Papadopoulos and M.~Worek,
%``Dominant QCD Backgrounds in Higgs Boson Analyses at the LHC: A Study of pp
%-> t anti-t + 2 jets at Next-To-Leading Order,''
1002.4009 [hep-ph].
%%CITATION = ARXIV:1002.4009;%%

%+% 5 refs
\bibitem{MCFM}
J.~M.~Campbell and R.~K.~Ellis,
%``Next-to-leading order corrections to W + 2jet and Z + 2jet production  at
%hadron colliders,''
Phys.\ Rev.\  D {\bf 65}, 113007 (2002)
[hep-ph/0202176].
%%CITATION = PHRVA,D65,113007;%%

%+% 3 refs
\bibitem{CS}
S.~Catani and M.~H.~Seymour,
%``The Dipole Formalism for the Calculation of QCD Jet Cross Sections at
%Next-to-Leading Order,''
Phys.\ Lett.\  B {\bf 378}, 287 (1996)
[hep-ph/9602277];
%%CITATION = PHLTA,B378,287;%%
%
% S.~Catani and M.~H.~Seymour,
%``A general algorithm for calculating jet cross sections in NLO QCD,''
Nucl.\ Phys.\  B {\bf 485}, 291 (1997)
[Erratum-ibid.\  B {\bf 510}, 503 (1998)]
[hep-ph/9605323].
%%CITATION = NUPHA,B485,291;%%

%+% 2 refs
\bibitem{AutomatedAmegic}
T.~Gleisberg and F.~Krauss,
%``Automating dipole subtraction for QCD NLO calculations,''
Eur.\ Phys.\ J.\  C {\bf 53}, 501 (2008)
[0709.2881 [hep-ph]].
%%CITATION = EPHJA,C53,501;%%

%+% 1 ref
\bibitem{FKS}
S.~Frixione, Z.~Kunszt and A.~Signer,
%``Three jet cross-sections to next-to-leading order,''
Nucl.\ Phys.\  B {\bf 467}, 399 (1996)
[hep-ph/9512328].
%%CITATION = NUPHA,B467,399;%%

%+% 1 ref
\bibitem{AutomatedSubtractionOther}
M.~H.~Seymour and C.~Tevlin,
%``TeVJet: A general framework for the calculation of jet observables in NLO
%QCD,''
0803.2231 [hep-ph];\\
%%CITATION = ARXIV:0803.2231;%%
%
K.~Hasegawa, S.~Moch and P.~Uwer,
%``Automating dipole subtraction,''
Nucl.\ Phys.\ Proc.\ Suppl.\  {\bf 183}, 268 (2008)
[0807.3701 [hep-ph]];\\
%%CITATION = NUPHZ,183,268;%%
%
R.~Frederix, T.~Gehrmann and N.~Greiner,
%``Automation of the Dipole Subtraction Method in MadGraph/MadEvent,''
JHEP {\bf 0809}, 122 (2008)
[0808.2128 [hep-ph]];\\
%%CITATION = JHEPA,0809,122;%%
%
M.~Czakon, C.~G.~Papadopoulos and M.~Worek,
%``Polarizing the Dipoles,''
JHEP {\bf 0908}, 085 (2009)
[0905.0883 [hep-ph]];\\
%%CITATION = JHEPA,0908,085;%%
%
R.~Frederix, S.~Frixione, F.~Maltoni and T.~Stelzer,
%``Automation of next-to-leading order computations in QCD: the FKS
%subtraction,''
JHEP {\bf 0910}, 003 (2009)
[0908.4272 [hep-ph]].
%%CITATION = JHEPA,0910,003;%%

%+% 2 refs
\bibitem{AntennaIntegrator}
A.~van Hameren and C.~G.~Papadopoulos,
%``A hierarchical phase space generator for QCD antenna structures,''
Eur.\ Phys.\ J.\  C {\bf 25}, 563 (2002)
[hep-ph/0204055].
%%CITATION = EPHJA,C25,563;%%

%+% 2 refs
\bibitem{GleisbergIntegrator}
T.~Gleisberg, S.~H\"oche and F.~Krauss,
%``How to calculate colourful cross sections efficiently,''
0808.3672 [hep-ph].
%%CITATION = ARXIV:0808.3672;%%

%+% 2 refs
\bibitem{Radcor09BH}
C.~F.~Berger {\it et al.},
%``Next-to-Leading Order Jet Physics with BlackHat,''
PoS {\bf RADCOR2009}, 065 (2009)
[0912.4927 [hep-ph]].
%%CITATION = ARXIV:0912.4927;%%

%+% 1 ref
\bibitem{DynamicalScaleChoice}
S.~Frixione,
%``A Next-to-leading order calculation of the cross-section for the
% production of W+ W- pairs in hadronic collisions,''
Nucl.\ Phys.\  B {\bf 410}, 280 (1993);\\
%%CITATION = NUPHA,B410,280;%%
%
U.~Baur, T.~Han and J.~Ohnemus,
%``QCD corrections and nonstandard three vector boson couplings in $W^{+}
%W^{-}$ production at hadron colliders,''
Phys.\ Rev.\  D {\bf 53}, 1098 (1996)
[hep-ph/9507336];
%%CITATION = PHRVA,D53,1098;%%
%
%U.~Baur, T.~Han and J.~Ohnemus,
%``QCD corrections and anomalous couplings in $Z \gamma$ production at hadron
%colliders,''
Phys.\ Rev.\  D {\bf 57}, 2823 (1998)
[hep-ph/9710416];\\
%%CITATION = PHRVA,D57,2823;%%
%
L.~J.~Dixon, Z.~Kunszt and A.~Signer,
%``Vector boson pair production in hadronic collisions at order $\alpha_s$ :
%Lepton correlations and anomalous couplings,''
Phys.\ Rev.\  D {\bf 60}, 114037 (1999)
[hep-ph/9907305];\\
%%CITATION = PHRVA,D60,114037;%%
%
G.~Bozzi, B.~J\"ager, C.~Oleari and D.~Zeppenfeld,
%``Next-to-leading order QCD corrections to W+Z and W-Z production via
%vector-boson fusion,''
Phys.\ Rev.\  D {\bf 75}, 073004 (2007)
[hep-ph/0701105].
%%CITATION = PHRVA,D75,073004;%%

%+% 2 refs
\bibitem{EarlyWplus2MP}
M.~L.~Mangano and S.~J.~Parke,
%``$W$ Boson Plus Two Jet Production at the Tevatron,''
Phys.\ Rev.\  D {\bf 41}, 59 (1990).
%%CITATION = PHRVA,D41,59;%%

%+% 4 refs
\bibitem{Bauer}
C.~W.~Bauer and B.~O.~Lange,
%``Scale setting and resummation of logarithms in pp -> V + jets,''
0905.4739 [hep-ph].
%%CITATION = ARXIV:0905.4739;%%

\bibitem{deFV}
D.~de Florian and W.~Vogelsang,
%``Resummed cross-section for jet production at hadron colliders,''
Phys.\ Rev.\  D {\bf 76}, 074031 (2007)
[0704.1677 [hep-ph]].
%%CITATION = PHRVA,D76,074031;%%

\bibitem{RSS}
M.~Rubin, G.~P.~Salam and S.~Sapeta,
%``Giant QCD K-factors beyond NLO,''
1006.2144 [hep-ph].
%%CITATION = ARXIV:1006.2144;%%

%+% 6 refs
\bibitem{Jetography}
G.~P.~Salam,
%``Towards Jetography,''
Eur.\ Phys.\ J.\  C {\bf 67}, 637 (2010)
[0906.1833 [hep-ph]].
%%CITATION = EPHJA,C67,637;%%

%+% 5 refs
\bibitem{SISCONE}
G.~P.~Salam and G.~Soyez,
%``A practical Seedless Infrared-Safe Cone jet algorithm,''
JHEP {\bf 0705}, 086 (2007)
[0704.0292 [hep-ph]].
%%CITATION = JHEPA,0705,086;%%

%+% 2 refs
\bibitem{antikT}
M.~Cacciari, G.~P.~Salam and G.~Soyez,
%``The anti-k_t jet clustering algorithm,''
JHEP {\bf 0804}, 063 (2008)
[0802.1189 [hep-ph]].
%%CITATION = JHEPA,0804,063;%%

%+% 2 refs
\bibitem{KTAlgorithm}
S.~Catani, Y.~L.~Dokshitzer, M.~H.~Seymour and B.~R.~Webber,
%``Longitudinally invariant $k_T$ clustering algorithms for hadron hadron
% collisions,''
Nucl.\ Phys.\  B {\bf 406}, 187 (1993).
%%CITATION = NUPHA,B406,187;%%

%+% 3 refs
\bibitem{KTES}
S.~D.~Ellis and D.~E.~Soper,
%``Successive combination jet algorithm for hadron collisions,''
Phys.\ Rev.\  D {\bf 48}, 3160 (1993)
[hep-ph/9305266].
%%CITATION = PHRVA,D48,3160;%%

%+% 1 ref
\bibitem{Midpoint}
R.~Akers {\it et al.}  [OPAL Collaboration],
%``QCD studies using a cone based jet finding algorithm for e+ e- collisions
%at LEP,''
Z.\ Phys.\  C {\bf 63}, 197 (1994);\\
%%CITATION = ZEPYA,C63,197;%%
M.~H.~Seymour,
%``Jet shapes in hadron collisions: Higher orders, resummation and
%hadronization,''
Nucl.\ Phys.\  B {\bf 513}, 269 (1998)
[hep-ph/9707338].
%%CITATION = NUPHA,B513,269;%%

%+% 2 refs
\bibitem{D0jetAlgorithm}
G.C. Blazey, {\it et al.}, in {\it Proceedings of the Workshop: QCD and
Weak Boson Physics in Run II}, eds. U.~Baur, R.~K.~Ellis and
D.~Zeppenfeld, Fermilab-Pub-00/297 (2000).

%+% 2 refs
\bibitem{CDFMidpoint}
A.~Abulencia {\it et al.}  [CDF Collaboration],
%``Measurement of the inclusive jet cross section in $p\bar{p}$ interactions
%at $\sqrt{s} =$ 1.96-TeV using a cone-based jet algorithm,''
Phys.\ Rev.\  D {\bf 74}, 071103(R) (2006)
[hep-ex/0512020].
%%CITATION = PHRVA,D74,071103;%%

%+% 3 refs
\bibitem{EHKetal}
S.~D.~Ellis, J.~Huston, K.~Hatakeyama {\it et al.},
%``Jets in hadron-hadron collisions,''
Prog.\ Part.\ Nucl.\ Phys.\  {\bf 60}, 484-551 (2008).
[0712.2447 [hep-ph]].

%+% 3 refs
\bibitem{BerendsRatio}
% S.~M.~Berman, J.~D.~Bjorken and J.~B.~Kogut,
%  %``Inclusive Processes At High Transverse Momentum,''
%  Phys.\ Rev.\  D {\bf 4}, 3388 (1971);\\
%  %%CITATION = PHRVA,D4,3388;%%
S.~D.~Ellis, R.~Kleiss and W.~J.~Stirling,
%``W's, Z's And Jets,''
Phys.\ Lett.\  B {\bf 154}, 435 (1985);\\
%%CITATION = PHLTA,B154,435;%%
%
F.~A.~Berends, W.~T.~Giele, H.~Kuijf, R.~Kleiss and W.~J.~Stirling,
%``Multi - Jet Production In W, Z Events At P Anti-P Colliders,''
Phys.\ Lett.\  B {\bf 224}, 237 (1989);\\
%%CITATION = PHLTA,B224,237;%%
%
F.~A.~Berends, H.~Kuijf, B.~Tausk and W.~T.~Giele,
%``On the production of a W and jets at hadron colliders,''
Nucl.\ Phys.\  B {\bf 357}, 32 (1991).
%%CITATION = NUPHA,B357,32;%%

%+% 3 refs
\bibitem{AbouzaidFrisch}
 E.~Abouzaid and H.~J.~Frisch,
 %``The Ratio of $W$ + $N$ jets to $Z^0 / \gamma^{*}$ + $N$ jets versus $N$ as
 % a test precision test of the standard model,''
 Phys.\ Rev.\  D {\bf 68}, 033014 (2003)
[hep-ph/0303088].
 %%CITATION = PHRVA,D68,033014;%%

%+% 1 ref
\bibitem{IntegralsExplicit}
G.~'t Hooft and M.~J.~G.~Veltman,
%``Scalar One Loop Integrals,''
Nucl.\ Phys.\  B {\bf 153}, 365 (1979);\\
%%CITATION = NUPHA,B153,365;%%
%
G.~J.~van Oldenborgh and J.~A.~M.~Vermaseren,
%``New Algorithms for One Loop Integrals,''
Z.\ Phys.\  C {\bf 46}, 425 (1990);\\
%%CITATION = ZEPYA,C46,425;%%
%
W.~Beenakker and A.~Denner,
%``INFRARED DIVERGENT SCALAR BOX INTEGRALS WITH APPLICATIONS IN THE
%ELECTROWEAK STANDARD MODEL,''
Nucl.\ Phys.\  B {\bf 338}, 349 (1990);\\
%%CITATION = NUPHA,B338,349;%%
%
A.~Denner, U.~Nierste and R.~Scharf,
%``A Compact expression for the scalar one loop four point function,''
Nucl.\ Phys.\  B {\bf 367}, 637 (1991);\\
%%CITATION = NUPHA,B367,637;%%
%
Z.~Bern, L.~J.~Dixon and D.~A.~Kosower,
%``Dimensionally regulated pentagon integrals,''
Nucl.\ Phys.\  B {\bf 412} (1994) 751
[hep-ph/9306240];\\
%%CITATION = NUPHA,B412,751;%%
%
T.~Hahn and M.~P\'erez-Victoria,
%``Automatized one-loop calculations in four and D dimensions,''
Comput.\ Phys.\ Commun.\  {\bf 118}, 153 (1999)
[hep-ph/9807565];\\
%%CITATION = CPHCB,118,153;%%
%
R.~K.~Ellis and G.~Zanderighi,
%``Scalar one-loop integrals for QCD,''
JHEP {\bf 0802}, 002 (2008)
[0712.1851 [hep-ph]].
%%CITATION = JHEPA,0802,002;%%

%+% 1 ref
\bibitem{Badger}
S.~D.~Badger,
%``Direct Extraction Of One Loop Rational Terms,''
JHEP {\bf 0901}, 049 (2009)
[0806.4600 [hep-ph]].
%%CITATION = JHEPA,0901,049;%%

%+% 1 ref
\bibitem{DdimUnitarity}
Z.~Bern and A.~G.~Morgan,
%``Massive Loop Amplitudes from Unitarity,''
Nucl.\ Phys.\  B {\bf 467}, 479 (1996)
[hep-ph/9511336];\\
%%CITATION = NUPHA,B467,479;%%
Z.~Bern, L.~J.~Dixon, D.~C.~Dunbar and D.~A.~Kosower,
%``One-loop self-dual and N = 4 superYang-Mills,''
Phys.\ Lett.\  B {\bf 394}, 105 (1997)
[hep-th/9611127];\\
%
%%CITATION = PHLTA,B394,105;%%
C.~Anastasiou, R.~Britto, B.~Feng, Z.~Kunszt and P.~Mastrolia,
%``D-dimensional unitarity cut method,''
Phys.\ Lett.\  B {\bf 645}, 213 (2007)
[hep-ph/0609191];\\
%%CITATION = PHLTA,B645,213;%%
%
R.~Britto and B.~Feng,
%``Integral Coefficients for One-Loop Amplitudes,''
JHEP {\bf 0802}, 095 (2008)
[0711.4284 [hep-ph]].
%%CITATION = JHEPA,0802,095;%%

%+% 1 ref
\bibitem{qqggg}
Z.~Bern, L.~J.~Dixon and D.~A.~Kosower,
%``One Loop Corrections To Two Quark Three Gluon Amplitudes,''
Nucl.\ Phys.\  B {\bf 437}, 259 (1995)
[hep-ph/9409393].
%%CITATION = NUPHA,B437,259;%%

%+% 1 ref
\bibitem{ZoltanColor}
W.~Giele, Z.~Kunszt and J.~Winter,
%``Efficient Color-Dressed Calculation of Virtual Corrections,''
0911.1962 [hep-ph].
%%CITATION = ARXIV:0911.1962;%%

%+% 1 ref
\bibitem{QD}
Y.~Hida, X.~ S.~Li and D.~H.~Bailey,
% Quad-Double Arithmetic: Algorithms, Implementation, and Application,
http://crd.lbl.gov/\~{}dhbailey/mpdist, report LBNL-46996.

%+% 1 ref
\bibitem{MultiChannel}
R.~Kleiss and R.~Pittau,
%``Weight optimization in multichannel Monte Carlo,''
Comput.\ Phys.\ Commun.\  {\bf 83}, 141 (1994)
[hep-ph/9405257].
%%CITATION = CPHCB,83,141;%%

%+% 1 ref
\bibitem{CTEQ6M}
J.~Pumplin {\it et al.},
% D.~R.~Stump, J.~Huston, H.~L.~Lai, P.~M.~Nadolsky and W.~K.~Tung,
%``New generation of parton distributions with uncertainties from global QCD
%analysis,''
JHEP {\bf 0207}, 012 (2002)
[hep-ph/0201195].
%%CITATION = JHEPA,0207,012;%%

%+% 4 refs
\bibitem{DMS}
M.~Dasgupta, L.~Magnea and G.~P.~Salam,
%``Non-perturbative QCD effects in jets at hadron colliders,''
JHEP {\bf 0802}, 055 (2008)
[0712.3014 [hep-ph]].
%%CITATION = JHEPA,0802,055;%%

%+% 2 refs
\bibitem{HustonLONLOJets}
J.~Huston,
%``LO, NLO, LO* and jet algorithms,''
PoS {\bf RADCOR2009}, 079 (2010)
[1001.2581 [hep-ph]].
%%CITATION = POSCI,RADCOR2009,079;%%

%+% 1 ref
\bibitem{JETCLU}
F.~Abe {\it et al.}  [CDF Collaboration],
%``The Topology of three jet events in $\bar{p}p$ collisions at $\sqrt{s} =
%1.8$ TeV,''
Phys.\ Rev.\  D {\bf 45}, 1448 (1992).
%%CITATION = PHRVA,D45,1448;%%

%+% 1 ref
\bibitem{ESW}
R.~K.~Ellis, W.~J.~Stirling and B.~R. Webber,
{\it QCD and Collider Physics} (Cambridge University Press, 1996).

%+% 2 refs
\bibitem{ADMP}
C.~Anastasiou, L.~J.~Dixon, K.~Melnikov and F.~Petriello,
%``High precision QCD at hadron colliders: Electroweak gauge boson rapidity
%distributions at NNLO,''
Phys.\ Rev.\  D {\bf 69}, 094008 (2004)
[hep-ph/0312266].
%%CITATION = PHRVA,D69,094008;%%

%+% 1 ref
\bibitem{OtherBerendsRatio}
V.~M.~Abazov {\it et al.}  [D0 Collaboration],
%``$t \bar{t}$ production cross-section in $p \bar{p}$ collisions at
%$\sqrt{s}$ = 1.8-TeV,''
Phys.\ Rev.\  D {\bf 67}, 012004 (2003)
[hep-ex/0205019];\\
%%CITATION = PHRVA,D67,012004;%%
%
W.~Wagner [for the CDF Collaboration],
%``Top quark cross-section measurements at the Tevatron,''
Eur.\ Phys.\ J.\  C {\bf 33}, S238 (2004)
[hep-ex/0312008];\\
%%CITATION = EPHJA,C33,S238;%%
%
D.~Cho, UMI-31-69569-MC, Fermilab-thesis-2005-31.
%``Measurement of the top - anti-top Production Cross Section at $s^{(/1/2)}$
%= 1.96-TeV in the $e +$ jets Final State of $p\bar{p}$ Collisions at the
%Tevatron,''
%%CITATION = UMI-31-69569-MC;%%

%+% 1 ref
\bibitem{MCNLO}
S.~Frixione and B.~R.~Webber,
%``Matching NLO QCD computations and parton shower simulations,''
JHEP {\bf 0206}, 029 (2002)
[hep-ph/0204244]; \\
%%CITATION = HEP-PH 0204244;%%
%
S.~Frixione, P.~Nason and B.~R.~Webber,
%``Matching NLO QCD and parton showers in heavy flavour production,''
JHEP {\bf 0308}, 007 (2003)
[hep-ph/0305252].
%%CITATION = HEP-PH 0305252;%%

%+% 1 ref
\bibitem{POWHEG}
P.~Nason,
%``A new method for combining NLO QCD with shower Monte Carlo algorithms,''
JHEP {\bf 0411}, 040 (2004)
[hep-ph/0409146];\\
%%CITATION = JHEPA,0411,040;%%
%
S.~Frixione, P.~Nason and C.~Oleari,
%``Matching NLO QCD computations with Parton Shower simulations: the POWHEG
%method,''
JHEP {\bf 0711}, 070 (2007)
[0709.2092 [hep-ph]];\\
%%CITATION = JHEPA,0711,070;%%
%
S.~Alioli, P.~Nason, C.~Oleari and E.~Re,
%``NLO vector-boson production matched with shower in POWHEG,''
JHEP {\bf 0807}, 060 (2008)
[0805.4802 [hep-ph]];
%%CITATION = JHEPA,0807,060;%%
%S.~Alioli, P.~Nason, C.~Oleari and E.~Re,
%``A general framework for implementing NLO calculations in shower Monte Carlo
%programs: the POWHEG BOX,''
JHEP {\bf 1006}, 043 (2010)
[1002.2581 [hep-ph]].
%%CITATION = JHEPA,1006,043;%%

%+% 2 refs
\bibitem{RikkertRadcor}
R.~Frederix, in {\it Proceedings of the 9th international Symposium on 
Radiative Corrections (RADCOR 2009)},
PoS {\bf RADCOR2009}, 066 (2009).

%+% 1 ref
\bibitem{BinothInterface}
T.~Binoth {\it et al.},
%``A proposal for a standard interface between Monte Carlo tools and one-loop
%programs,''
1001.1307 [hep-ph].
%%CITATION = ARXIV:1001.1307;%%

\bibitem{SHERPAWebPage}
\SHERPA{} home page, http://www.sherpa-mc.de/

%+% 1 ref
\bibitem{HV}
G.~'t Hooft and M.~Veltman,
%``Regularization And Renormalization Of Gauge Fields,''
Nucl.\ Phys.\ B {\bf 44}, 189 (1972).
%%CITATION = NUPHA,B44,189;%%

\end{thebibliography}
\end{document}